\documentclass[twocolumn,astrosymb]{aastex631}

\submitjournal{ApJ}

\shorttitle{Distinguishing X-ray Stars vs. AGN through ML}
\shortauthors{Hebbar \& Heinke}

\graphicspath{{./}{Figures/}}

\usepackage{graphicx}	
\usepackage{multirow}
\usepackage{ragged2e}
\usepackage{listings}

\usepackage[caption=false]{subfig}

\begin{document}


\title{Machine learning applied to X-ray spectra: separating stars in Orion nebula cluster from active galactic nuclei in CDFS}

\author[0000-0002-4961-4690]{Pavan R. Hebbar}
\affiliation{Department of Physics, CCIS 4-181, University of Alberta, Edmonton, AB T6G 2E1, Canada}

\author[0000-0003-3944-6109]{Craig O. Heinke}
\affiliation{Department of Physics, CCIS 4-181, University of Alberta, Edmonton, AB T6G 2E1, Canada}

\correspondingauthor{Pavan R. Hebbar}
\email{hebbar@ualberta.ca}

\begin{abstract}

Modern X-ray telescopes have detected hundreds of thousands of X-ray sources in the universe. However, current methods to classify these sources using the X-ray data themselves suffer problems --- detailed X-ray spectroscopy of individual sources is too time-consuming, while hardness ratios often lack accuracy, and can be difficult to use effectively. These methods fail to use the power of X-ray CCD detectors to identify X-ray emission lines and distinguish line-dominated spectra (from chromospherically active stars, supernova remnants etc.) from continuum-dominated ones (e.g. compact objects or active galactic nuclei [AGN]). In this paper, we probe the use of artificial neural networks (ANN) in differentiating  {\it Chandra} spectra of young stars in the {\it Chandra} Orion Ultradeep Project (COUP) survey from AGN in the {\it Chandra} Deep Field South (CDFS) survey. We use these surveys to generate 100,000 artificial spectra of stars and AGN, and train our ANN models to separate the two kinds of spectra. We find that our methods reach an accuracy of $\sim$ 92\% in classifying simulated spectra of moderate-brightness objects in typical exposures, but their performance decreases on the observed COUP and CDFS spectra ($\sim 91 \%$), due in large part to the relatively high background of these long-exposure datasets. We also investigate the performance of our methods with changing properties of the spectra such as the net source counts, the relative contribution of background, the absorption column of the sources, etc. We conclude that these methods have substantial promise for application to large X-ray surveys.

\end{abstract}


\section{Introduction} \label{sec:intro}

The X-ray sky consists of a variety of extremely hot objects ($kT \sim 1$ keV). These sources include chromospherically active stars \citep[e.g.,][]{Gudel_2004, Preibisch_2005}, supernova remnants \citep[SNRs; e.g.,][]{Vink_2011}, isolated neutron stars \citep[NSs; e.g.,][]{Kaspi_2006, Pavlov_2002}, X-ray binaries
\citep[XRBs; e.g.,][]{Remillard_2006, Campana_1998}, and active galactic nuclei \citep[AGN; e.g.,][]{Netzer_2015, Padovani_2017}. These different kinds of X-ray sources emit radiation through distinct physical processes --- active stars and SNRs emit X-rays predominantly from thermal bremsstrahlung and line radiation, young isolated NSs show thermal blackbody-like emission, NSs with strong magnetic fields can also produce synchrotron radiation, and X-ray binaries and AGN cool via inverse Compton scattering \citep[refer][for detailed review of the emission processes]{Bradt_2014}.

X-ray sources, especially compact objects, host exotic environments with extremely hot temperatures, high densities and immensely strong gravitational and magnetic fields \citep[see][for detailed reviews]{Remillard_2006, Lattimer_2007, Turner_2009}. The surfaces of young NSs can have temperatures $T \sim 10^6$ K, the centres of NSs can reach densities  $\rho \sim 10^{14}-10^{15}$ g/cm$^{-3}$, magnetars can host magnetic fields of $10^{14} - 10^{15}$ G \citep{Olausen_2014}, and black holes (BHs) exert gravitational fields that test the limits of modern physics. Thus, studying these compact objects can help us understand important physics such as nuclear forces, interaction of matter with strong magnetic fields, etc. X-ray emission is universal from these sources due to their temperatures and the high-energy processes involved in emitting radiation, thus making X-ray surveys ideal to search for and study compact objects. However, our knowledge of many of these sources has been constrained by the small numbers of identified objects. 

Over the last decades, we have launched several powerful X-ray telescopes to observe these high energy X-ray sources, including the {\it Chandra X-ray observatory} (CXO or {\it Chandra}, for short), the {\it XMM-Newton} telescope, {\it eROSITA} instrument etc. These instruments record the position, time of arrival and energy of each X-ray photon detected. These telescopes have detected hundreds of thousands of X-ray sources in the sky --- e.g. the {\it Chandra} Source Catalog (CSC) has $\sim$300,000 unique sources \citep{Evans_2010}, the XMM-Newton Serendipitous Source Catalog has detected $\sim$600,000 X-ray sources \citep{Webb_2020}, and the eROSITA Final Equatorial Depth Survey (eFEDS) survey has detected $\sim$30,000 sources \citep{Brunner_2021}, with several million X-ray sources expected in the final eROSITA survey \citep{Predehl_2021}. However, most of these sources haven't been studied in detail and we do not know the type of the X-ray source. Detailed individualized X-ray spectroscopy to model the X-ray emission of individual X-ray sources, understand their properties, and detect compact objects can be time-consuming for these large catalogs. 
Automated spectroscopic catalogs \citep[e.g.][]{Corral15} can be useful, though these also have limits (the selection of models, time and storage required for the fitting).
Hardness ratios, that compare the number of X-ray photons in the soft X-ray band (say, 0.5--2 keV) and the hard X-ray band (say, 2--6 keV), can also be used to estimate the properties of the X-ray source \citep{Yokogawa_2000,Prestwich_2003,Brassington_2008}, and quantile analysis provides an alternative with substantial benefits \citep{Hong_2004}. However, hardness ratios do not have strong discriminatory power for faint (and sometimes even moderately bright) sources, which can lead to inaccurate classification of X-ray sources \citep[e.g.][]{Hebbar_2019}. Thus, we need efficient ways to identify the
X-ray sources detected in large X-ray surveys.

\begin{figure}
    \centering
    \includegraphics[width=0.99\columnwidth]{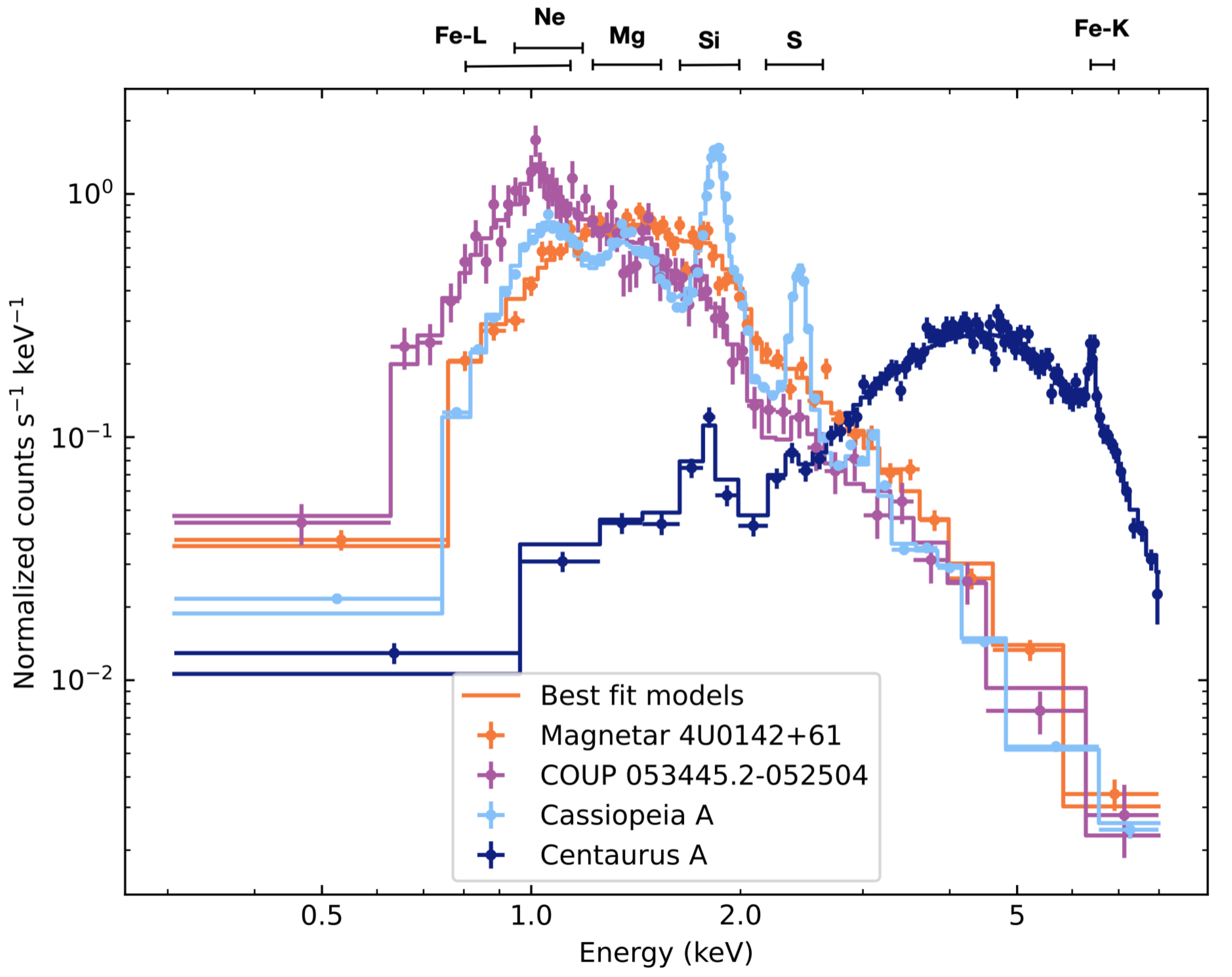}
    \figcaption{Example spectra of an isolated neutron star (4U 0142+61), active star (COUP 053445.2-052504), SNR (Cassiopeia A), and AGN (Centaurus A). We also show the approximate positions of  Fe-L, Ne-K, Mg-K, Si-K, S-K, and Fe-K emission lines. The spectra of active stars and SNRs are dominated by these lines while that of AGN and NSs are largely continuous. AGN can sometimes show an Fe-K line. \label{fig:typical_spectra}}
\end{figure}

CCD detectors that are used in most X-ray telescopes have an energy resolution of $\Delta E \sim$ 0.1--0.2 keV \citep{Fano_1947,Struder_2001,Predehl_2021}. Such moderate-resolution X-ray spectra are capable of detecting and resolving emission and absorption lines from elements such as neon (Ne), magnesium (Mg), silicon (Si), and iron (Fe). The coronae of active stars emit significant X-ray energy through the Ne-K \& Fe-L lines. (Both Ne \& Fe emit at $\sim 1$ keV. 
The dominant line among them depends on conditions in the stellar coronae. The comparison between the relative strengths of these lines is not important for the moderate resolution spectra used here.) The spectra of SNRs also have bright emission lines from Mg, Si, S, etc. Most AGN have a continuum dominated inverse Compton power-law spectrum from the hot corona. Some AGN  also have additional soft energy components and fluorescent emission lines from the surrounding colder gas. But these line features are usually much fainter than the continuum component, except for the Fe-K line at $\sim 6.4$ keV (which can be redshifted) \citep[][]{Matt_1997, Garcia_2013}. NSs typically have  continuum X-ray spectra with no emission lines. Typical spectra of these sources are shown in Fig.~\ref{fig:typical_spectra}. Thus, an ability to identify these emission lines and distinguish between continuum- and line-dominated X-ray spectra will allow us to understand the nature of the X-ray source.

The large datasets used in astronomy, minimal privacy and ethical concerns in sharing data, etc. make astronomy research a great potential application for machine learning (ML). Machine learning algorithms have been used for calculating photometric redshifts \citep[e.g.,][]{Carrasco_2013}, classification of galaxies and identification of AGN using optical observations \citep[e.g.,][etc.]{Cavouti_2014, Rozo_2016, Chattopadhyay_2019}, classification of supernovae \citep[][]{Moller_2020} and modelling their X-ray spectra\citep[][]{Parker_2022, Matzeu_2022}, identification of exoplanets \citep[][]{Shallue_2018}, etc. Recently \citet{Yang_2021}, \citet{Schneider}, and \citet{Tranin22} have used ML algorithms to identify the multiwavelength counterparts of X-ray sources in CSC v2.0, eFEDS, and XMM-Newton data, respectively, using  properties such as angular separation, X-ray flux, X-ray hardness ratios, Gaia magnitude and colors, etc.,  and classify the X-ray sources. While such methods work efficiently for sources far from the Galactic plane, it can be 
very time-consuming 
to find multiwavelength counterparts of X-ray sources in highly absorbed or crowded regions such as the Galactic bulge and globular clusters. We also anticipate that combining information from the X-ray spectra themselves along with multiwavelength information will provide more accurate source identification.

\subsection{Artificial neural networks}
Artificial neural networks (ANNs) are supervised learning algorithms that use hidden layers to learn non-linear representations of the data. In a classical feed-forward ANN, a weight matrix is applied on the input data that transforms the dimensionality of input. Then, we apply non-linear activation functions on this transformed input to calculate the values of nodes in the first hidden layer. This process can be repeated on the first hidden layer to calculate the next layer, and so on, and finally, the output layer. Typical activation functions include Rectified Linear Unit (ReLU) and its variants, sigmoid function, softmax function, etc. The choice of activation functions depends on the problem and nature of output needed (check \citealt{Bishop_2007} for detailed reference). In particular, using the sigmoid and the softmax functions that give values between 0 and 1 on the output layer gives a probabilistic interpretation for binary and multi-class classification.

As non-linear classifiers, the ANNs can learn complex relationships in the data and thus lead to better accuracy of classification results. The probabilistic nature of the output  allows us to change the threshold based on our requirement. ANNs can also be modified for semi-supervised training through auto-encoders, and use physical expressions for better interpretation of the results and the training process. Thus ANNs can be adapted to a wide variety of problems. However, ANNs need a large dataset for training and are prone to over-fitting. Thus, it is important that we use proper regularization to avoid spurious weights and test the trained model with a separate dataset that is not used for training (a test set).

In this paper, we aim to develop ANN algorithms that distinguish  different kinds of X-ray sources based on their moderate-resolution X-ray spectra from CCDs. We test if ANNs can differentiate young stars from AGN, in simulated and real Chandra X-ray Observatory data. We describe our methods of data generation and the setup of our analysis in \S\ref{sec:data_analysis}. We show the results of our classification and investigate the robustness of our ANN model in \S\ref{sec:results}. We discuss the implications of our results and its broader application in \S\ref{sec:discussions}. We summarize the results of our analysis and discuss future prospects in \S\ref{sec:conclusion}.

\section{Data Acquisition and Analysis}
\label{sec:data_analysis}

We use ANNs as a supervised training algorithm to test their feasibility. Training and testing a complex ANN requires a large dataset of labelled sources. Thus, we rely on XSPEC \texttt{fakeit} simulations of the spectra of stars and AGN for the training. The {\it Chandra} Orion Ultradeep Project \citep[COUP,][]{Getman_2005} and the {\it Chandra} Deep Field South \citep[CDFS, ][]{Giacconi_2002} surveys are ideal for extracting the properties of stars and AGN for the simulated spectra. 
Most of the sources in the COUP survey of a star forming region are young stars, while the majority of the sources detected in the CDFS are AGN. 
All the COUP observations were conducted in January 2003, thus removing the effects of the changing soft energy response of {\it Chandra} ACIS. The CDFS observations were taken during the years 2000, 2007 and 2010. We only consider the observations in year 2000 for our analysis as they are the ones used in the CDFS AGN spectral properties catalog \citep[][]{Tozi_2006}. This also ensures a better soft energy response for our AGN data. 
With a few hundred sources detected in each of the COUP and CDFS surveys, they  provide us samples of observed spectra to test our results.

We use the X-ray sources in the {\it Chandra} Orion Ultradeep Project \citep[COUP,][]{Getman_2005} survey for our ensemble of young stars. The COUP survey used the ACIS-I instrument of {\it Chandra} for an exposure time of 838 ks and detected 1616 X-ray point sources, that are mostly active young stars, and modeled them using one or two-component thermal plasma spectra. We 
utilize their catalog of objects identified as stars, and 
do not consider sources that are marked uncertain. 
The COUP catalog specifies the best-fitting parameter values of these models for each COUP spectrum. We use the distribution of these properties to extract the general properties of X-rays from young stars. Since we are interested in getting the distribution of the properties of the COUP spectra, we also omit sources with marginal (null hypothesis property from $\chi^2$ is between 0.05 and 0.005) and poor (null hypothesis probability from $\chi^2$ is less than 0.005) fits. These poor fits could be due to contamination from surrounding sources, high absorption leading to poor fitting of lines, spectra more complicated than that of thermal-equilibrium plasma, etc. Using the membership information of the COUP catalog provided by \citet{Getman_2005b}, we only select sources within the Orion Nebula Cluster. This gives us a set of 1045 sources. We also remove sources with flags for: deviations in the emission lines (presence of narrow spectral features not explained by the model, probably due to different elemental abundance; 62 such sources), soft or hard excess (mostly due to poor subtraction of a non-uniform background around weak sources; 214 such sources), confusion from nearby sources (54 such sources), two components with different absorption column densities (6 such sources) or a poor fit (poor $\chi^2$ statistics and/or based on visual examination; 89 such sources). Note that these are only removed
for the creation of 
the distribution of spectral properties, not for testing our trained ML model.

\begin{figure*}
    \centering
    \subfloat[Distribution of absorption column densities, $N_H$. \label{fig:star_nh}]{%
    \includegraphics[width=0.9\columnwidth]{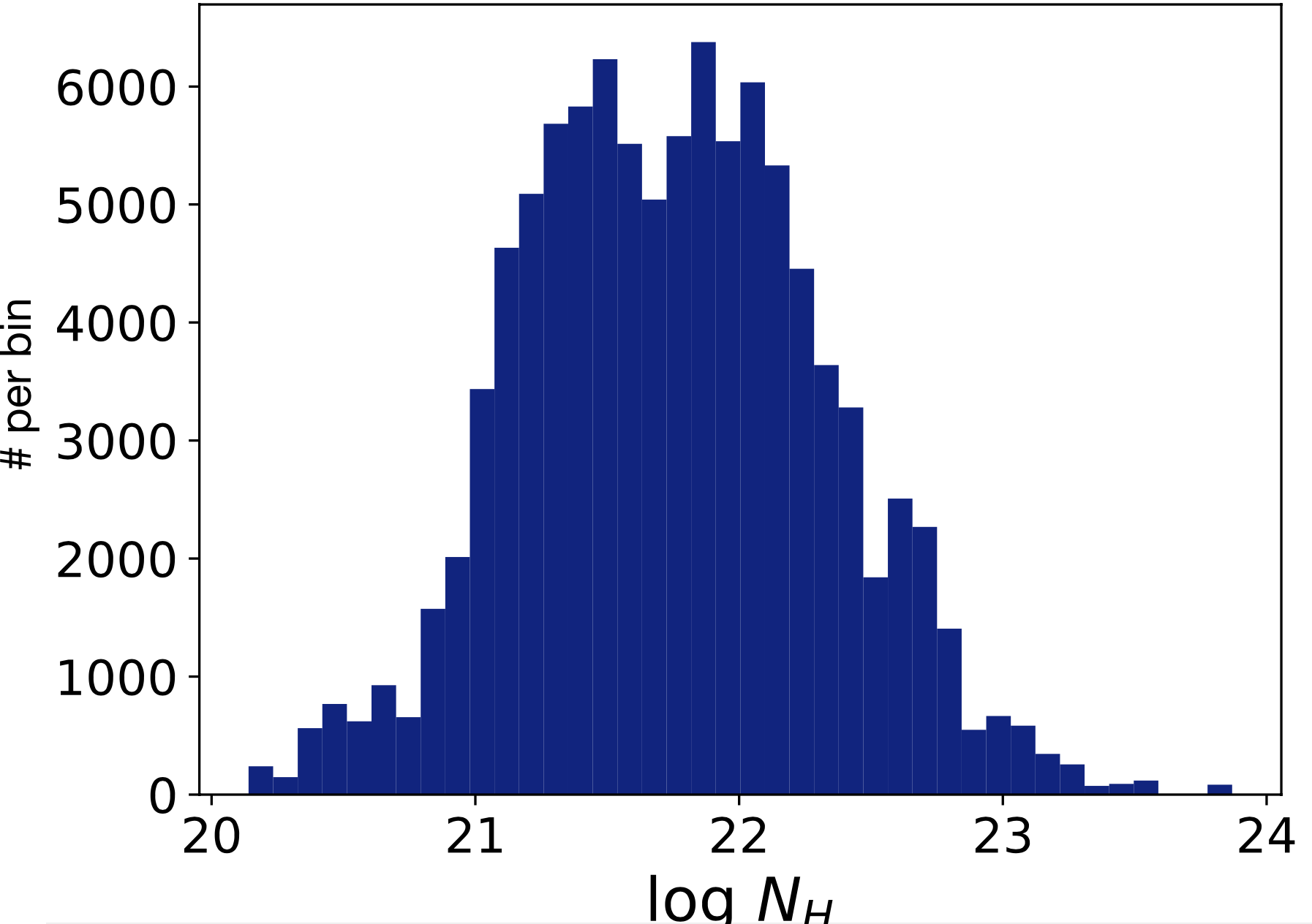}} \hfill
    \subfloat[Distribution of temperatures, $kT1$. \label{fig:star_kT1}]{%
    \includegraphics[width=0.9\columnwidth]{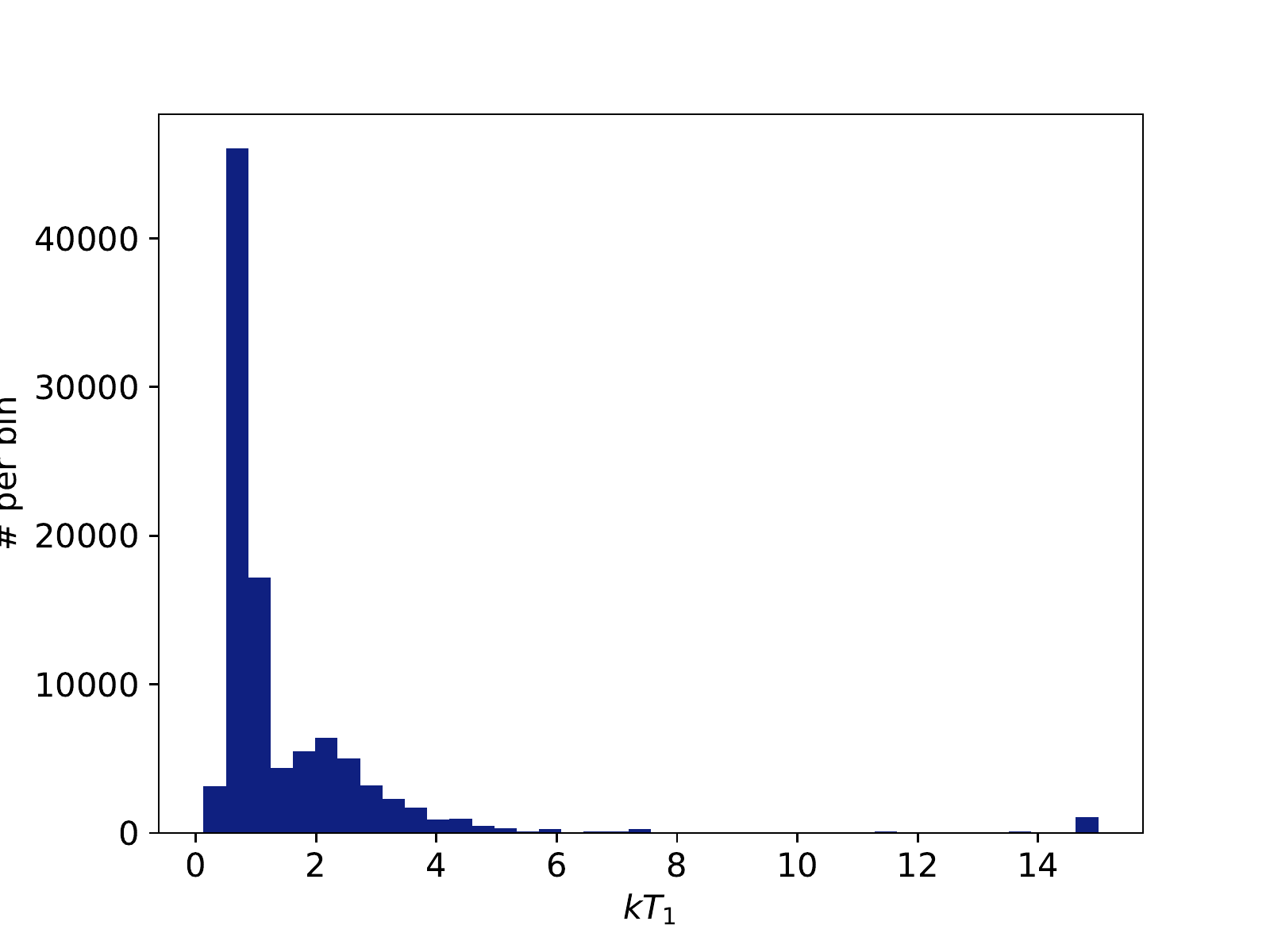}}
    
    \subfloat[2D histogram of $kT1$, and $kT2$ used to simulate stars with two-temperature plasma models ($kT1$ refers to colder plasma).\label{fig:star_kT1_kT2}]{%
    \includegraphics[width=0.9\columnwidth]{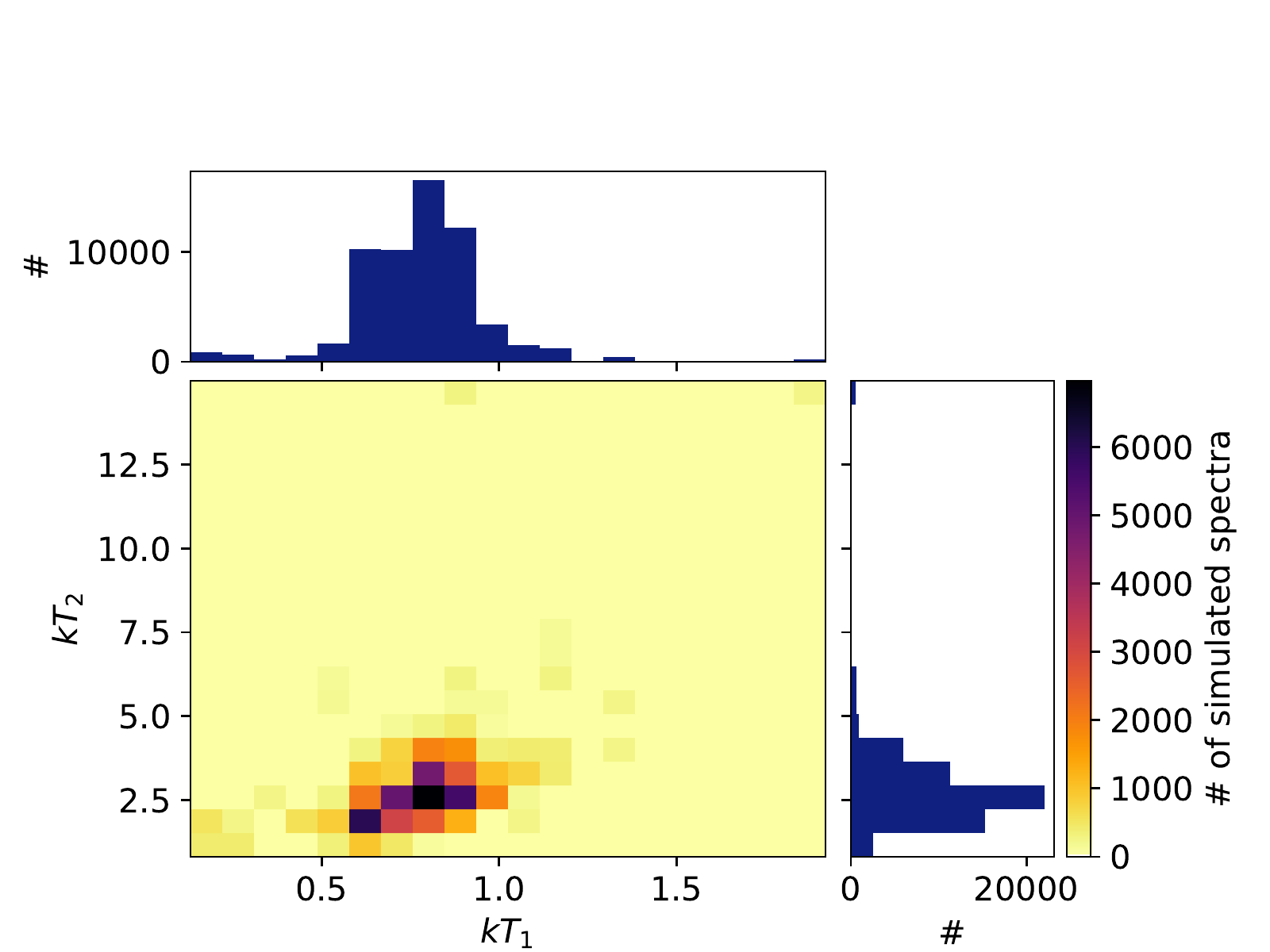}}\hfill
    \subfloat[2D histogram of emission measures used to simulate stars with two-temperature plasma models. \label{fig:star_em1_em2}]{%
    \includegraphics[width=0.9\columnwidth]{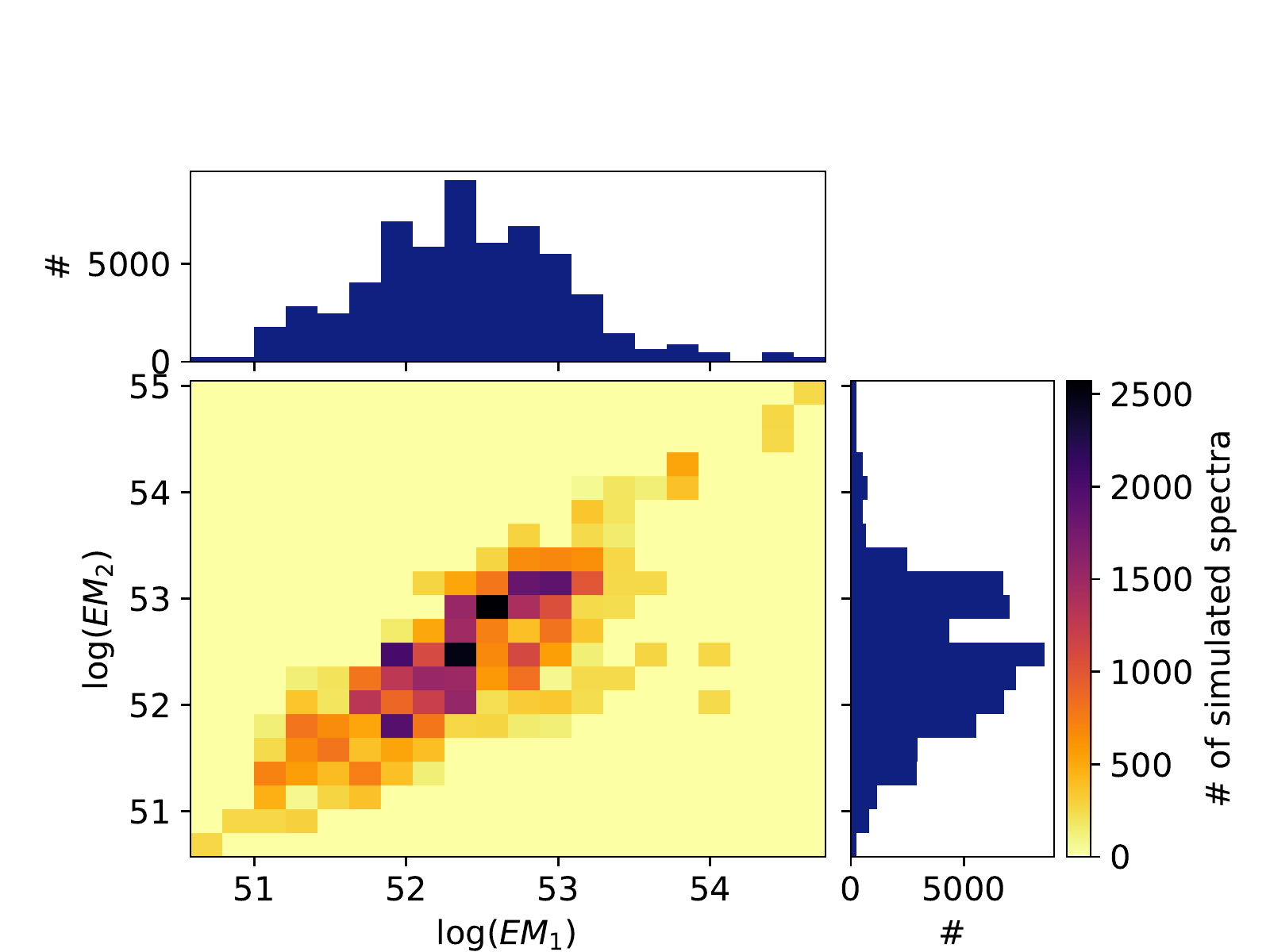}}
    \figcaption{Distribution of the parameters used to simulate the spectra of stars for training our ANN. Since the Orion nebula is close to the galactic plane, $N_H$ is high. Most stars have $kT < 2$ keV. The few sources with $kT \approx$ 14 keV are likely to be heavily absorbed stars where emission lines could not be fit (and a higher temperature plasma could fit the relatively hard spectra). We see some correlation between $kT1$ and $kT2$, and a strong correlation between $EM_1$ and $EM_2$. Since we match the fluxes of stars and AGN, specifying the emission measure values only contributes to maintaining the ratio of emission from the two plasmas.}
    \label{fig:star_props}
\end{figure*}

This filtering
leaves us with 679 sources. Of these 679 sources, 406 sources were modeled with single plasma models and 273 sources were fit with two-temperature plasma models. We use the distributions of $N_H$ (absorption column densities), $kT_1$, $kT_2$ (temperatures of plasma [stellar atmosphere in this case]), and the emission measures of the components of these COUP X-ray spectra models to simulate spectra of 100,000 young stars in our sample. We fix the abundance values to 0.3 times the solar values in accordance with \citet{Getman_2005}. We show the distribution of parameters used to simulate our active star spectra in Figs.~\ref{fig:star_props}.  From Figs.~\ref{fig:star_nh} \& \ref{fig:star_kT1}, we notice that 90\% of the stars have $\log N_H$ (in cm$^{-2}$) $\in (20.82, 22.68)$ and $kT_1 \in (0.43, 3.3)$  keV. We use the \texttt{tbabs*apec} and \texttt{tbabs*(apec+apec)} models in XSPEC v12.11.1 and the \texttt{fakeit} command to simulate  spectra with an exposure time of 1 megasecond (Ms). (The \texttt{tbabs} component is used to model the absorption with \citet{Wilms_200} abundances and the \texttt{apec} models X-ray emission from plasma in thermal equilibrium). \citet{Tsujimoto_2005} detected a 6.4 keV Fe fluorescence line in seven of the COUP stars. However, this fluorescent emission is
only present in young stellar objects with significant disks, and is weak (usually $<$150 eV, and weaker than the nearby 6.7 keV Fe line). 
Therefore, we do not add this line to our simulated stellar spectra.

The properties of the AGN X-ray spectra are selected from the {\it Chandra} Deep Field South AGN spectral properties catalog \citep[CDFSAGNCXO,][]{Tozi_2006}. This catalog 
lists 
the properties of 321 high redshift AGN in the 1 Ms {\it Chandra} Deep Field South (CDFS) survey. Most of these sources were fit with an absorbed power-law model (Compton thin or `C-thin' model). For 8 sources, the fit included an additional soft power-law component (same slope as the absorbed power-law but no absorption; Soft component or `Soft-C' model), and 14 AGN were fit with a reflection-dominated model (Compton thick or the `C-thick' model). \citet{Tozi_2006} fixed the Galactic absorption (from our Galaxy) to $N_{H, Gal}$ = 8 $\times$ 10$^{19}$ cm$^{-2}$. Accordingly, we use a C-thin model for 93\%, C-thick model for 4\%, and Soft-C model for 3\% of the 100,000 artificial AGN spectra that we generated.  For models where the intrinsic absorption of the host galaxy could not be constrained [i.e. where  \citet{Tozi_2006} reported intrinsic absorption column density $N_{H, int} = 0$], we use a value of $10^{19}$ cm$^{-2}$.$\ $ Note that effects from Galactic absorption and the response of {\it Chandra} dominate the soft spectra in these AGN. 

\begin{figure*}
    \centering
    \subfloat[Distribution of power-law indices for AGN distribution. Most spectra have $\Gamma \sim 1.8$. \label{fig:cdfs_gamma_hist}]{%
    \includegraphics[width=0.9\columnwidth]{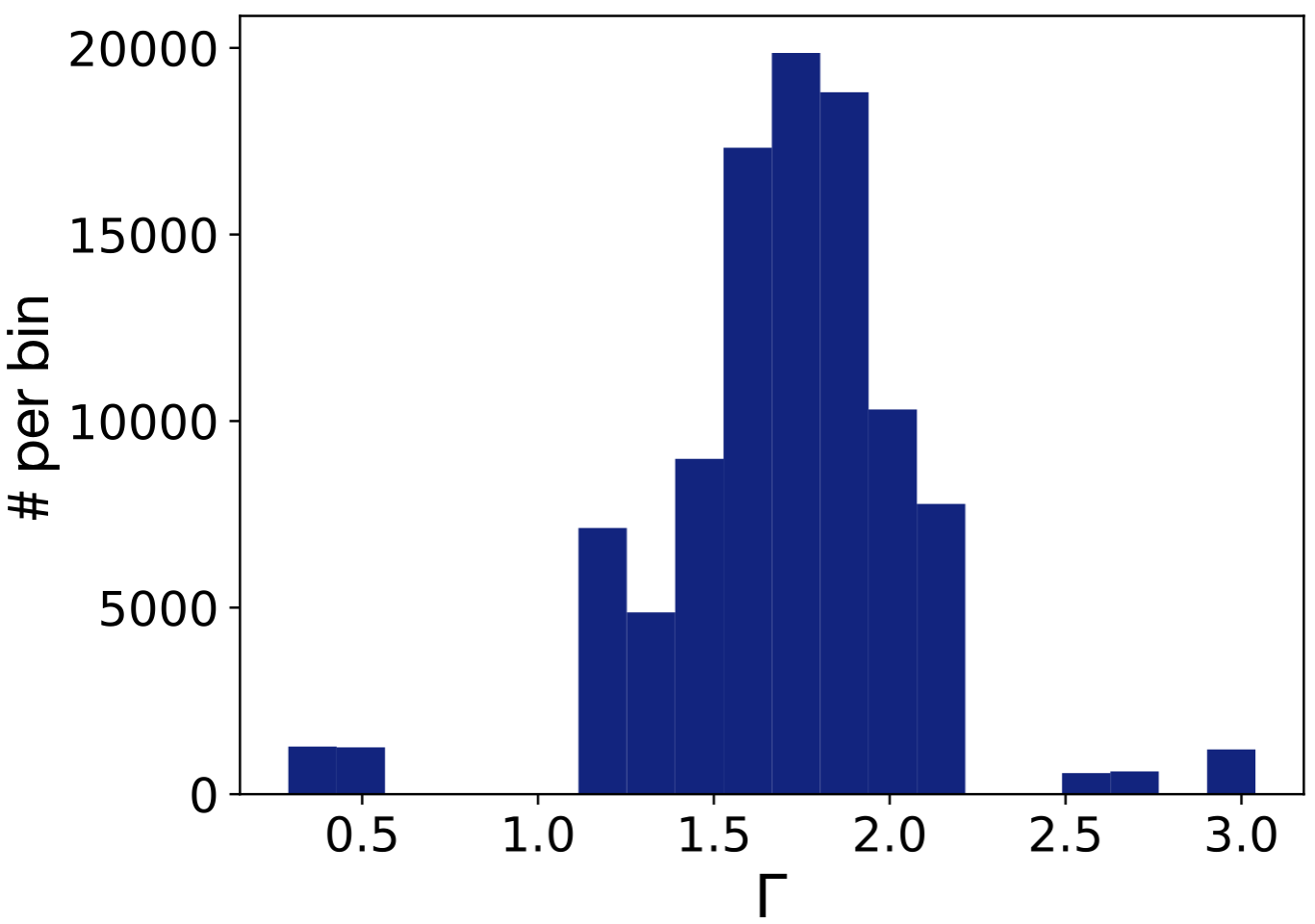}} \hfill
    \subfloat[2D histogram of absorption column density, $N_H$ and redshift, $z$. For sources where \citet{Tozi_2006} could not constrain the $N_H$, we use a fixed $N_H =10^{19}$ cm$^{-2}$. AGN with higher redshift usually have higher absorption. \label{fig:cdfs_nh_z_hist}]{%
    \includegraphics[width=0.9\columnwidth]{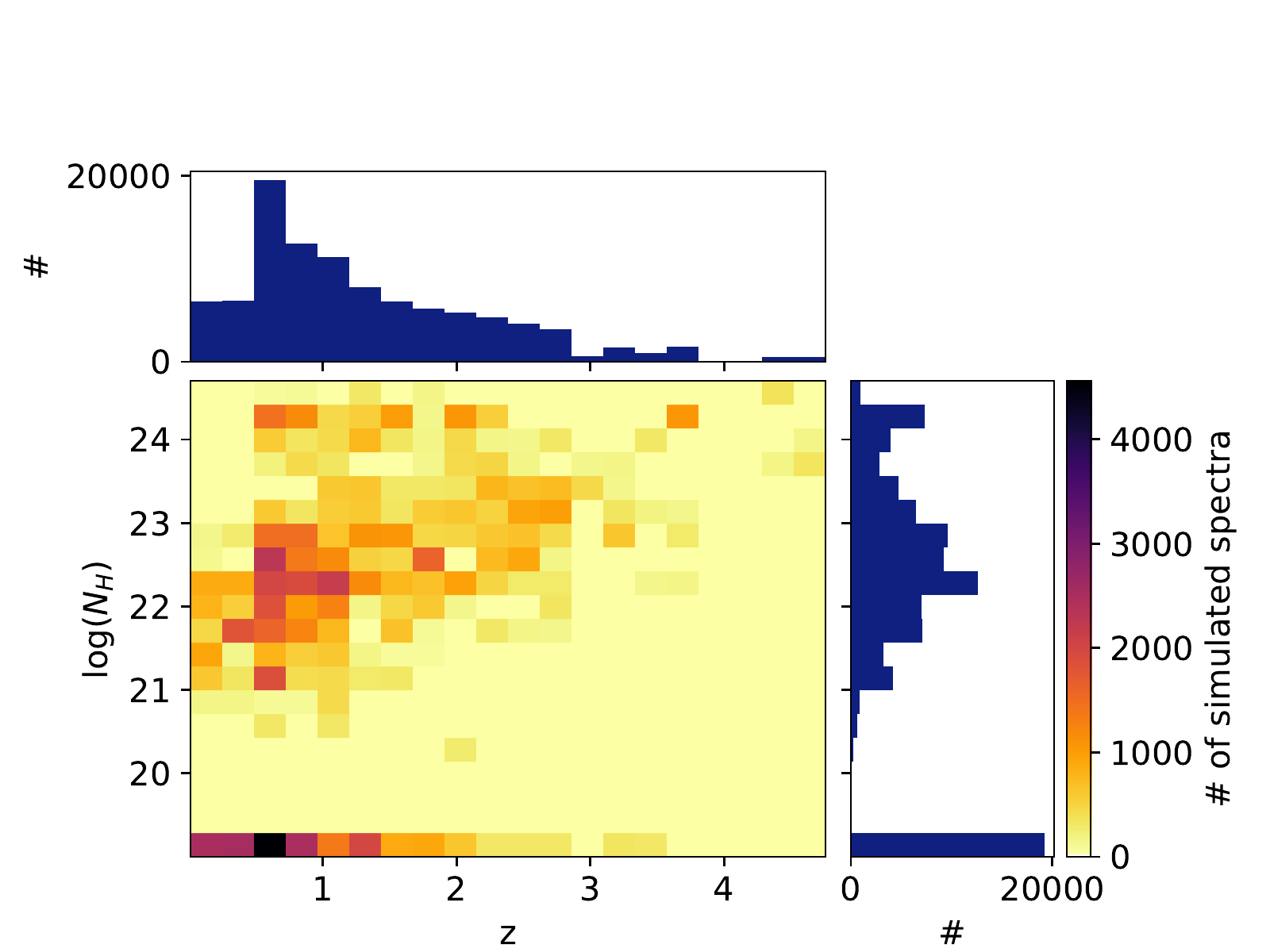}}
    
    \subfloat[2D histogram of hard (2--10 keV) and soft (0.5--2.0 keV) flux used for simulating our AGN spectra. The two fluxes generally show a positive correlation except at low values of soft flux, which could be due to high absorption in these sources. \label{fig:cdfs_fsoft_hard_hist}]{%
    \includegraphics[width=0.9\columnwidth]{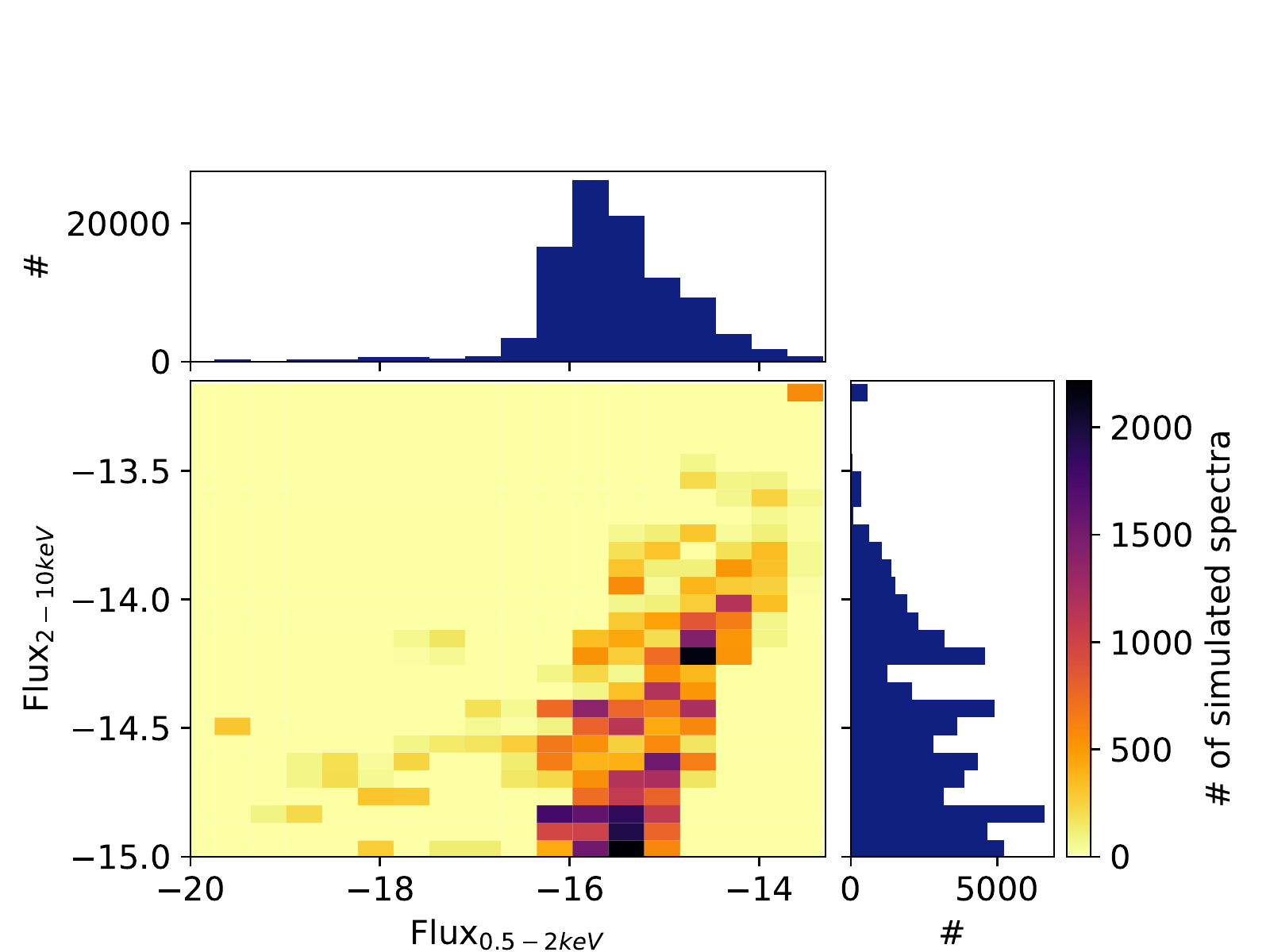}} \hfill
    \subfloat[Positions of the Fe-K emission line in the AGN spectra after accounting for the redshift. This figure only considers the AGN spectra in which an Fe-K emission was added (50\% of all AGN spectra). We assume a uniform distribution of equivalent widths from 100--3000 eV for these spectra. \label{fig:fe_pos}]{%
    \includegraphics[width=0.9\columnwidth]{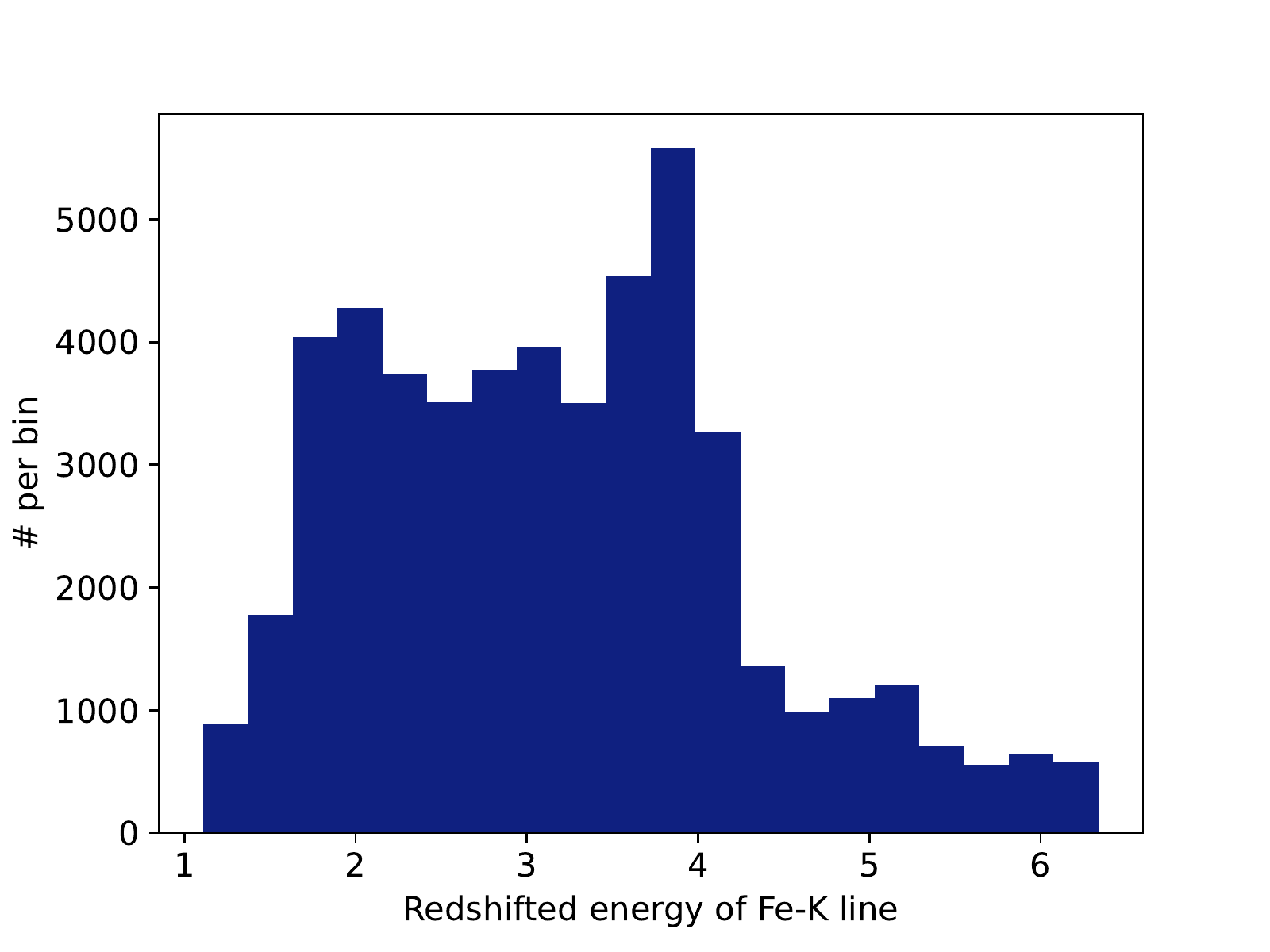}}
    \figcaption{Properties of AGN sample used for training and testing our ANN.}
    \label{fig:agn_prop}
\end{figure*}

Fe line emission has been detected in 34\% of  well-detected CDFSAGNCXO AGN with spectroscopic redshifts, with equivalent widths between 100--3000 eV \citep{Liu17}. Since we desire our ML algorithms to account for the possibility of Fe-K emission in the AGN spectra, we introduce an Fe-K line in 50\% of our simulated spectra of AGN. This may or may not be representative of the AGN population in the CDFS sample or across the universe (it is unclear, since our modelled equivalent width distribution reaches to small values that may not be detected in e.g. the CDFS sample), but this ensures that AGN with and without lines are  represented in our sample.  Our goal is that our model can identify AGN with or without the Fe-K line. The equivalent width of the Fe-K emission line was assumed to follow a uniform distribution between 100-3000 eV. The position of the Fe-K line is calculated based on the chosen redshift for each AGN, and a rest-frame energy of 6.4 keV.
We show the properties of our AGN sample in Figs.~\ref{fig:agn_prop}. We notice that 90\% of the sources have $\log N_H$ (in cm$^{-2}$) $\in (19.05, 24.2)$. Among sources where $N_H$ could be properly constrained, the range is (21.37, 24.2), $z \in (0.01, 2.56)$ and $\Gamma \in (1.2, 2.2)$. We use the XSPEC models \texttt{cflux*tbabs*ztbabs*pegpwrlw}, \texttt{tbabs*pegpwrlw + cflux*ztbabs*pegpwrlw}, and \texttt{cflux*tbabs*pexrav} for the C-thin, Soft-C and C-thick AGN respectively (we use \texttt{cflux} on the absorbed flux, as  the CDFSAGNCXO catalog only reports the absorbed flux values) to generate an artificial AGN spectra sample with an exposure time of 1 Ms. The \texttt{tbabs} and \texttt{ztbabs} are used to model the absorption from the Galaxy (fixed to $8 \times 10^{-19}$ cm$^{-2}$ in the direction of the CDFS sample) and the redshifted intrinsic absorption in the host galaxy. The \texttt{pegpwrlw} component models the power-law emission from the AGN. We add an additional \texttt{tbabs*pegpwrlw} component to  model the soft component in Soft-C AGN. In these AGN, we use the same power-law indices for the two components, in accordance with the spectral fitting in \citet{Tozi_2006}. We use \texttt{pexrav} model to generate the reflected AGN spectra in C-thick AGN. We add an additional \texttt{tbabs*ztbabs*gaussian} component for AGN with an Fe-K line.

\begin{figure}
    \centering
    \includegraphics[width=0.9\columnwidth]{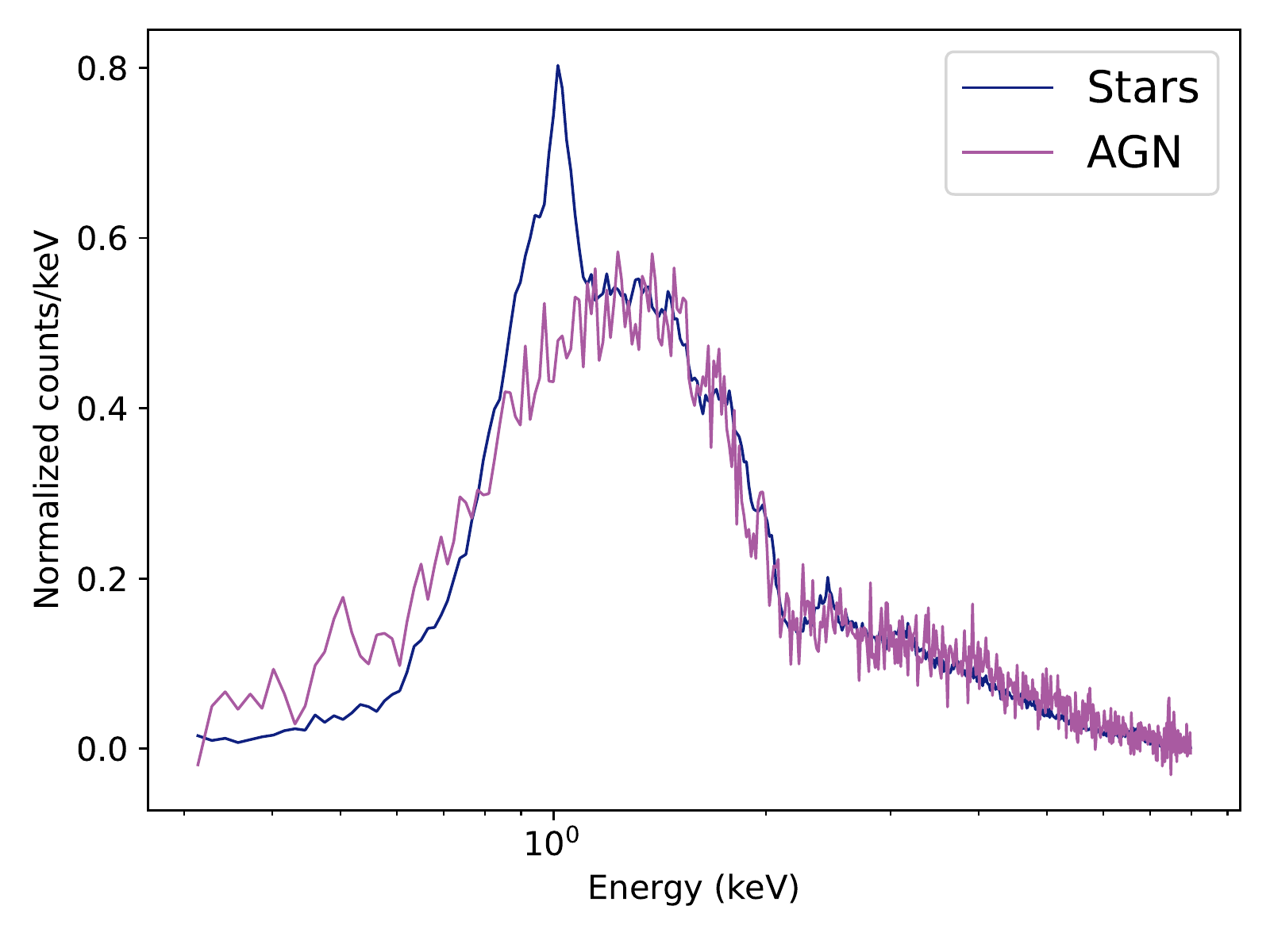}
    \figcaption{Normalized (divided by total counts) mean spectra of the observed COUP stars (solid blue) and CDFS AGN (solid purple). For the purpose of plotting, we only show the mean spectra calculated from sources with net counts more than 100. From the figure, we see that the spectra of stars show a prominent Ne-K/Fe-L emission line at $\sim$1 keV along with slight hints of Mg-K ($\sim$1.3 keV), Si-K ($\sim$1.8 keV), and S-K ($\sim$2.3 keV), while the AGN show no such lines. Since the CDFS AGN have lower flux than the COUP stars in general, we notice that the spectra of AGN have more noise (both background and Poisson) than the stars.}
    \label{fig:obs_meanspectra}
\end{figure}

\begin{figure}
    \centering
    \includegraphics[width=0.9\columnwidth]{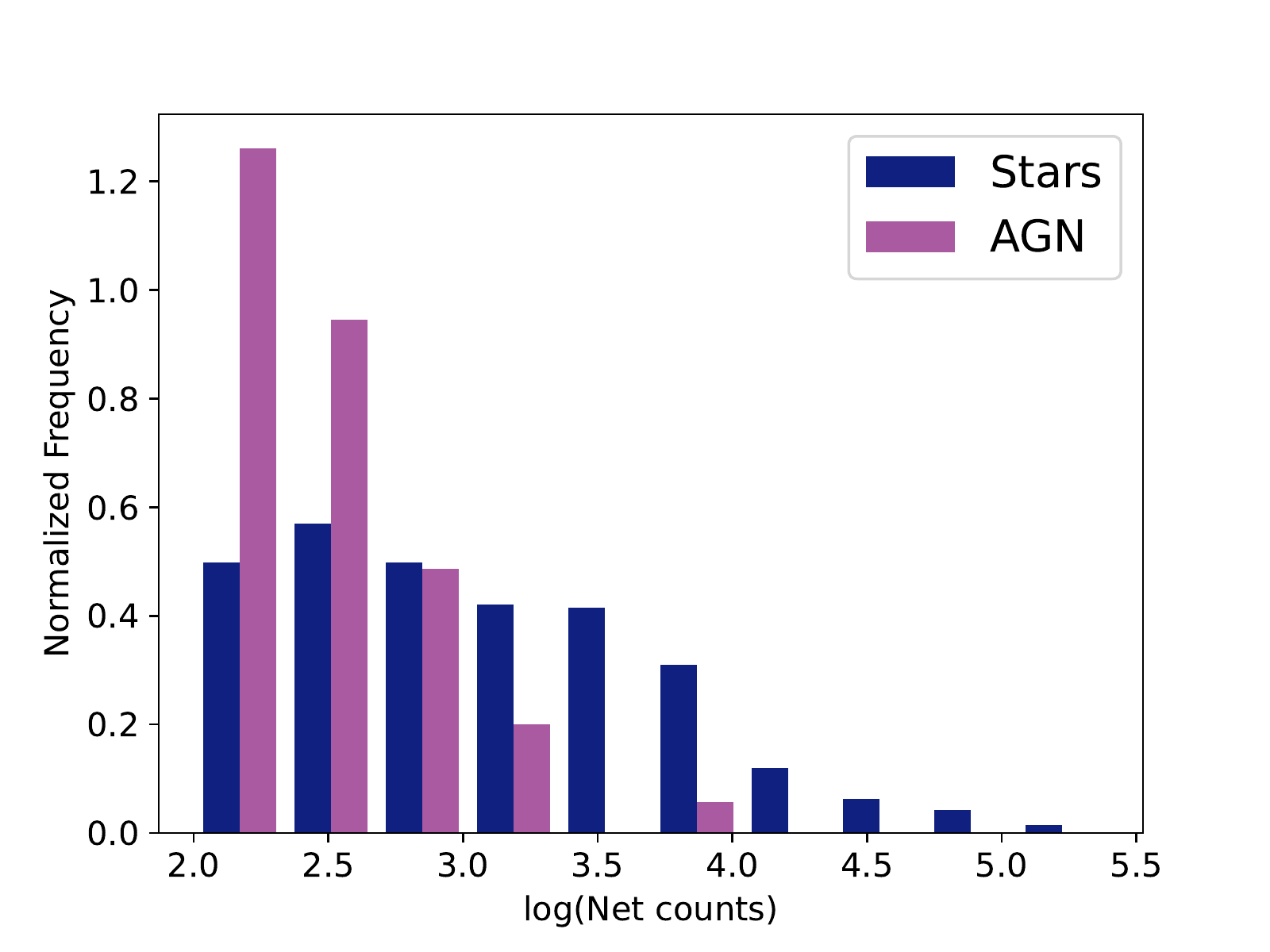}
    \figcaption{Histogram of the net counts of the observed COUP stars (solid blue) and CDFS AGN (solid purple). We have normalized the histograms such that the area under both histograms is unity (to avoid distortion due to more COUP sources than CDFS AGN). We notice that the COUP stars are typically $\sim$ 10 times brighter than the CDFS AGN. \label{fig:obs_netcounts}}
\end{figure}

\begin{figure}
    \centering
    \includegraphics[width=0.9\columnwidth]{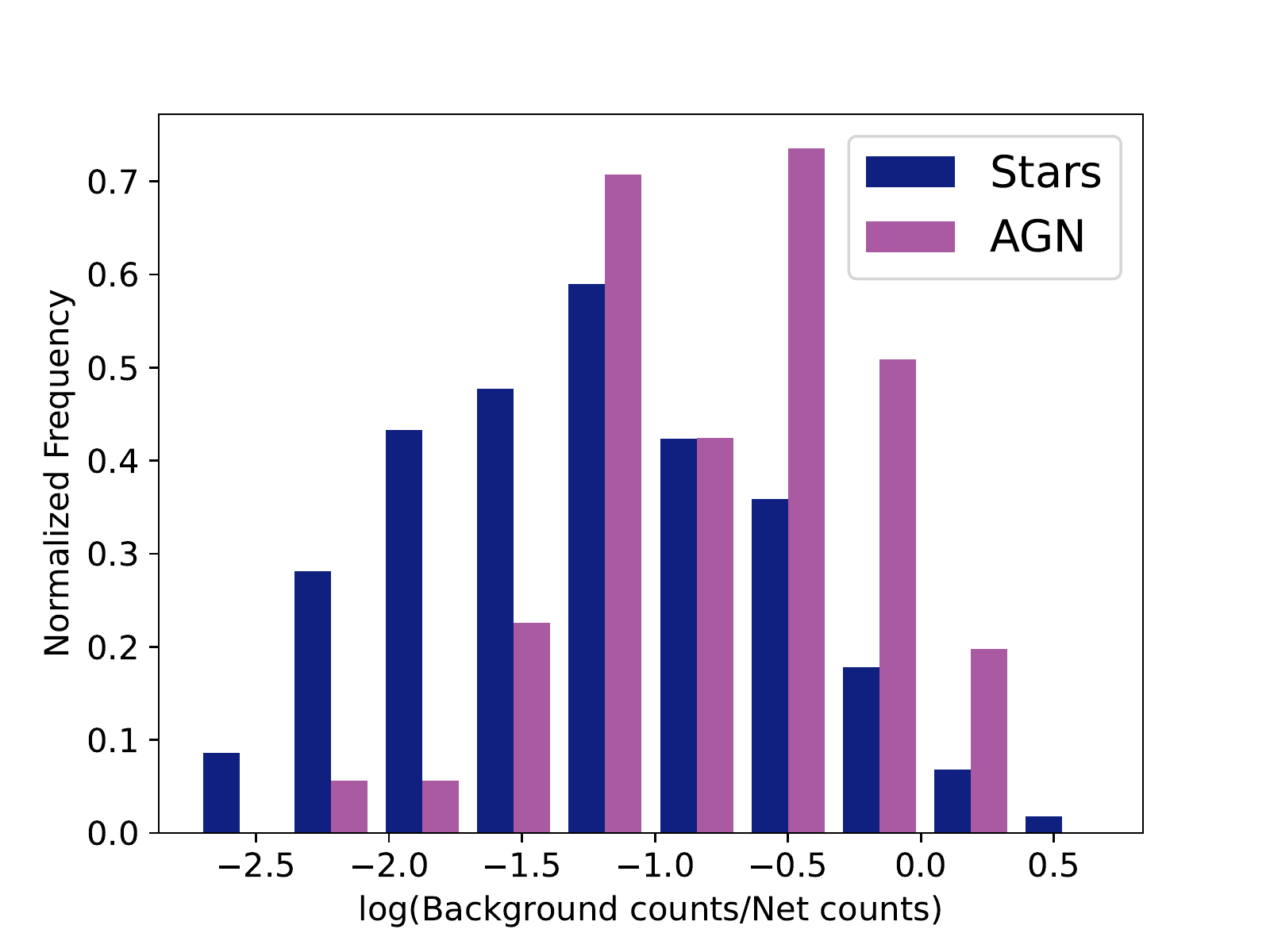}
    \figcaption{Normalized histogram of the ratio of background-to-net counts for the observed COUP stars and CDFS AGN. Since CDFS AGN are mostly faint,  most of them have a very high background contribution. \label{fig:obs_bgcounts}}
\end{figure}

We cross-match the X-ray positions of the COUP and CDFSAGNCXO sources with the CSC v2.0 catalog using a search radius of 3.0$\arcsec$, and download the spectra 
from
the COUP and CDFS observations. We are able to retrieve the spectra of 1373 stars out of the 1616 COUP sources, and 296 AGN out of the 321 CDFS sources. (We fail to extract all sources because the CSC pipeline masks the corners of the ACIS detectors, where the background is very high, to minimize the detection of erroneous sources.) The CDFS observations were taken in four epochs -- 1 Ms between 1999-2000, 1 Ms in 2007, 2 Ms in 2010, and 3 Ms in 2014. Since the soft energy response of {\it Chandra} ACIS has been degrading, we only use the 1999-2000 observations in this paper. These are also the observations used by \citet{Tozi_2006} to study the AGN properties. In order to generate the artificial spectra of our sample, we use the response matrix and effective area files of the cross-matched COUP and CDFS sources.
We show the mean spectra of these sources in Fig.~\ref{fig:obs_meanspectra}. We only used sources with net counts greater than 100 in  Fig.~\ref{fig:obs_meanspectra} to reduce the noise. We also show the distribution of the net counts and the background contribution to these sources in Figs.~\ref{fig:obs_netcounts} \& \ref{fig:obs_bgcounts}.

We note a few caveats of using the COUP and CDFS surveys. Both COUP and CDFS surveys have exposure time $\sim 1$Ms, much longer than the exposure times of typical Chandra observations. Thus, the net counts of sources in COUP are higher than those of typical stars, and the faint AGN in CDFS have larger background-to-net count ratios than most AGN detected. From Figs~\ref{fig:obs_netcounts}, we notice that the COUP stars are $\sim 10$ times brighter than CDFS AGN, implying that the spectra of our sample stars will have higher signal-to-noise ratio than our sample AGN. Therefore, we use the flux distribution from the CDFS AGN in the 0.5--10 keV regime to simulate the spectra of stars so that both AGN and stellar spectra have similar signal-to-noise ratios. We do this to check if our ML method is inherently better at picking AGN/stars. However, this doesn't imply that all our sources have the same noise. The $\sim 2$ orders of magnitude range in the flux of CDFS AGN ensures that we can check our performance with varying net-counts and background contribution.

We attempt to analyze how the background can affect the performance of our methods by simulating three kinds of spectra from AGN and stars --- without background, with the observed background, and with a reduced background rate (i.e., we increased the flux of the sources by a factor of 10 and reduced the exposure to 100 ks). We consider a reduced background because the targets we selected are unrepresentative of typical Chandra observations (as noted above),
so the reduced background analysis will be more helpful to estimate potential future performance.
We also notice the COUP sources are $\sim$10 times brighter than the AGN in the CDFS survey. Therefore, we use the fluxes from the CDFS AGN in the 0.5--10 keV regime to simulate the spectra of stars so that both AGN and stellar spectra have similar signal-to-noise ratios.

\subsection{Setting up the ANN}
\begin{figure}
    \centering
    \includegraphics[width=\columnwidth]{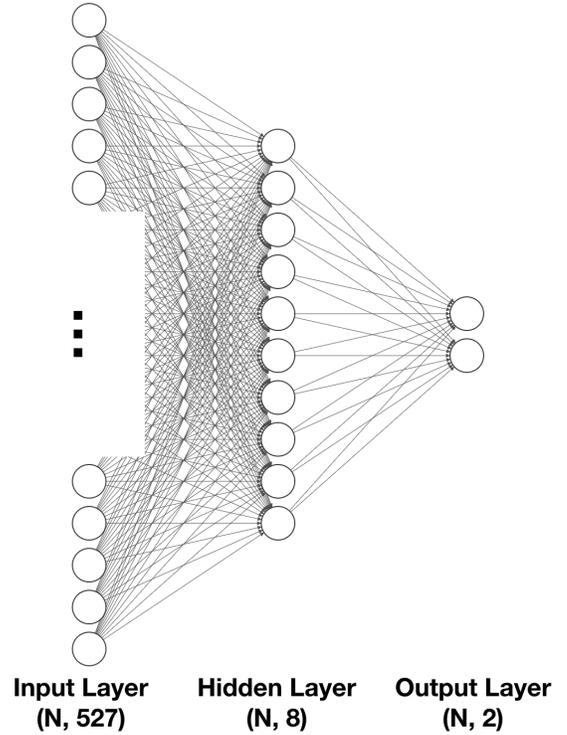}
    \figcaption{ Architecture of the ANN model. The input layer consists of 527 nodes corresponding to the number of channels in the {\it Chandra} ACIS-S spectra in the energy range 0.3--8.0 keV. We use one hidden layer with 10 nodes which are fully connected to the input layer. The hidden layer uses a ReLU activation and a L1 regularization with $\lambda = 0.001$. These hidden nodes are then used to calculate the values in the two output nodes that use a softmax activation function and signify the probability of the source being an active star or an AGN. \label{fig:ann_arch}}
\end{figure}

We incorporate a sequential artificial neural network (ANN) model using Tensorflow \citep[][]{tensorflow2015-whitepaper} for our primary analysis. In this model, we add one hidden layer with ten nodes (we also tried other configurations, but found no significant increase in the accuracy with a higher number of nodes/layers) and a Rectified Linear Unit (ReLU) activation function (we show the neural network architecture in Fig.~\ref{fig:ann_arch}). The output layer has sigmoid activation ensuring the probabilistic interpretation of the output values. As the problem of classifying the X-ray spectra is essentially a Bernoulli problem (i.e. an X-ray spectrum could be from an AGN with probability {\it p} and star with probability {\it 1-p}), we use a binary cross-entropy loss  for training. The binary cross-entropy loss is defined by,
\begin{equation}
    H_p (D) = -\frac{1}{N} \sum_{i=1}^{N} y_i \log p(y_i)  + (1 - y_i)\log (1-p(y_i)),
\end{equation}
where $H_{p}(D)$ signifies loss calculated over dataset D, N is the number of  input data, $y_i \in \{0,1\}$ (we chose 0 for stars and 1 for AGN), and $p(y_i)$ is the probability that $y_i = 1$. We use the Adam optimization algorithm \citep[][]{Kingma_2014} for training the model since it results in less computation time and faster convergence. We fix the L1 regularization value, $\lambda = 0.001$, to avoid over-fitting of the data (we test for various values).
We consider the 0.3--8.0 keV energy interval while training our ANN classification model, to limit the background contribution (not to be confused with the 0.5--10 keV flux values reported in \citealt{Tozi_2006} and used in our simulations). In order to limit the Poisson noise, we only use simulated spectra with greater than 100 net counts in the 0.3--10 keV energy range for training (this net count criteria is not for simulating sources but only for training and testing ANN).
We use 80\% of the spectra for training the model and use the remaining 20\% as our test set. We perform 10-fold cross-validation while training the model, to ensure that we do not overfit the data. We then apply the trained ANN model to classify spectra of the test set and evaluate our performance. We perform 20 such iterations to study the behaviour of our ANN model.

For several ML applications, the input is standardized by subtracting each feature by the mean of that feature across all classes, and dividing the difference by the standard deviation of that feature. In general, this allows for better fitting of the weights. We also perform ANN classification with standardized input, but the improvement in classification accuracy is $\lesssim 1\%$. In astrophysics, having unequal number of sources in each source class is common, which can lead to the mean being biased to one particular class.
Therefore, we choose to present results that do not use 
standardized inputs.

We analyze the results of our classification using the following metrics (for the purpose of this paper, we consider AGN as positives and stars as negatives):
\begin{itemize}
    \item {\bf Accuracy:} The total fraction of correct predictions across the entire set.
    \begin{equation}
        \mathrm{Accuracy} = \frac{\mathrm{TP + TN}}{\mathrm{TP + TN + FP + FN}}
    \end{equation}
    \item {\bf Recall:} The ratio of true positives to the total number of positives in the dataset. In our case, this signifies the fraction of AGN identified correctly.
    \begin{equation}
        \mathrm{Recall} = \frac{\mathrm{TP}}{\mathrm{TP + FN}}
    \end{equation}
    \item {\bf True Negative Rate (TNR):} The ratio of true negatives to the total number of negatives in the dataset, i.e. the fraction of stars identified correctly.
    \begin{equation}
        \mathrm{TNR} = \frac{\mathrm{TN}}{\mathrm{TN + FP}}
    \end{equation}
    \item {\bf Precision:} The ratio of true positives to the number of samples that have been classified as positives. In our case, it is the fraction of true AGN among the spectra classified as AGN.
    \begin{equation}
        \mathrm{Precision} = \frac{\mathrm{TP}}{\mathrm{TP + FP}}
    \end{equation}
    \item {\bf Negative Predictive Value (NPV):} The ratio of true negatives to the number of samples that have been classified as negative, i.e. the fraction of true stars among the spectra that have been identified as stars.
    \begin{equation}
        \mathrm{NPV} = \frac{\mathrm{TN}}{\mathrm{TN + FN}}
    \end{equation}
    
\end{itemize}
In the above equations TP, TN, FP, and FN stand for the true positives (AGN identified correctly), true negatives (stars identified correctly), false positives (stars identified as AGN), and false negatives (AGN identified as stars), respectively.

We also evaluate the performance of the trained ML model on the observed data. This set consists of all COUP stars (including sources that have been flagged for deviations in emission lines, poorly or marginally fit spectra, soft/hard excess, etc; only extragalactic sources are removed) and  CDFS AGN that have more than 100 counts. Since there are only 108 such CDFS AGN as compared to 679 COUP stars, we randomly select 108 COUP stars so that both classes are equally sampled. Note that we select a new random sample of COUP stars with each of the 20 iterations to avoid any hidden biases.

\section{Results}
\label{sec:results}

\begin{table*}
\caption{Performance of our machine learning model on different datasets}
\label{table:performance_metrics}
\hspace{-30pt}
\resizebox{\linewidth}{!}{
\begin{tabular}{ccccccc}
\hline
Type of Dataset          & Testing data      & Accuracy                     & Recall                  & TNR                         & Precision               & NPV                         \\ \hline
\multirow{2}{*}{No background} & Simulated spectra & $(88.9 \pm  0.9)$\% & $(83 \pm 5)$\% & $(92 \pm 3)$\% & $(86 \pm 3)$\% & $(91 \pm 2)$\% \\
                               & Observed spectra  & $(81 \pm 3)$\%  & $(67 \pm 7)$\% & $(93 \pm 5)$\%  & $(92 \pm 6)$\% & $(74 \pm 4)$\%                             \\ \hline
\multirow{2}{*}{Observed background} & Simulated spectra & $(88.7 \pm  0.6)$\% & $(86 \pm 3)$\% & $(91 \pm 2)$\% & $(87 \pm 2)$\% & $(90 \pm 2)$\% \\
                               & Observed spectra  & $(89 \pm 3)$\%  & $(84 \pm 5)$\% & $(95 \pm 2)$\%  & $(94 \pm 3)$\% & $(85 \pm 4)$\%                             \\ \hline
\multirow{2}{*}{Reduced background} & Simulated spectra & $(91.7 \pm  0.8)$\% & $(91 \pm 2)$\% & $(92 \pm 2)$\% & $(91 \pm 2)$\% & $(93.0 \pm 2)$\% \\
                               & Observed spectra  & $(91 \pm 2)$\%  & $(90 \pm 3)$\% & $(92 \pm 4)$\%  & $(91 \pm 3)$\% & $(91 \pm 3)$\%                             \\ \hline  
\multirow{2}{*}{Reduced background, Net Counts $>$ 1} & Simulated spectra & $(86 \pm  1)$\% & $(86 \pm 3)$\% & $(85 \pm 3)$\% & $(85 \pm 2)$\% & $(86 \pm 2)$\% \\
                               & Observed spectra  & $(82 \pm 3)$\%  & $(77 \pm 4)$\% & $(87 \pm 4)$\%  & $(86 \pm 4)$\% & $(80 \pm 3)$\%                             \\ \hline
\end{tabular}
}\\
Note: The error-bars indicated above correspond to one standard deviation
\end{table*}

\begin{figure}
    \centering
    \includegraphics[width=0.9\columnwidth]{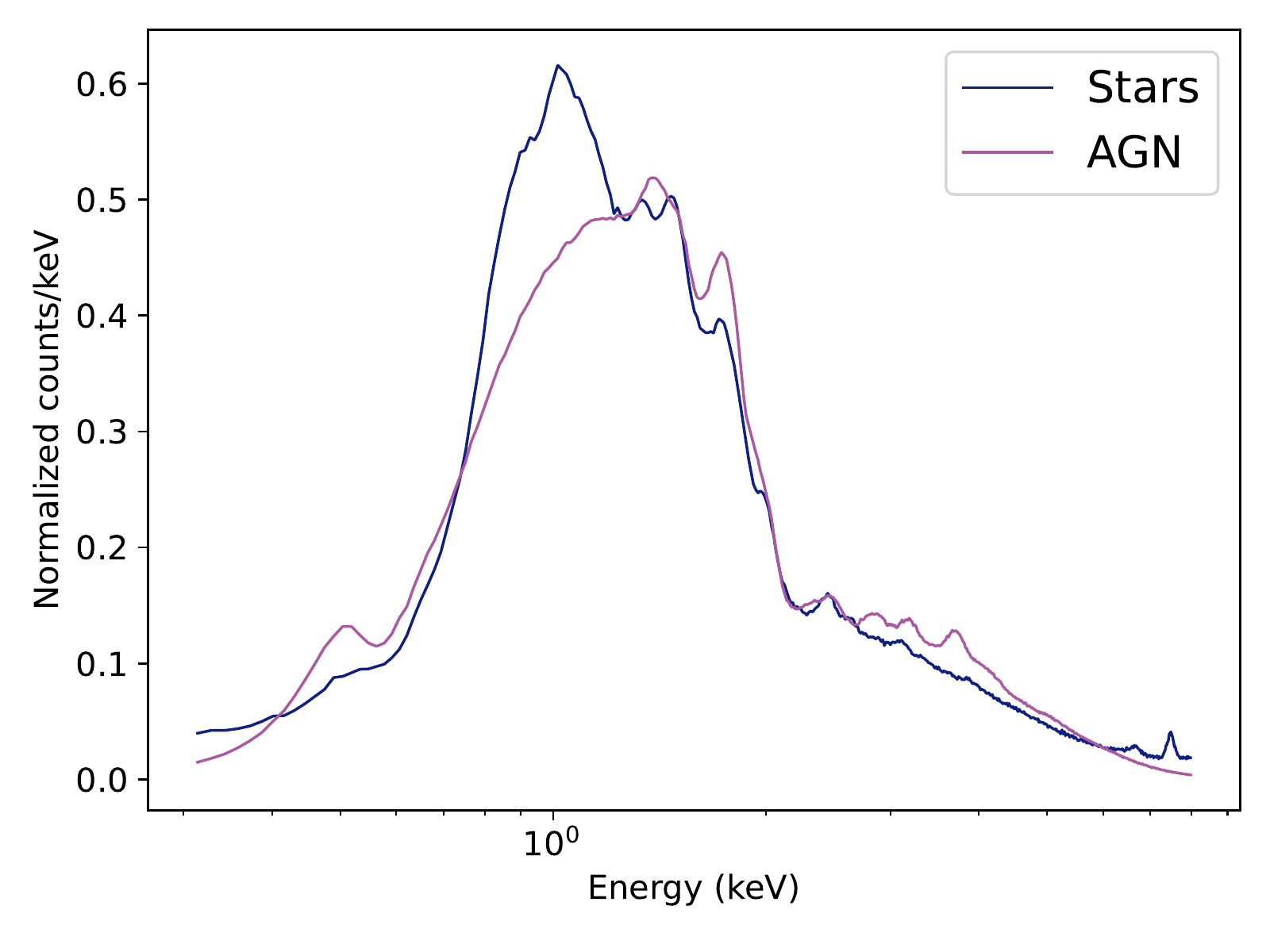}
    \figcaption{Normalized mean of the artificially generated AGN and star spectra simulated without including the background. These spectra are similar to that in Fig.~\ref{fig:obs_meanspectra}, but less noisy since they are the mean of 100,000 spectra. With less noise, the emission features in the spectra of stars are more clear. The emission-line-like features in the AGN spectra are due to a combination of the intrinsic absorption columns, redshifts, and the {\it Chandra} response. Red-shifted Fe-K emission lines cause the appearance of bumps in the mean AGN spectra in the 2--4 keV energy range.}
    \label{fig:nobg_meanspec}
\end{figure}

\begin{figure}
    \centering
    \includegraphics[width=0.9\columnwidth]{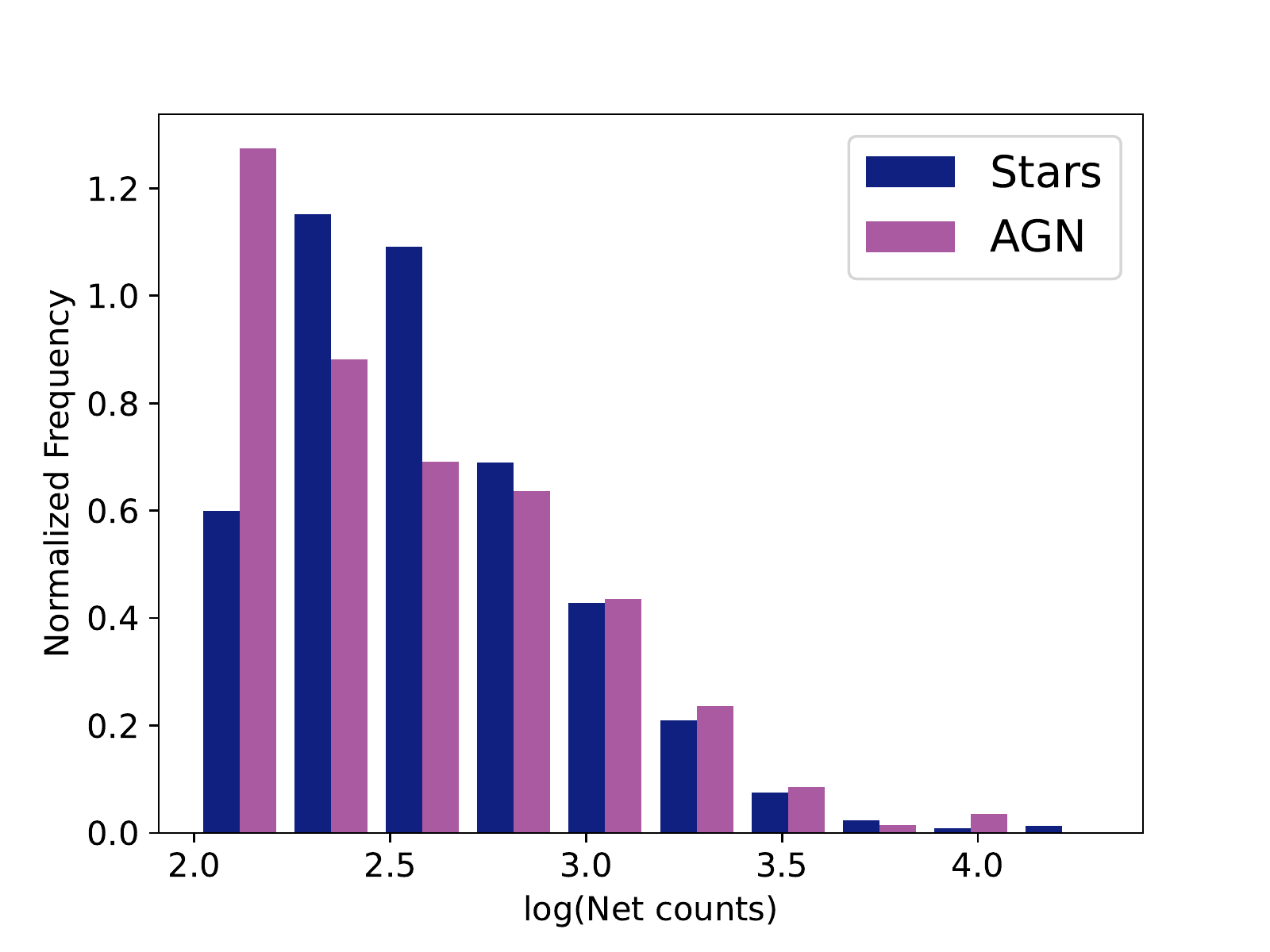}
    \figcaption{Distribution of total counts, in the 0.3--8.0 keV energy range, in the simulated sample of AGN and stars generated without including the background. Despite equating the 0.5-10 keV flux of simulated stars and AGN, stars have higher X-ray photon counts since they have softer spectra than AGN. \label{fig:nobg_netcounts}}
\end{figure}

We first analyze the spectra of AGN and stars without including the background. We show the mean spectra of stars and AGN in Fig.~\ref{fig:nobg_meanspec}. From the figure, we notice that AGN in general have harder spectra than their stellar counterparts (note that some AGN in our sample also have soft spectra as seen from Fig.~\ref{fig:cdfs_gamma_hist}). In our case this is further amplified due to their higher (intrinsic) $N_H$ and the presence of redshifted Fe-K lines in the CDFS AGN spectra. We also notice that the spectra of stars have dominant Ne, Fe, Mg, Si, and S lines. The feature at $\sim$2 keV in the mean AGN spectrum is due to a combination of  $N_H$, $z$, position of the Fe-K line, and the response of the {\it Chandra} ACIS detector, and is not a real emission line. We show the distribution of the net counts in the $0.3-8.0$ keV energy range in  Fig.~\ref{fig:nobg_netcounts}. We see that the stars have higher count rates than AGN on average by a factor of $\sim$ 5, even though their fluxes in the 0.5-10 keV interval are similar. This is because of the softer spectra of stars in comparison to AGN (i.e. we match the X-ray fluxes of stars and AGN in the 0.5--10 keV range while simulating spectra, but consider 0.3--8.0 keV for the classification and analysis.).

Applying our ANN model to this dataset gives us an overall accuracy $\sim$89\%, recall $\sim$83\%, TNR $\sim$92\%, precision of $\sim$86\%, and NPV of $\sim$91\%.
However, when this trained model is used to classify the set of observed spectra, we get an accuracy of only $\sim$81\%. The decreased accuracy is due to the poor performance of the classification algorithms at low net counts, where the contribution of background cannot be ignored.

\begin{figure}
    \centering
    \includegraphics[width=0.9\columnwidth]{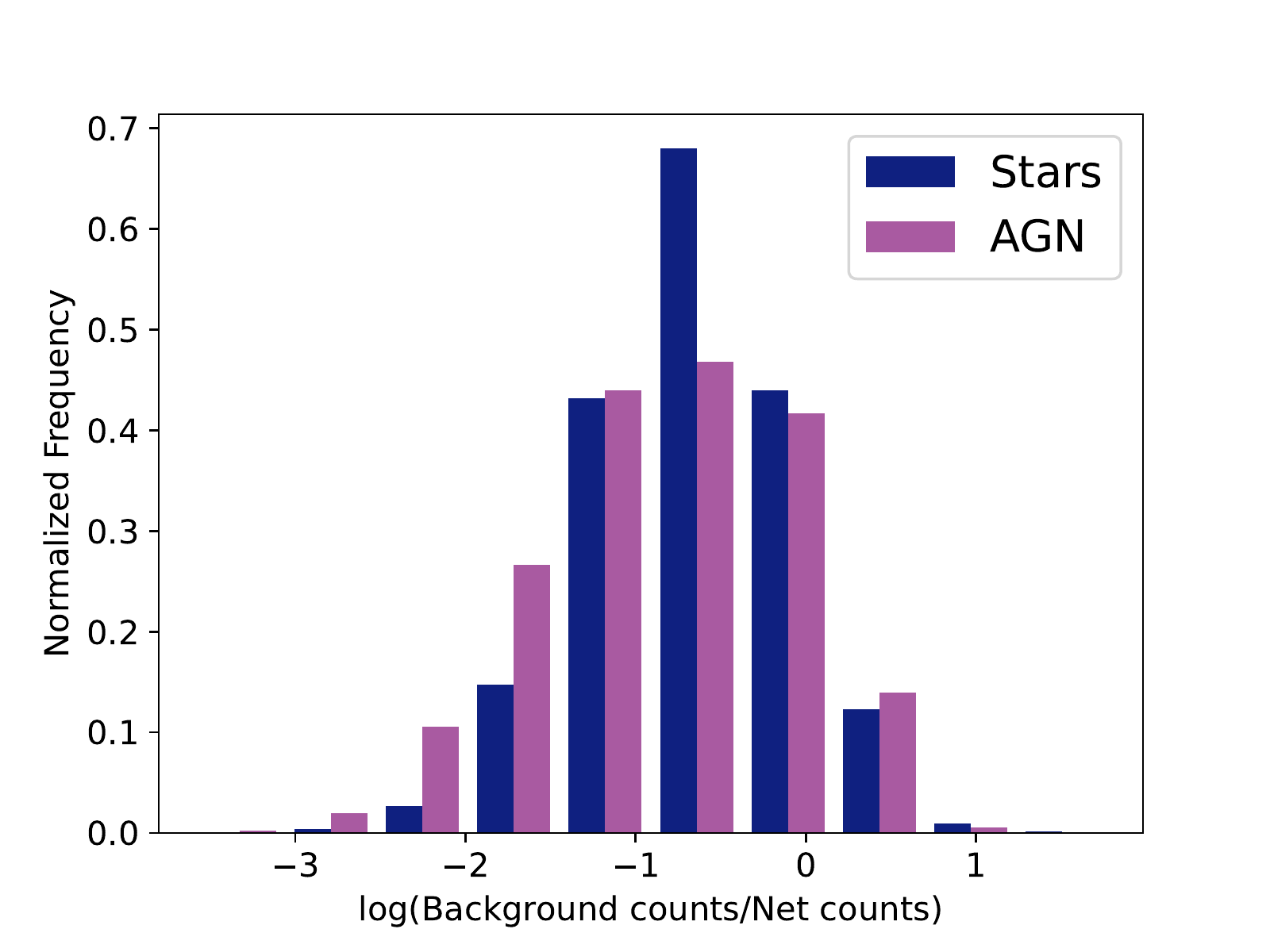}
    \figcaption{Distribution of background-to-net count ratio in the simulated AGN and star spectra when the observed background is included. The plot only represents spectra with net counts greater than 100 in the 0.3--8.0 keV energy range. Many of these spectra have a high contribution from the background as these sources are fainter than typical sources detected by {\it Chandra}. \label{fig:observedbg_bgcount}}
\end{figure}

Next, we explore the effect of background on our simulations. For this purpose, we simulate the spectra of AGN and stars using similar background levels as the COUP and CDFS sources. Fig.~\ref{fig:observedbg_bgcount} shows the histogram of the ratio of  background counts to net counts for the simulated sources in the 0.3--8.0 keV energy range. (The background counts have been scaled such that they correspond to the source extraction region). We notice that the stars and AGN in the COUP and CDFS catalogs have a high background, in general, with many sources having background counts $>$ 0.1 times their net counts. Our classification accuracy is $\sim$ 89\% on these simulated spectra.
When we apply the trained model to the observed spectra of CDFS and COUP sources, we get a a similar classification accuracy of $\sim$89\%. This performance is better than the ANN model trained on simulated spectra without background.

\begin{figure}
    \centering
    \includegraphics[width=0.9\columnwidth]{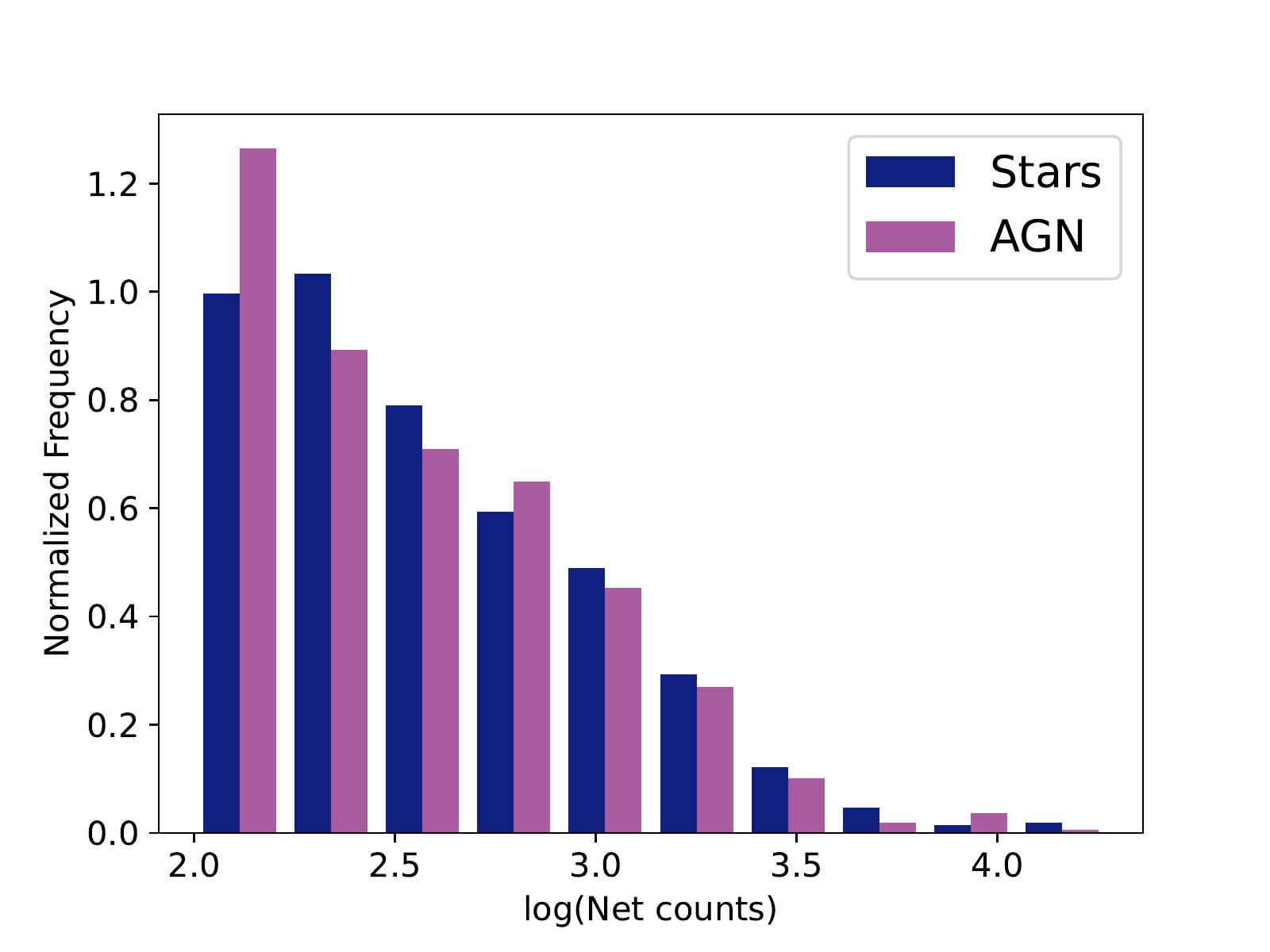}
    \figcaption{Distribution of net counts of star and AGN spectra, simulated with a reduction of the background contribution by a factor of 10. \label{fig:reducedbg_netcount}}
\end{figure}

\begin{figure}
    \centering
    \includegraphics[width=0.9\columnwidth]{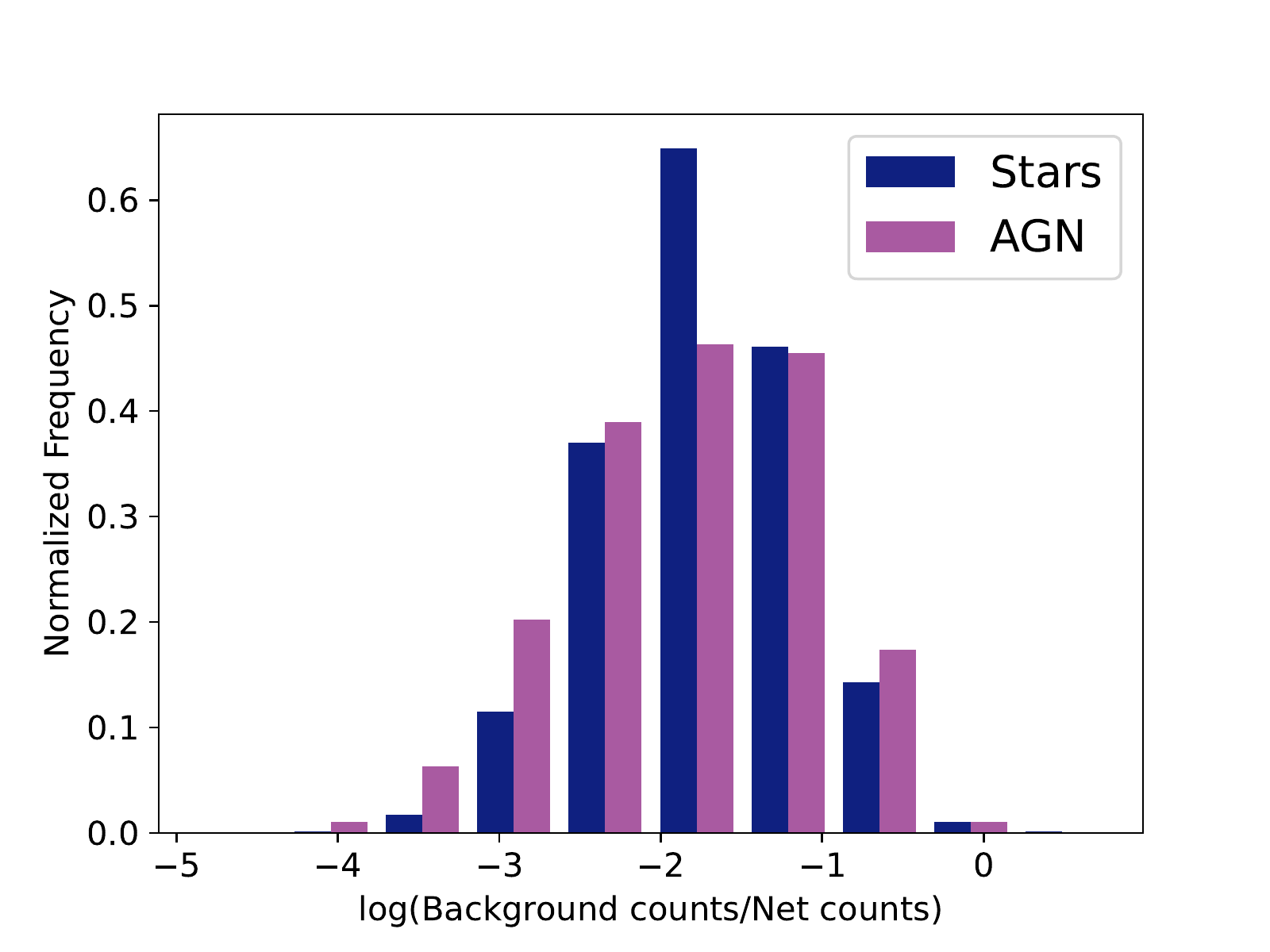}
    \figcaption{Distribution of the background-to-net count ratio in the simulated AGN and star spectra simulated with the reduced background. The y-axis is normalized such that the figure shows a density function. This distribution is more typical of the X-ray sources detected by {\it Chandra}. \label{fig:reducedbg_bgcounts}}
\end{figure}

The high background in our simulated flux is due to the low fluxes of the CDFS AGN (as well as the off-axis positions of a substantial portion of the sources). In general, most X-ray sources detected by {\it Chandra} have a lower background contribution than these AGN.
For instance, a typical Chandra observation is of order 30 ks, rather than the 1 Ms CDFS, reducing the typical background by a factor of 33. However, more Chandra sources with $>$100 counts can be found in longer observations; we use 100 ks observations for our standard as a compromise. 
To consider the performance of our classification algorithms for typical stars and AGN, we thus consider simulated spectra of stars and AGN where the background contribution has been reduced by a factor of 10.

We show the distribution of the net counts and the background-to-net count ratio in Figs.~\ref{fig:reducedbg_netcount} \& \ref{fig:reducedbg_bgcounts} respectively. These values of background are more typical of X-ray sources detected by {\it Chandra}. Applying our ANN model to these spectra gives an overall accuracy $\sim$ 92\%, recall of $\sim$ precision $\sim$91\%, and TNR and NPV$\sim$92.5\%.
Applying the trained model to observed spectra gives an accuracy of $\sim$91\%. One of the reasons why the model trained on reduced background performs better on the observed data, as compared to the model trained on the observed background, could be that the lower noise from the background contribution helps better constrain the values of the best fit model. We summarize the results of our classification on the simulated and the training data in Table~\ref{table:performance_metrics}. 

\begin{figure}
    \centering
    \includegraphics[width=\columnwidth]{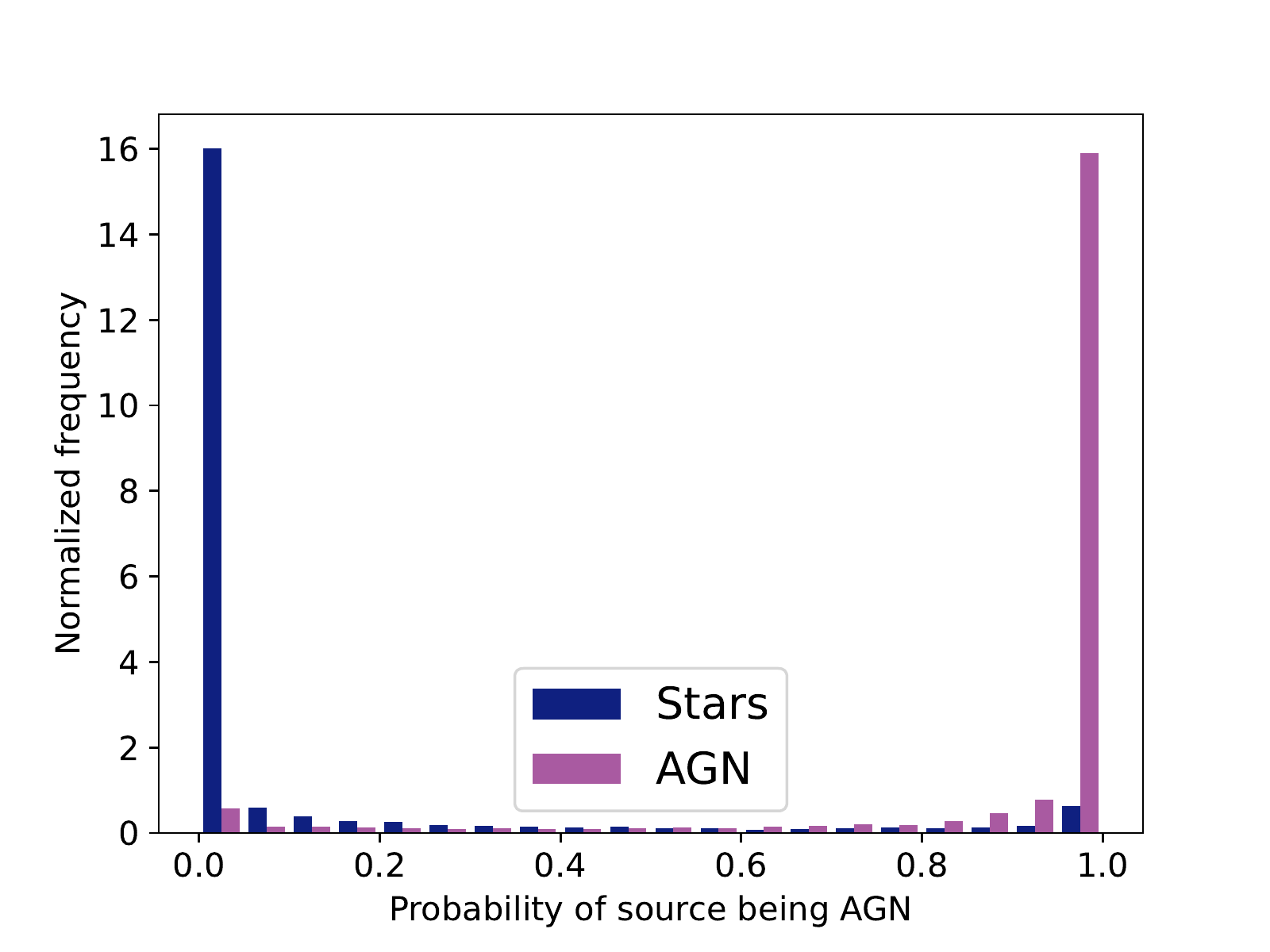}
    \figcaption{Distribution of probabilities of the source being AGN, as calculated by our trained model. Notice that most of the stars and AGN have been correctly identified with a certainty of more than 95\%. In this article we have used a cutoff of 50\% to differentiate stars and AGN. \label{fig:prob_hist}}
\end{figure}

We also look at the certainty with which our model identifies the AGN vs. stars. Fig.~\ref{fig:prob_hist} shows the distribution of probabilities of the input source being an AGN as estimated by our model. From the figure, we note that our model is able to distinguish most AGN and stars with very high confidence. We used a cutoff of 50\% to classify sources as AGN/stars in this paper. For situations where we need a purer sample of AGN, we can use a higher cutoff, or we can use a lower cutoff if we want to lower the number of AGN that could be identified as stars.

\subsection{Variation of performance with properties of sources.}

\begin{figure*}
    \centering
    \subfloat[Change in accuracy with net counts. The accuracy increases with net counts, due to the decrease in the Poisson noise, and is $>$90\% for net counts $>$300.]{%
    \includegraphics[width=0.3\textwidth]{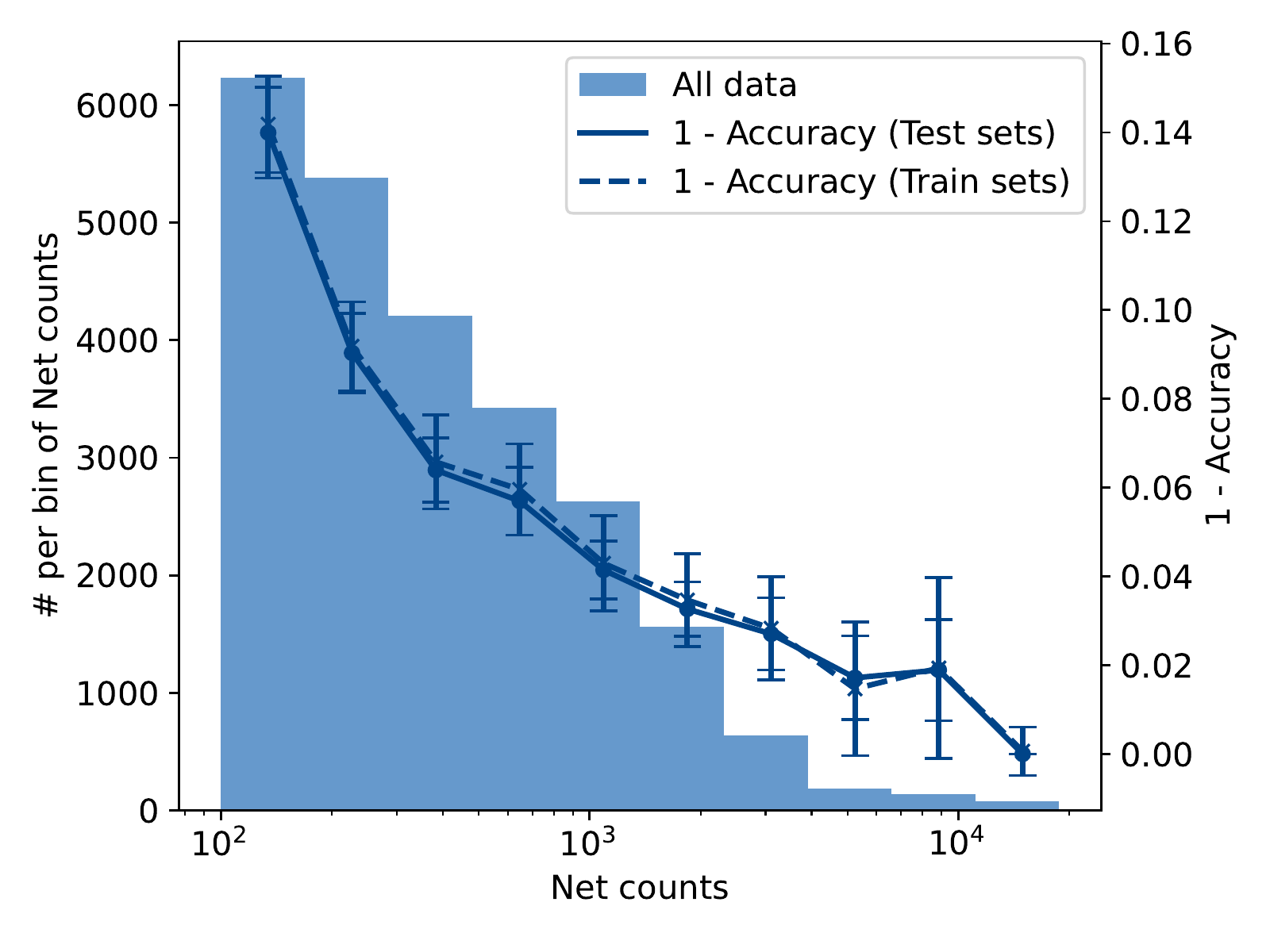}}\hfill
    \subfloat[Change in recall (fraction of AGN retrieved) and TNR (fraction of stars retrieved) with net counts in the spectra. These show similar trends as the accuracy.]{%
    \includegraphics[width=0.3\textwidth]{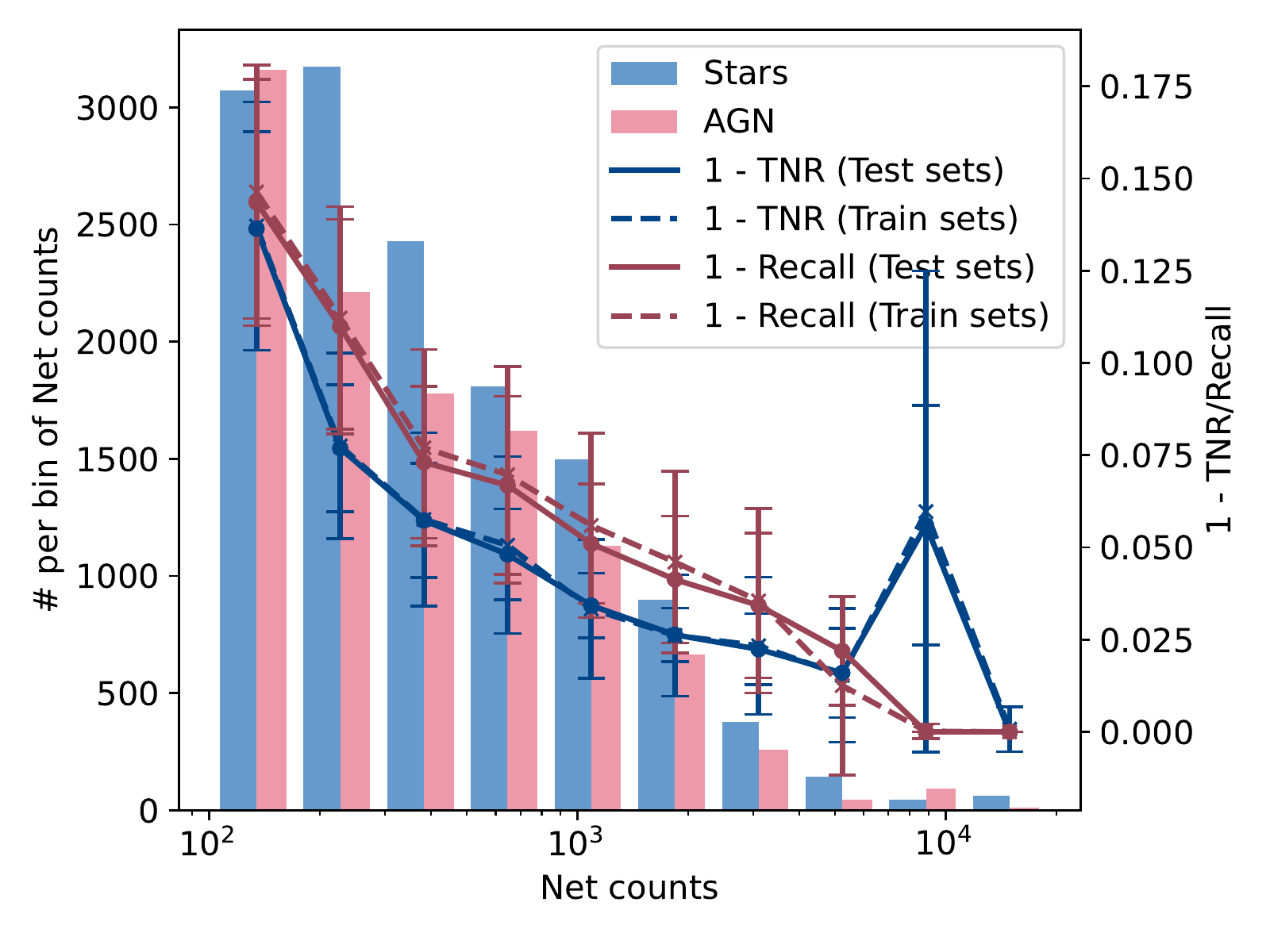}}\hfill
    \subfloat[Change in precision (fraction of true AGN among sources classified as AGN) and NPV (fraction of true stars among sources classified as stars) with net counts. Both show similar trends as the accuracy.]{%
    \includegraphics[width=0.3\textwidth]{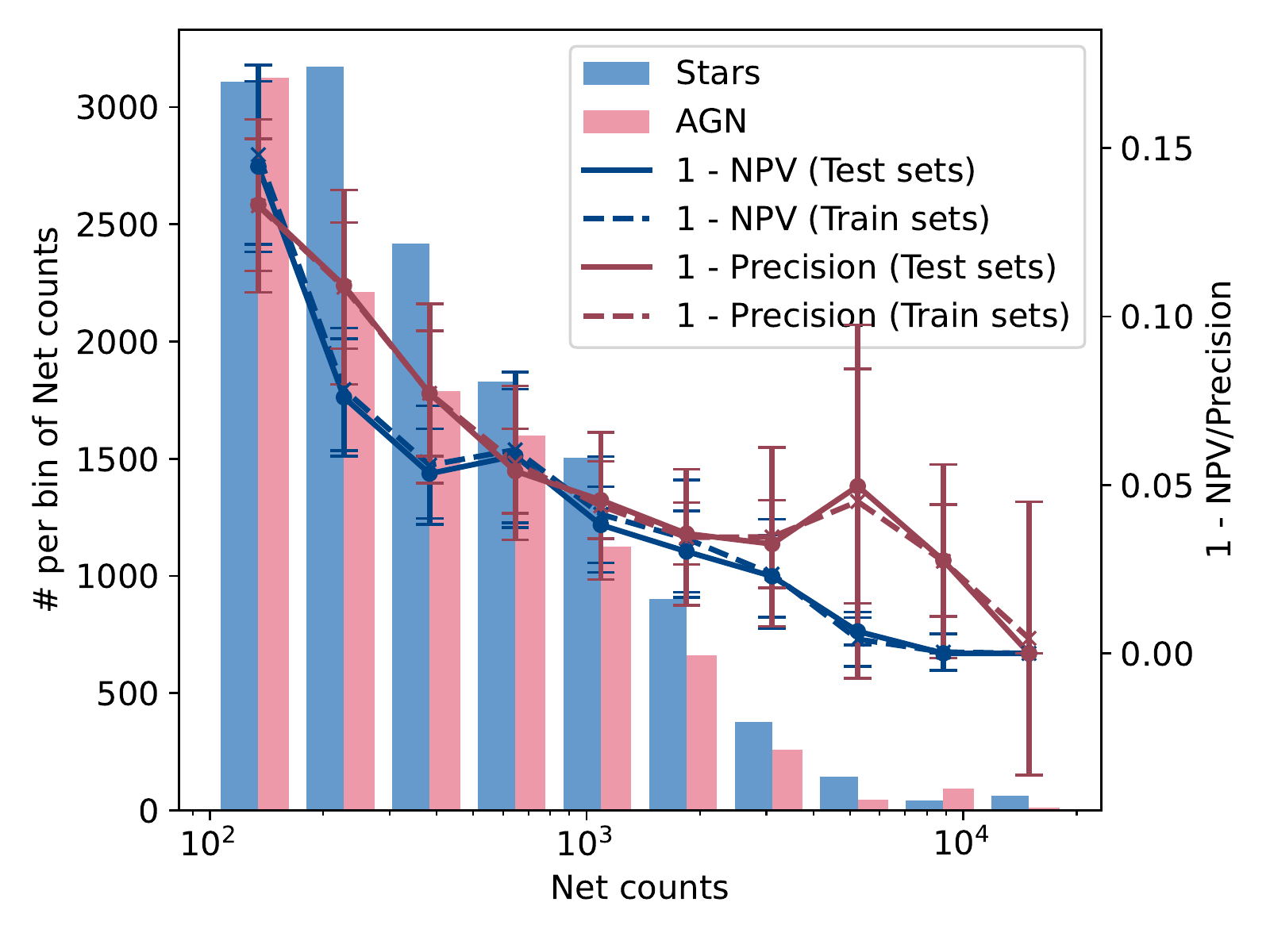}}
    
    \vspace{-10pt}
    
    \subfloat[Change in accuracy with the ratio of background-to-net counts. The accuracy decreases with increasing background contribution. Accuracy $>$ 90\%,  for background-to-net count ratio $<$ 0.08.]{%
    \includegraphics[width=0.3\textwidth]{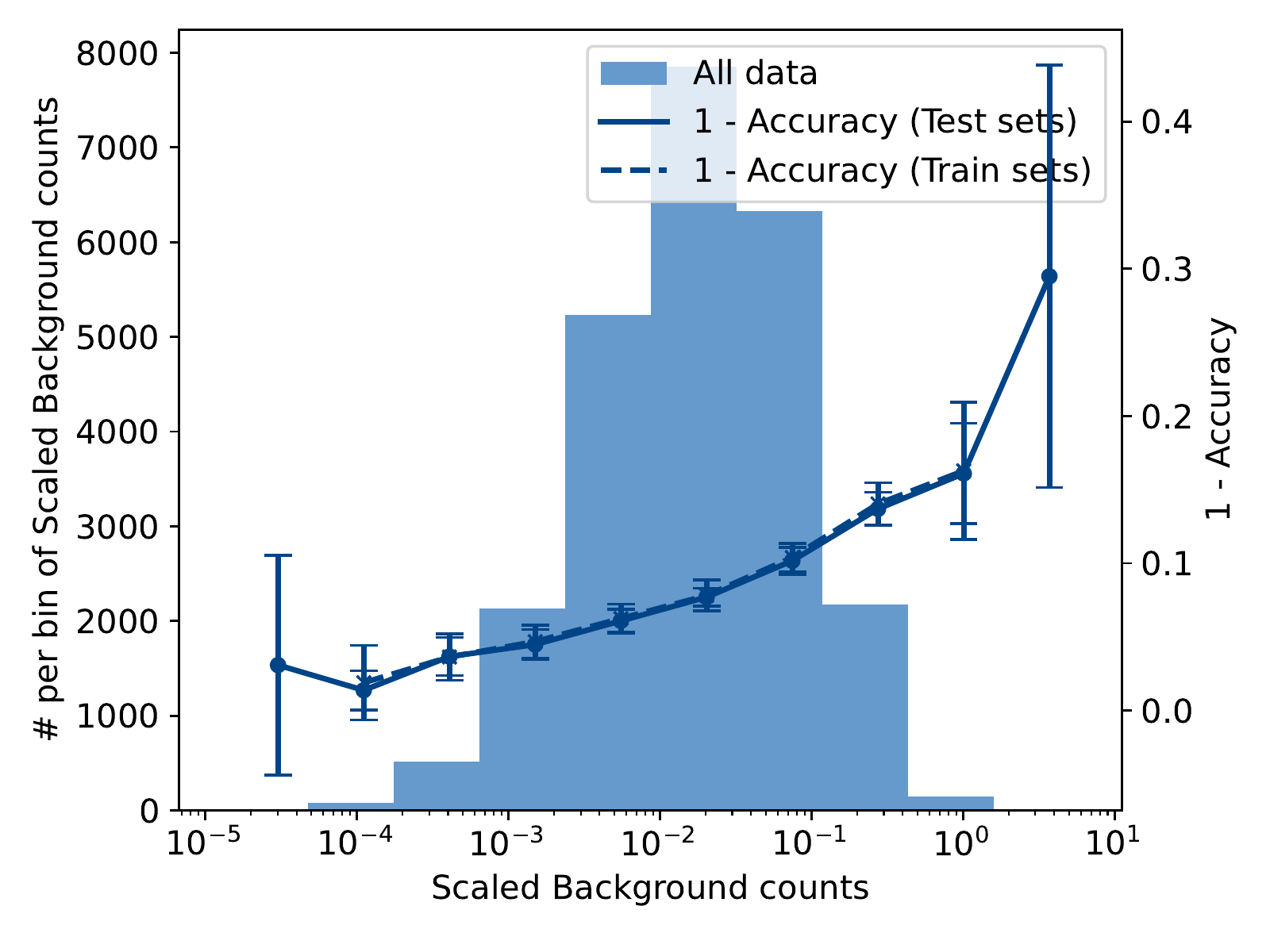}
    \label{subfig:bg_simulated}}\hfill%
    \subfloat[Change in recall and TNR with the changing background-to-net count ratio. These show similar trends as the accuracy.]{%
    \includegraphics[width=0.3\textwidth]{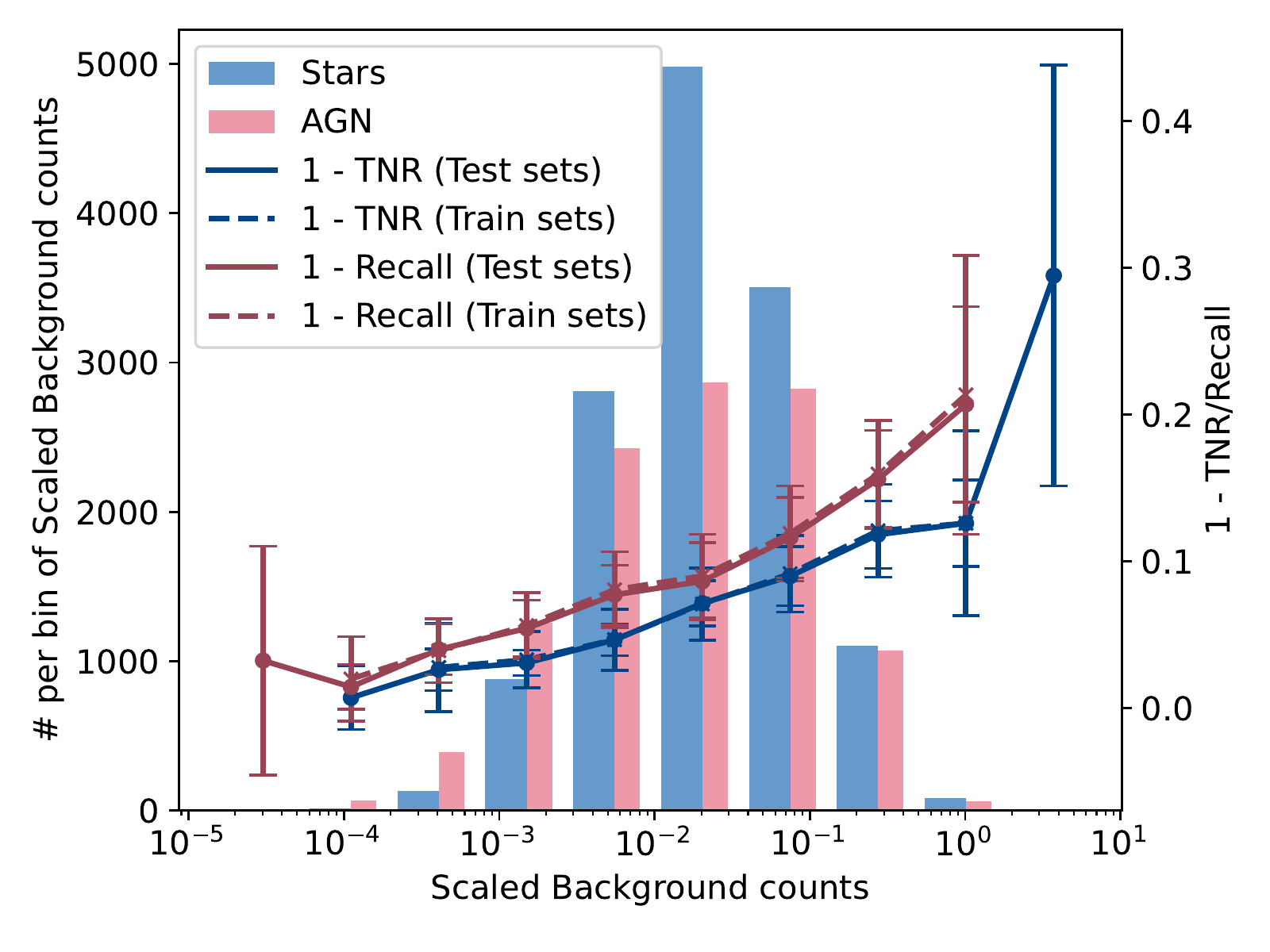}}\hfill
    \subfloat[Change in precision and NPV with the background-to-net count ratio. As background increases, more stars can be confused as AGN.]{%
    \includegraphics[width=0.3\textwidth]{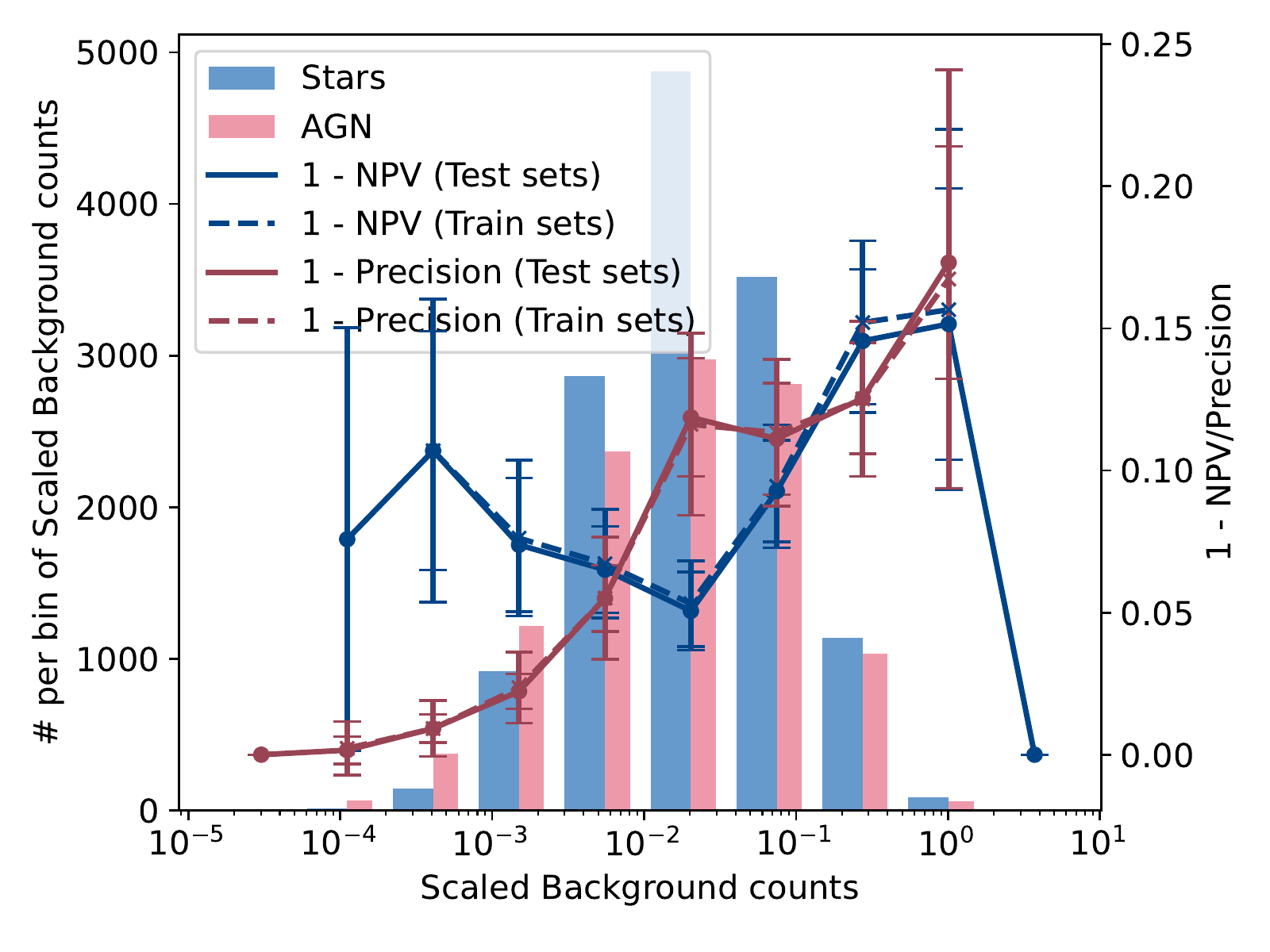}}
    
    \vspace{-10pt}
    
    \subfloat[Change in recall and TNR with  absorption column density. High absorption ($N_H > 10^{22}$ cm$^{-2}$) leads to very low flux $<$ 2 keV, leading to difficulties in the detection of Fe-L, Mg and Si lines, and thus to identifying stars. \label{subfig:perf_nh}]{%
    \includegraphics[width=0.3\textwidth]{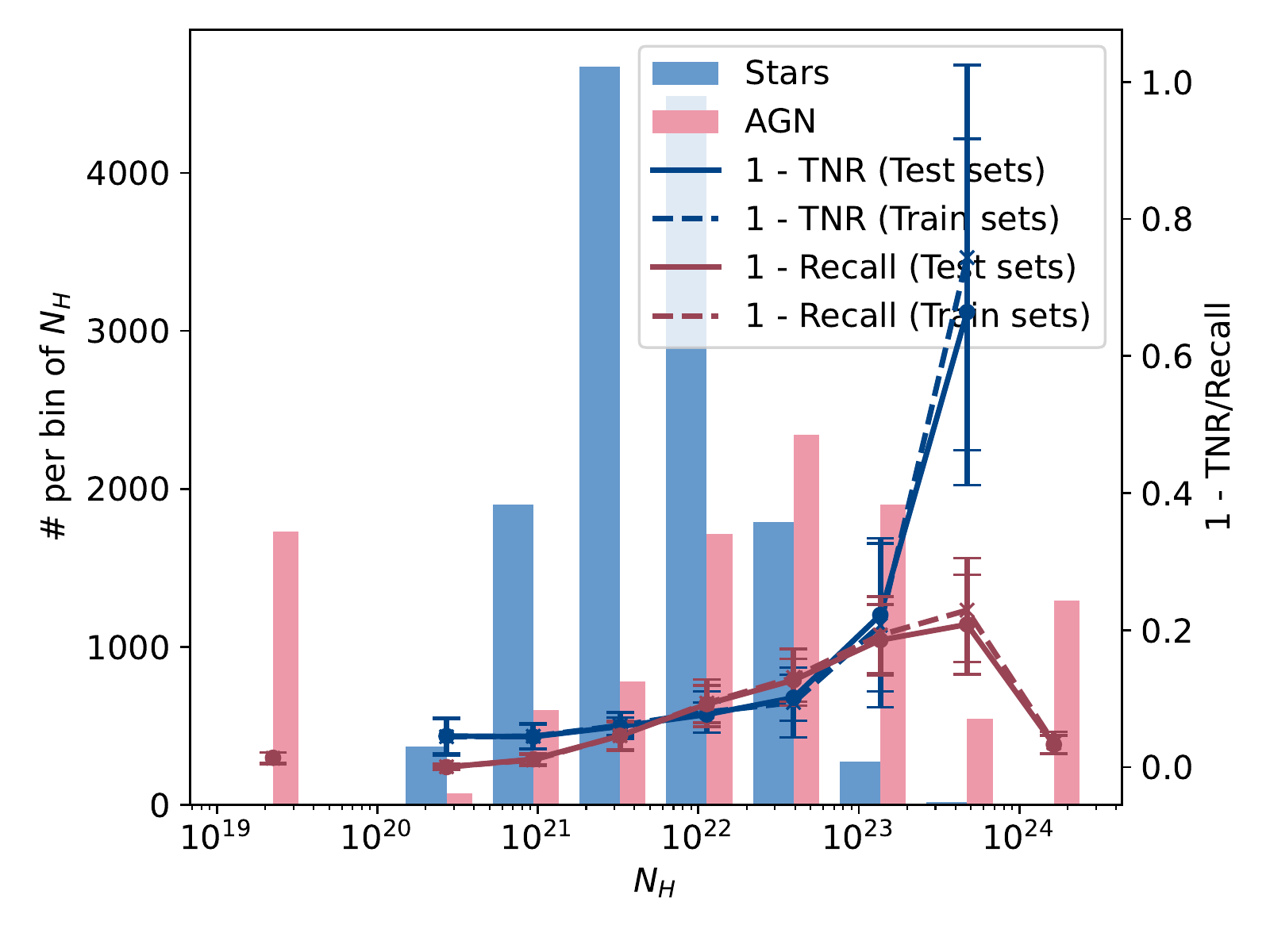}}\hfill
    \subfloat[Change in fraction of stars detected with the changing temperature of the thermal plasma used. At high temperatures, the atoms are fully ionized and the spectra are continuum dominated, making it harder to identify the spectra as stellar. \label{subfig:perf_kT}]{%
    \includegraphics[width=0.3\textwidth]{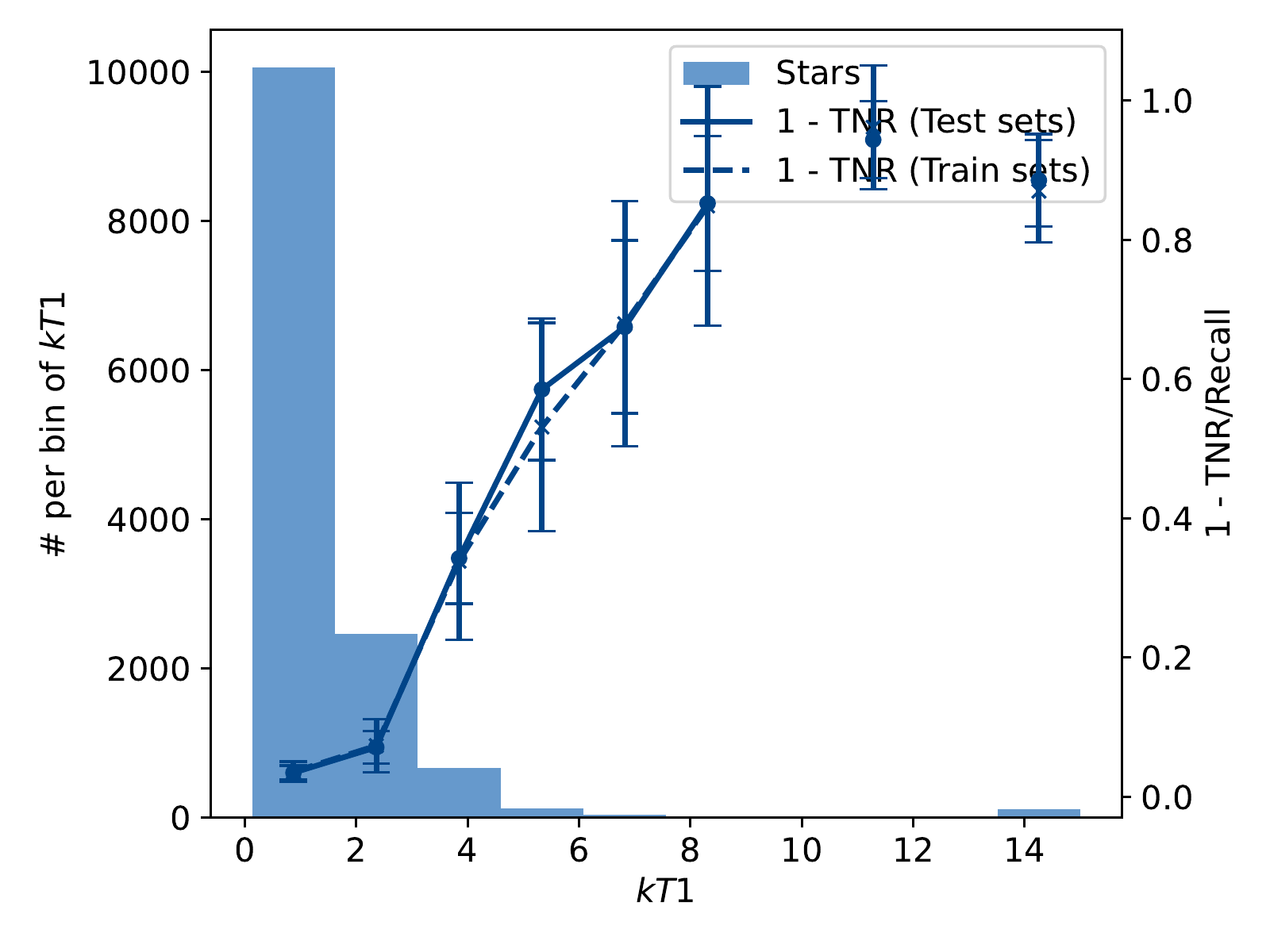}}\hfill
    \subfloat[Change in fraction of true AGN detected with the power-law index of AGN. Our model tends to select AGN with harder spectra. This is expected since most of the AGN have $\Gamma < 2$. \label{subfig:perf_gamma}]{%
    \includegraphics[width=0.3\textwidth]{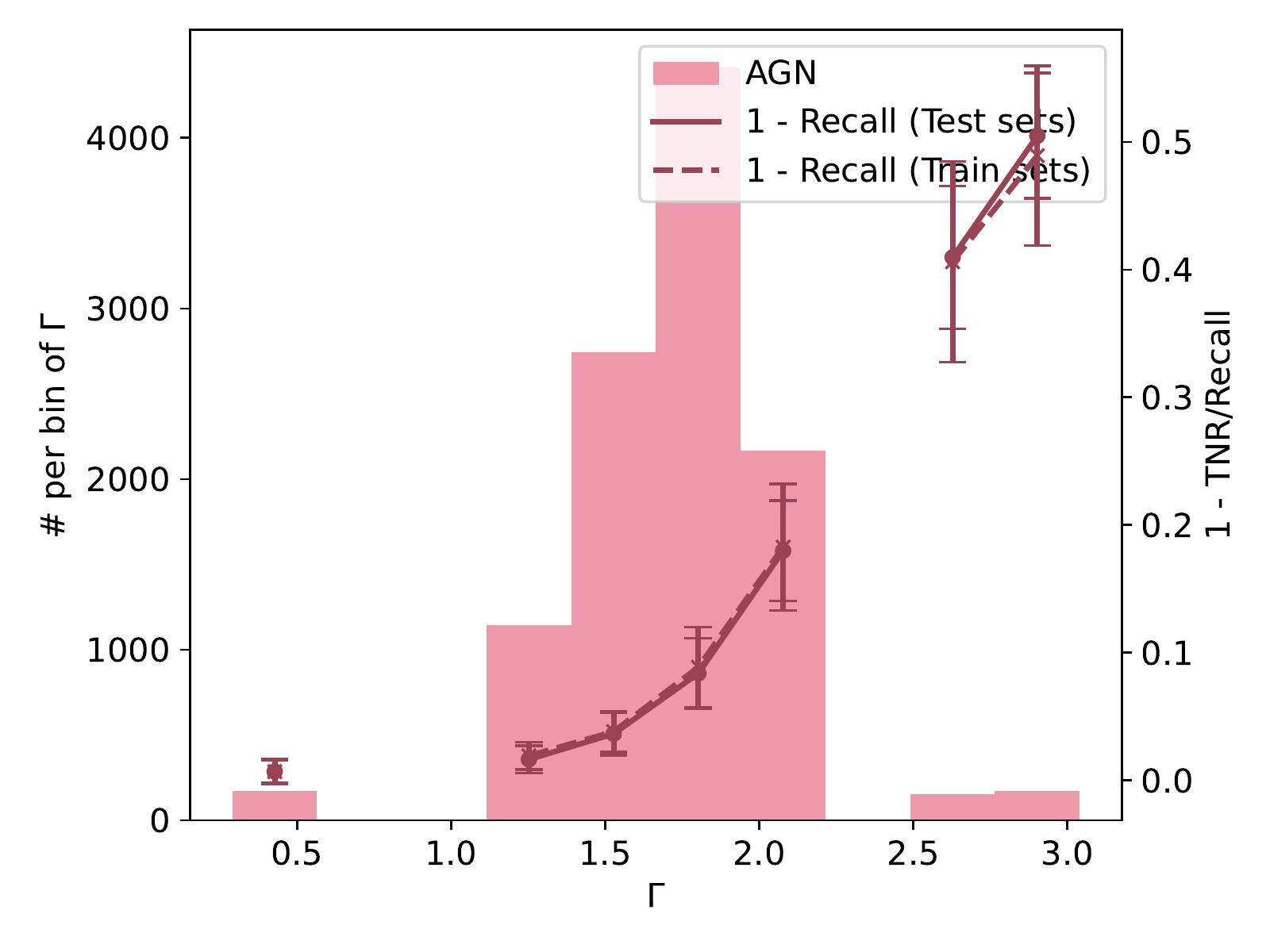}}
    
    \vspace{-20pt}
    
    \subfloat[Change in fraction of true AGN detected vs. the redshift. We notice that the recall doesn't change with redshift.]{%
    \includegraphics[width=0.3\textwidth]{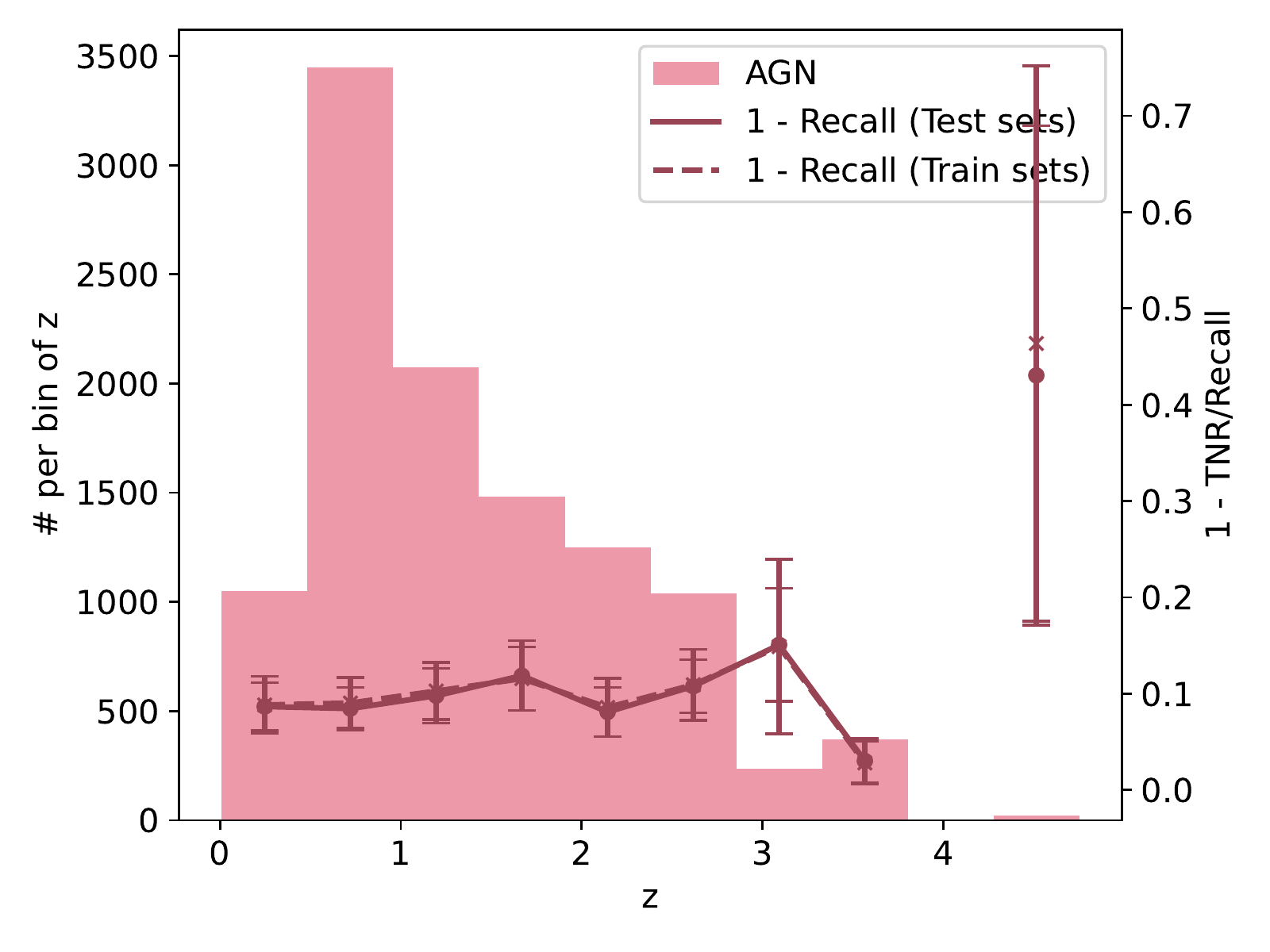}}\hfill
    \subfloat[Change in recall vs. spectral model used. Recall is better for `C-thick' and `Soft-C' models.]{%
    \includegraphics[width=0.3\textwidth]{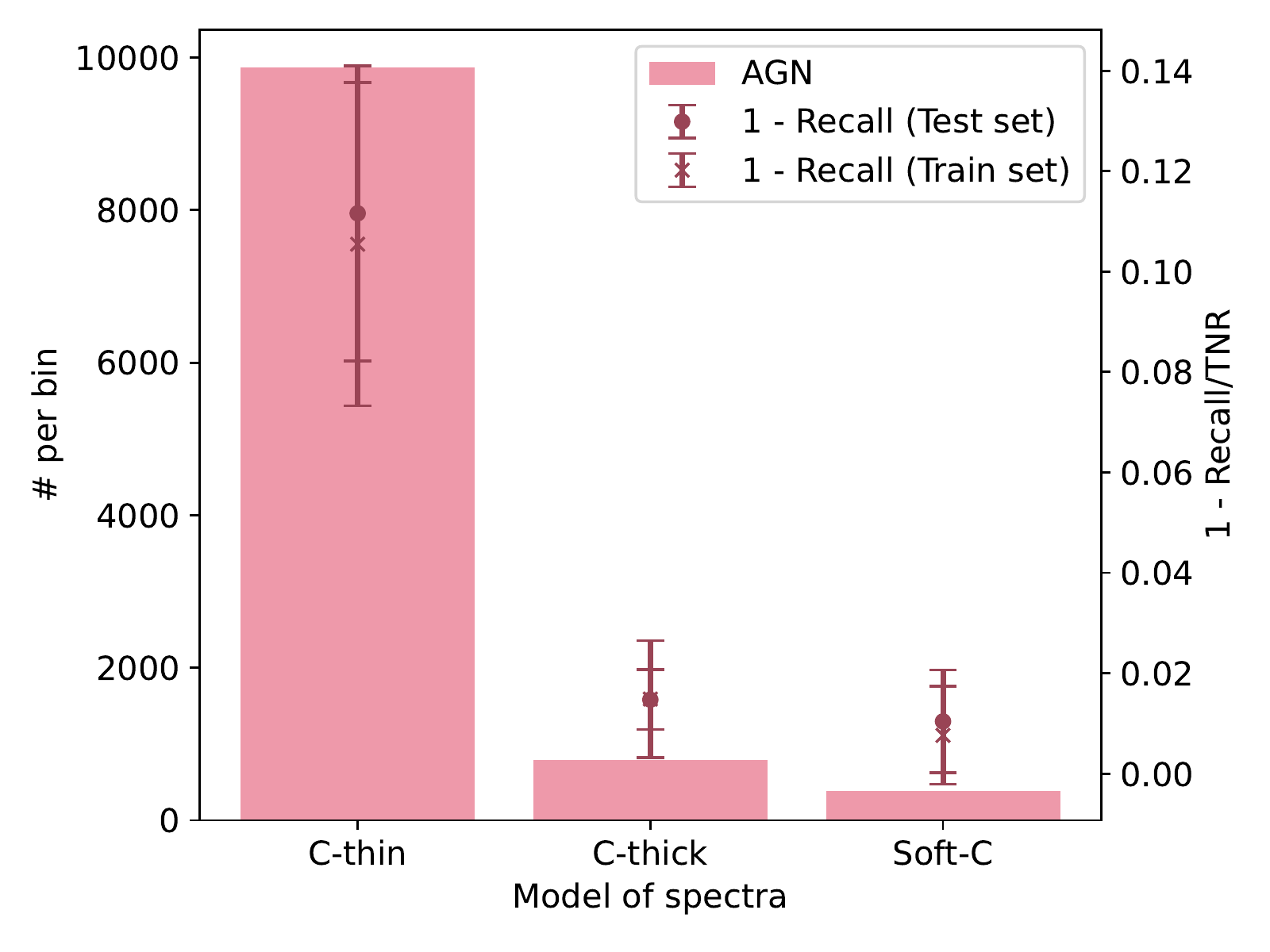}}\hfill
    \subfloat[Change in recall with equivalent width of the Fe-K line. Our model selects AGN with strong Fe-K emission preferentially. \label{subfig:perf_fe}]{%
    \includegraphics[width=0.3\textwidth]{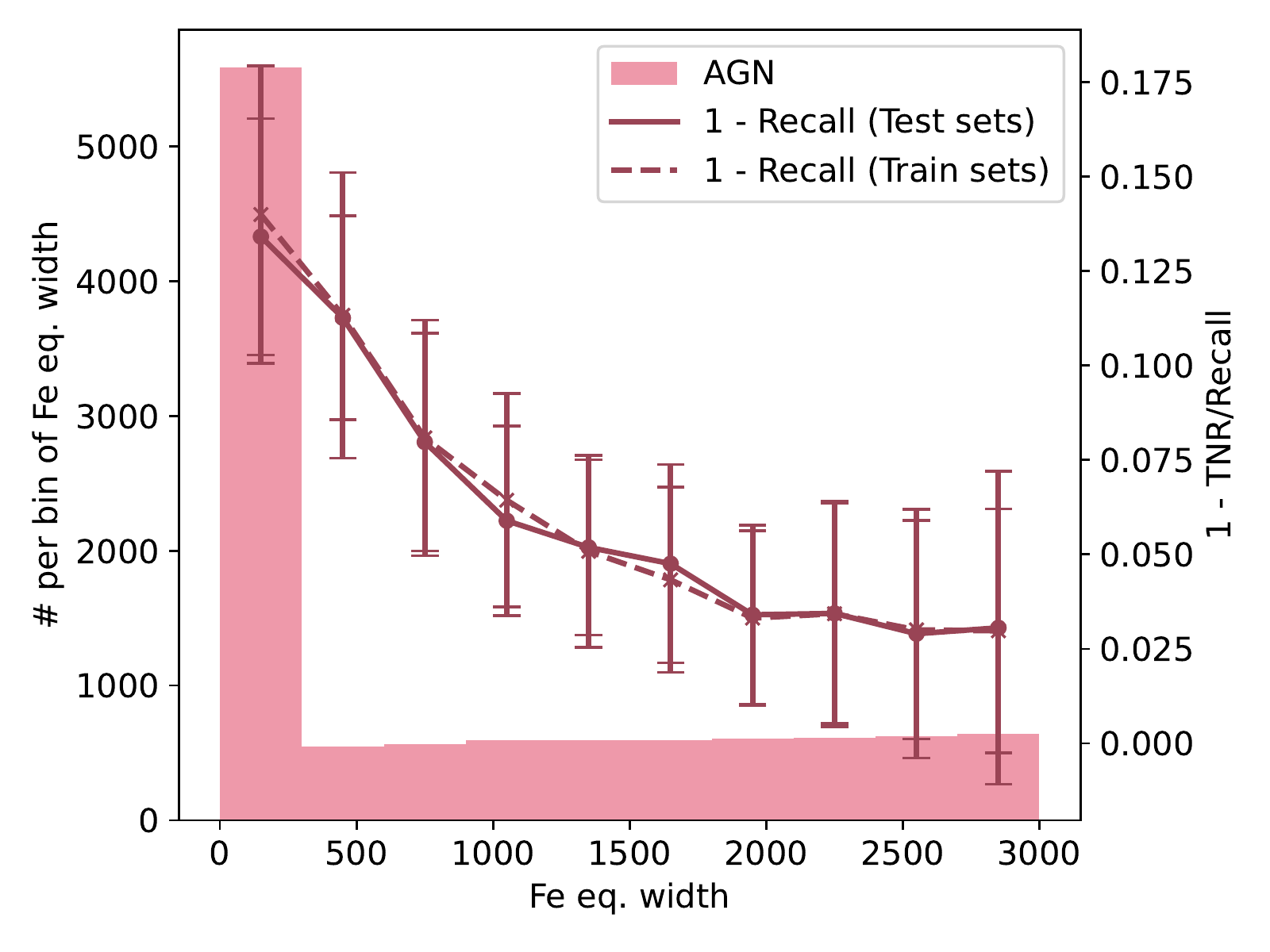}
    \label{subfig:fe_line}}
    \figcaption{Variation in performance of our ANN model for different properties of the sources. In each case the histogram represents the number of spectra in the given binning of the property. The y-axis on the right corresponds to the values of confusion matrix (1-Accuracy, 1-Recall/TNR etc.).
    \label{fig:reducedbg_prop}}
\end{figure*}

Since the model trained on the reduced background spectra, which represent typical {\it Chandra} sources, gives us the best results for both simulated and observed spectra, we use these to analyze if our classification model preferentially selects stars or AGN of  certain properties. We show the performance metrics with respect to different properties of stars and AGN in Fig.~\ref{fig:reducedbg_prop}.

We first analyze the performance with respect to the net counts and the background contribution (background-to-net count ratio). The performance improves as net counts increase.  As the net counts increase, the Poisson noise decreases, and hence the emission lines can be more easily identified. Similarly, as the background contribution decreases, the background noise can be distinguished from the true emission lines leading to better differentiation of stars and AGN.
In general, we get good results (better than 90\%) for sources with net counts greater than 
200 
and/or background-to-net count ratio smaller than 0.05.

We then study the performance with respect to the absorption column densities used for the AGN and star spectra. We see that the true negative rate (i.e., the fraction of stars detected correctly) decreases steeply for $N_H \gtrsim 10^{22}$ cm$^{-2}$. This is because high absorption column densities block the soft X-rays which are more important to the detection of the Ne, Fe-L, Mg and Si lines that identify the spectra as stars. We expect such high $N_H$ only for stars very close to the Galactic plane and those in dense nebulae. The detection of AGN
suffers only slightly with the increasing $N_H$. This is because as $N_H$ increases, the count rate at softer energies decreases, thus increasing the Poisson noise, which could be confused as lines. At $N_H \gtrsim 10^{24}$, the performance improves because of the reprocessed X-ray emission from the Compton thick model we use in this range.

The detection of stars also decreases with the increase in the plasma temperature used to model 
X-ray emission. For high temperatures ($kT > 2$ keV), the continuum emission 
dominates while lines fade,
making it 
harder to detect the line emission. However, we expect that most of the stars have $kT < 2$ keV (e.g. \citealt{Gudel_2004}). Higher temperatures are mostly due to improper fitting of highly absorbed sources.

The detection of AGN varies with the hardness of the X-ray emission. Our algorithm seems to detect harder AGN at a higher rate. This is expected since most CDFS AGN have $\Gamma = 1.75 \pm 0.02$. We notice that the performance is better on Compton thick AGN, and AGN with an additional soft component, as compared to Compton thin AGN. The detection fraction of AGN increases slightly with the redshift of AGN. The decrease in recall at $z > 4$ may be  due to very few data points in these bins.

We notice that our model detects AGN with an Fe-K line more efficiently than the ones without Fe-K emission. This can be deduced through the increasing recall as the equivalent width of the Fe-K emission line increases. The change in recall with the position of the Fe-K emission line seems to be consistent within the error-bars.

\subsection{Performance at low counts}

\begin{figure}
    \centering
    \includegraphics[width=0.8\columnwidth]{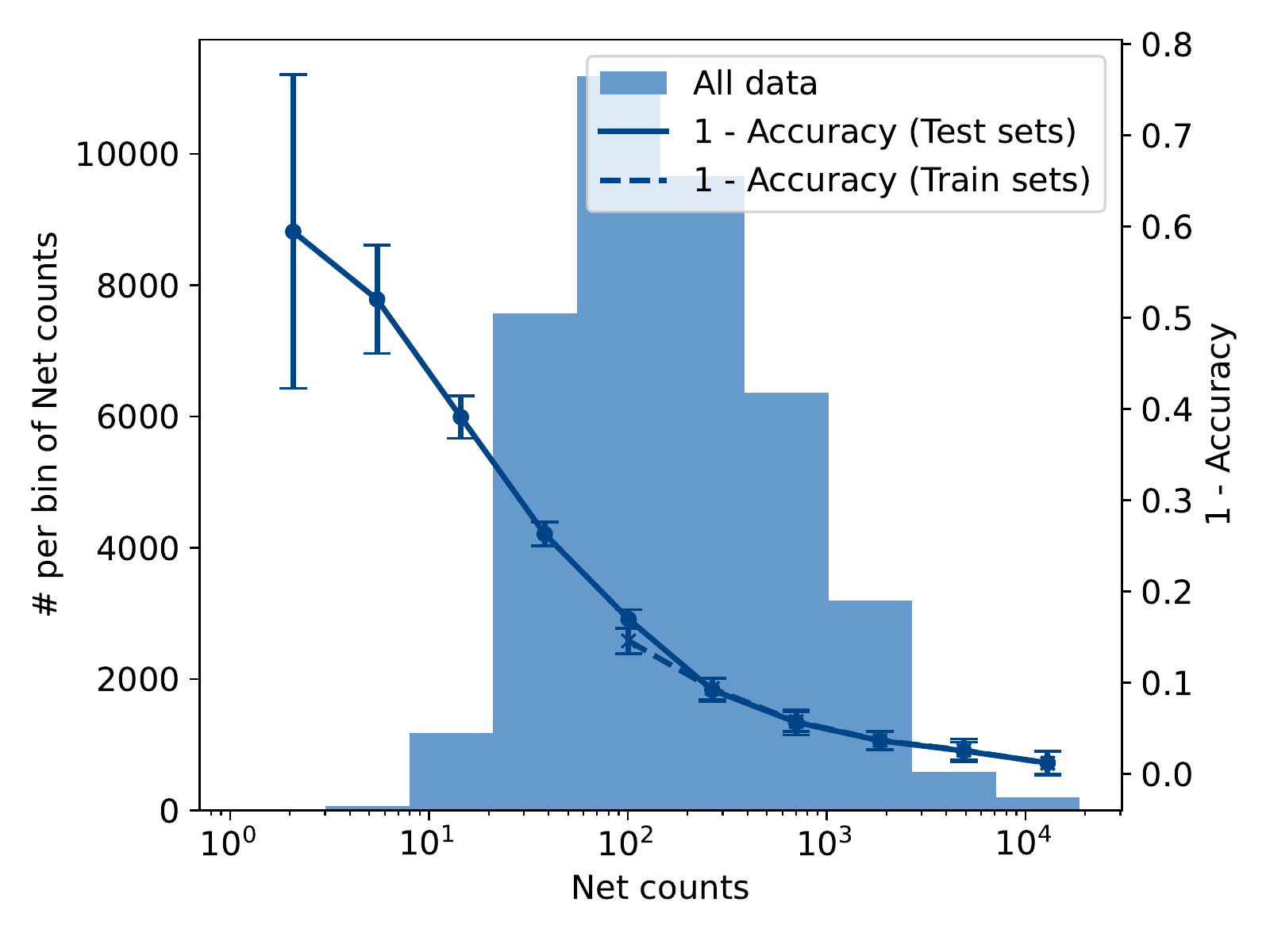}
    \figcaption{Change in the accuracy with net counts for simulated spectra with net counts $>$ 1.0. From the figure, we see that our algorithms does not perform well on sources with ~1--10 net counts (accuracy marginally better than 50\%). The accuracy is $\gtrsim70\%$ for sources with net counts $\gtrsim 20$ and $\gtrsim 80\%$ for net counts $\gtrsim 70$. \label{fig:lowcount_accuracy}}
\end{figure}

We also test the performance of the ML algorithms on sources with reduced background and net counts less than 100. For this purpose, we select all sources with net counts $>$ 1 in our simulated and observed dataset to train and test our performance. The overall accuracy is 86\% on the simulated sample and 82\% on the observed spectra. The recall, TNR, precision and NPV have similar values as the accuracy. We list the values of the performance metrics for the observed and simulated data in Table~\ref{table:performance_metrics}. We also show the variation in accuracy with net counts in Fig.~\ref{fig:lowcount_accuracy}. The accuracy is very poor ($\lesssim 50\%$) for net counts $\lesssim 8$. The accuracy is $\sim 60\%$, $\sim 74\%$, and $\sim 83\%$ for net counts $\sim 8-21$, $\sim 21-55$, and $\sim 55-146$, respectively.
For these low count sources, including the information from their positions in the sky, detection of multiwavelength counterparts and their properties along with the classification probabilities from X-ray spectra alone will help in improving the overall classification accuracy.

\subsection{Weights applied by ANN on energy bins}

\begin{figure}
    \centering
    \includegraphics[width=0.9\columnwidth]{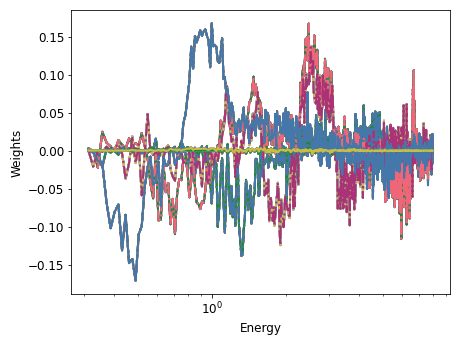}
    \figcaption{Weights to each energy channel in the first layer of a trained ANN model. We notice that the model assigns larger weights (magnitude only) to the spectra around the position of the Ne, Fe, Mg, and Si lines. The different curves correspond to the weights of ten nodes in our ANN. \label{fig:reducedbg_weights}}
\end{figure}

Since we use a simple ANN model with only one neural network layer, the weights assigned to the input layer can be interpreted as the relative importance of the energy channel to classifying the source as a star or an AGN. We show the weights assigned to each energy channel in Fig. \ref{fig:reducedbg_weights}. From this figure, we see that the X-ray spectra in the energy range $\sim$ 0.8--1.4 keV are strongly selected, indicating the preference given to the Ne, Fe, Mg, and Si lines. The bumps in the spectra at energy bins $\gtrsim$ 2 keV are probably due to the model searching for the redshifted Fe-K line. We notice that training the ANN model using only the spectra without an Fe-K line reduces the weights in this regime. Thus, our model indeed tries to identify emission lines in the spectra to classify the source, and does not use the hardness ratio of the spectra alone.

\subsection{Applying trained ANN model to observed data}
\begin{figure*}
    \centering
    \subfloat[Change in accuracy with the net counts in the spectra. The accuracy increases with net counts, due to the decrease in Poisson noise and in general, a smaller background contribution.]{
    \includegraphics[width=0.3\textwidth]{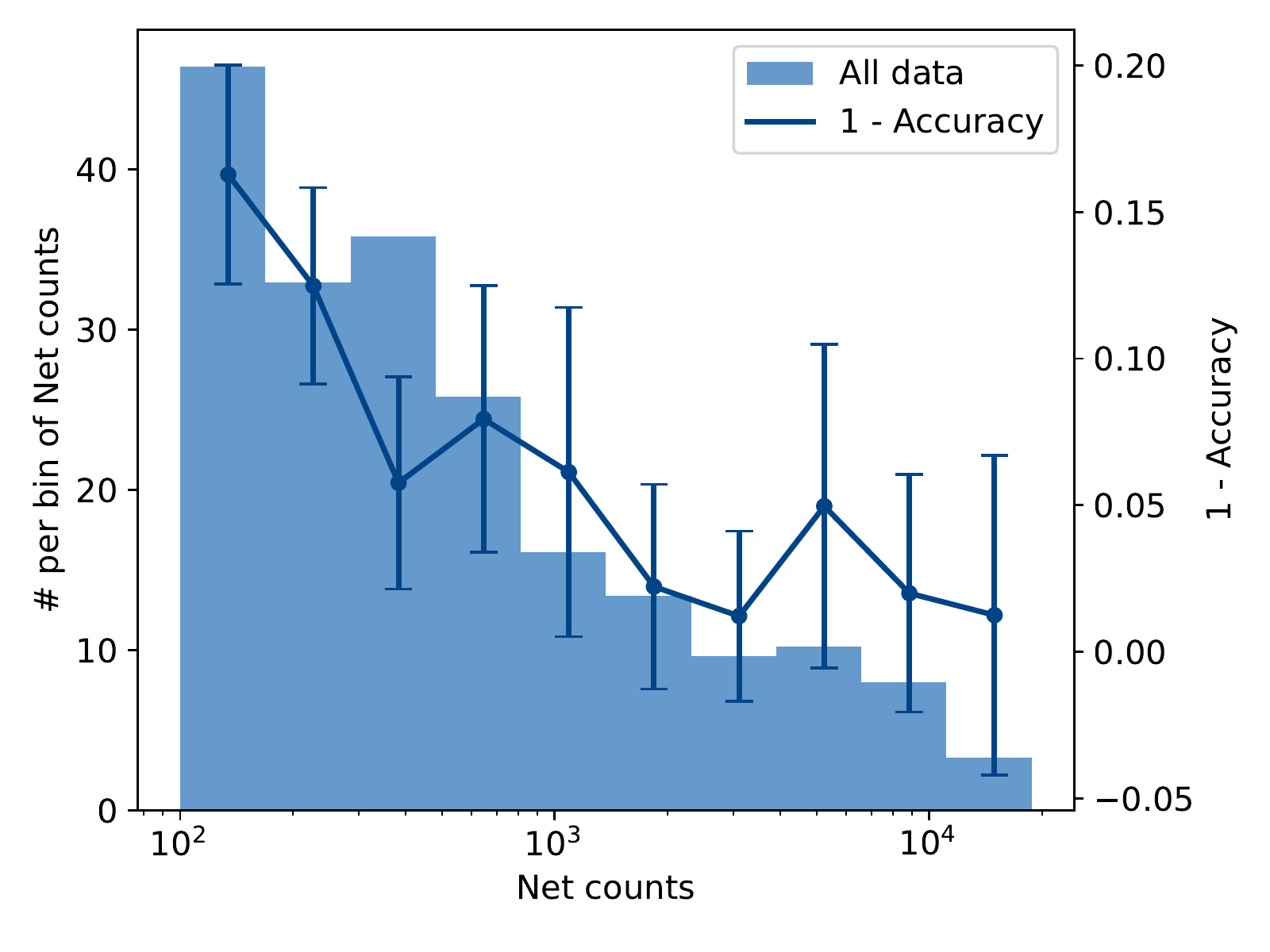}
    }\hfill
    \subfloat[Change in recall (fraction of AGN retrieved) and TNR (fraction of stars retrieved) with the net counts in the spectra. These show similar trends as the accuracy.]{
    \includegraphics[width=0.3\textwidth]{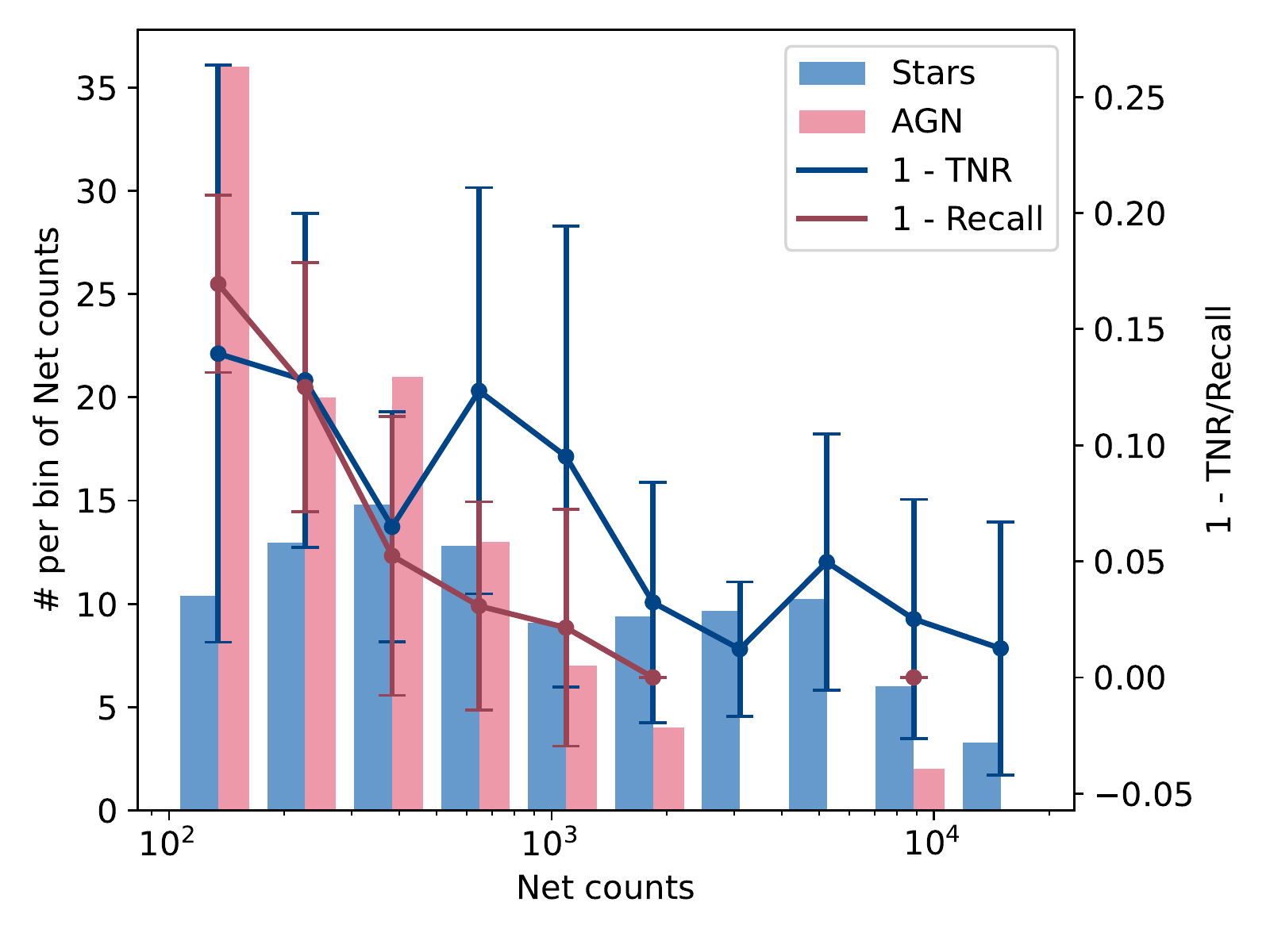}
    }\hfill
    \subfloat[Change in precision (fraction of true AGN among sources classified as AGN) and NPV (fraction of true stars among sources classified as stars) with the net counts. Both show similar trends as the accuracy.]{
    \includegraphics[width=0.3\textwidth]{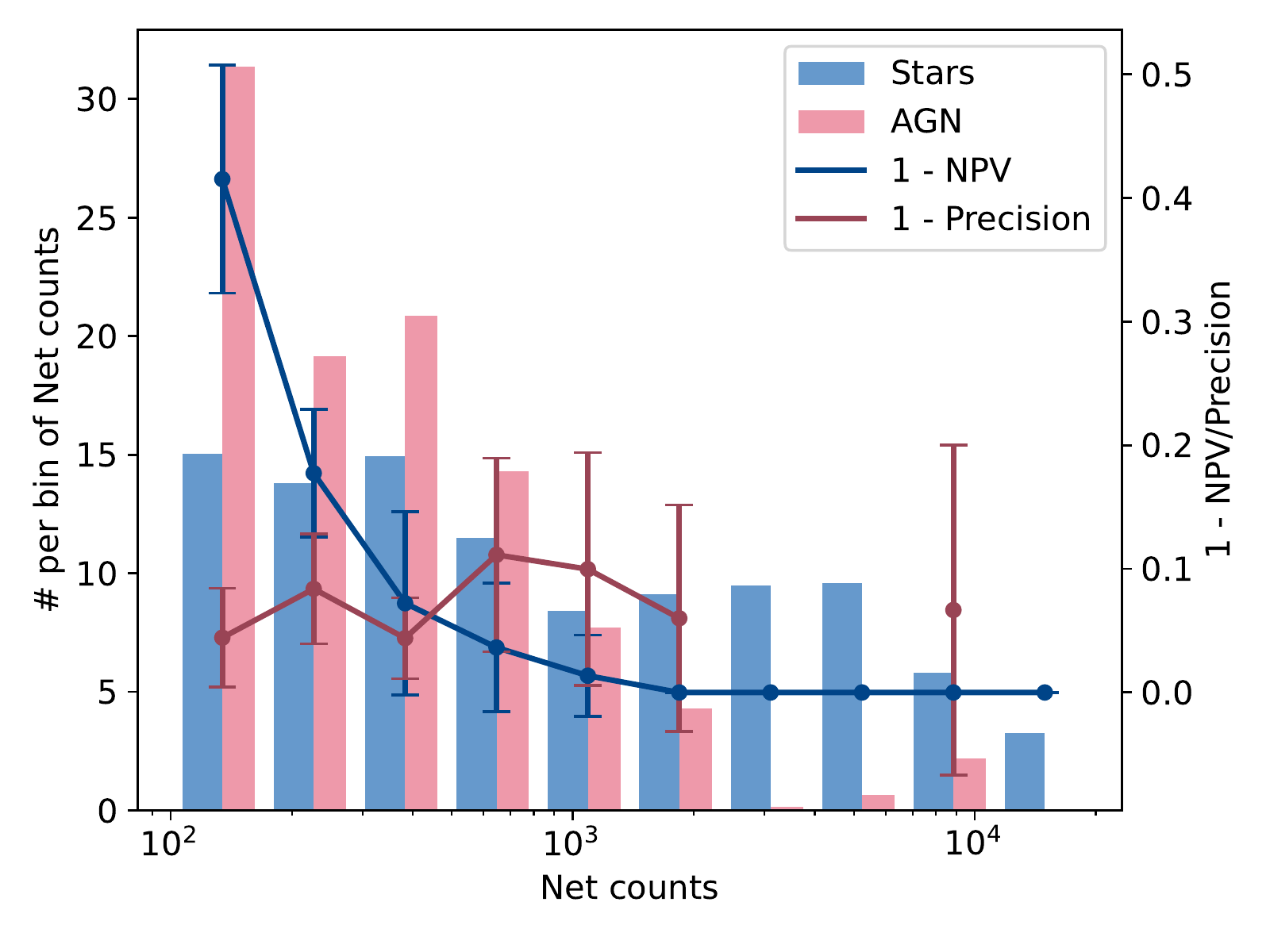}}

    \subfloat[Change in accuracy with the ratio of background-to-net counts. The accuracy decreases with the increase in the background contribution.]{
    \includegraphics[width=0.3\textwidth]{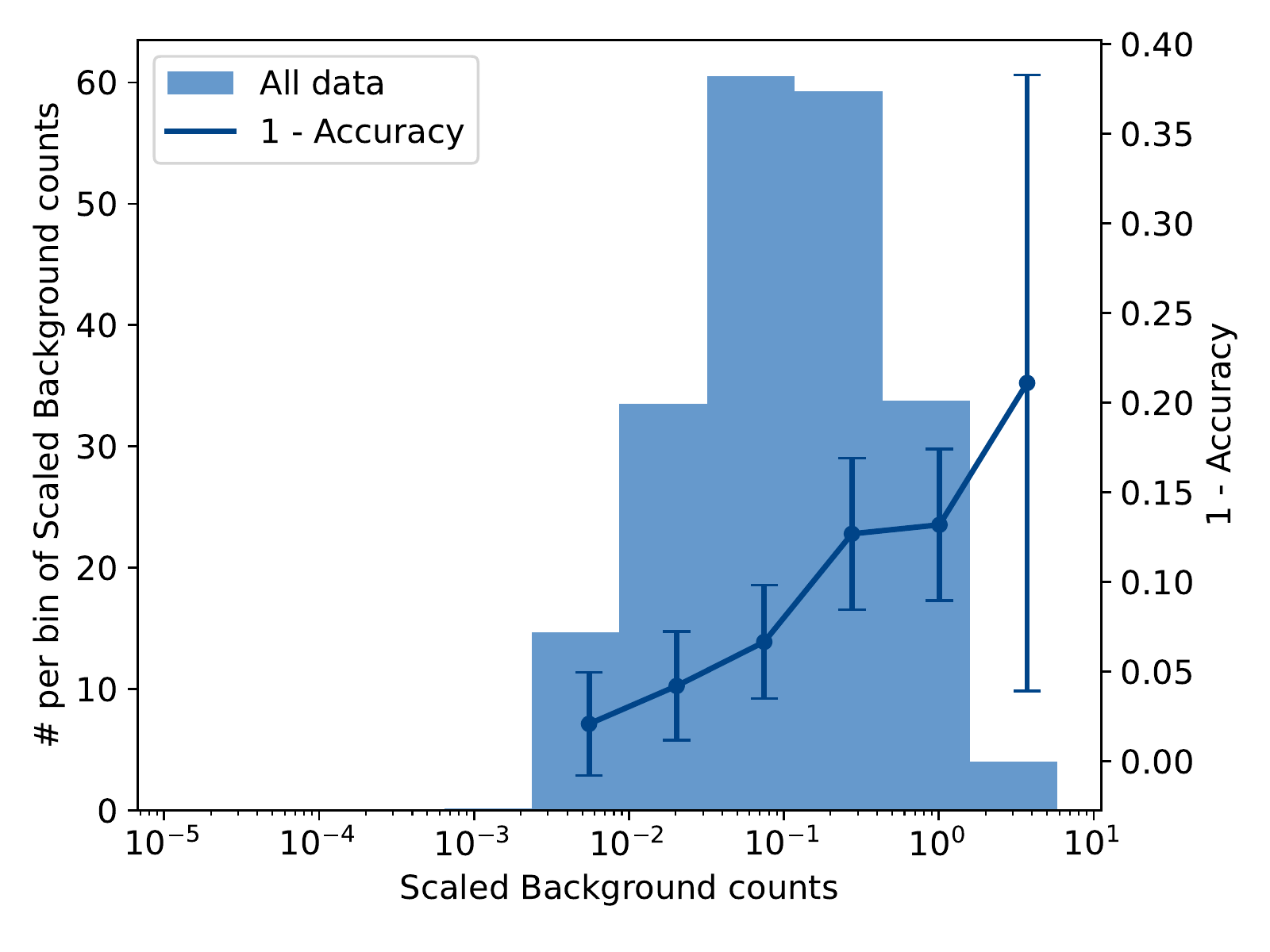}
    \label{subfig:bg}} \hfill
    \subfloat[Change in recall and TNR with the background-to-net count ratio. These show similar trends as the accuracy.]{
    \includegraphics[width=0.3\textwidth]{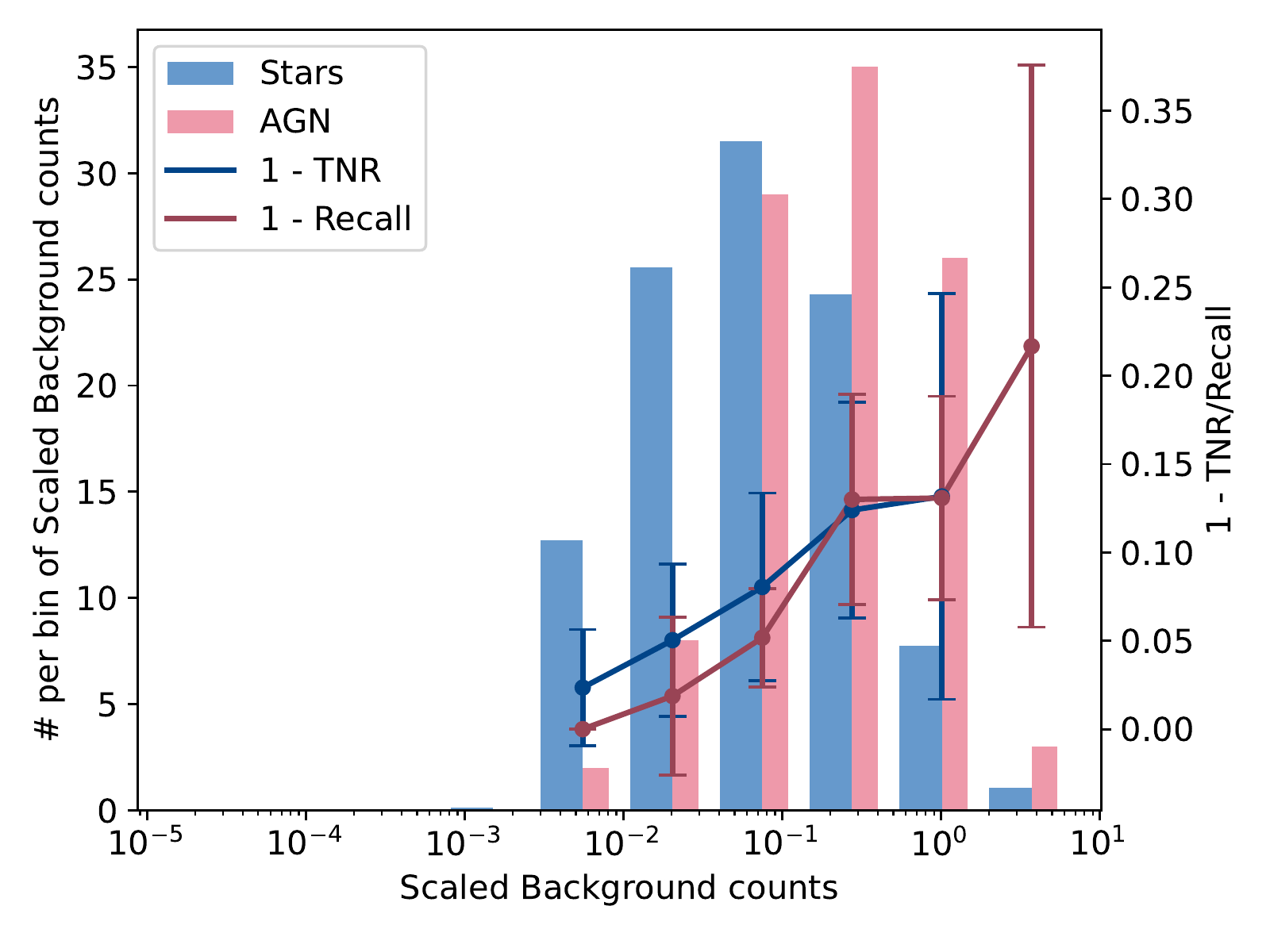}}\hfill
    \subfloat[Change in precision and NPV with the background-to-net count ratio. As background increases, more AGN can be confused as stars.]{
    \includegraphics[width=0.3\textwidth]{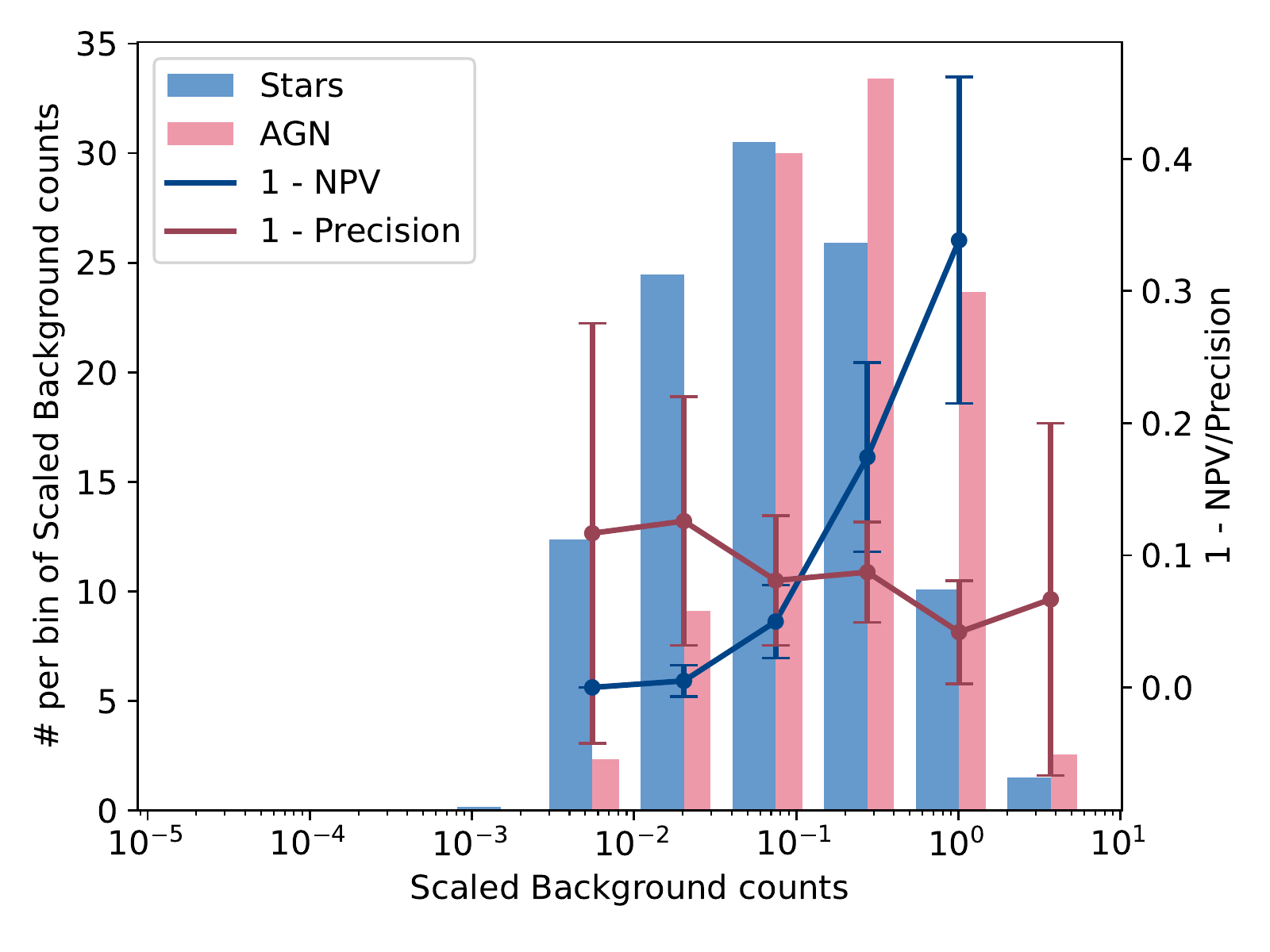}}
    \figcaption{Performance of the trained ML model with net counts, and with  background contribution, on the observed spectra. \label{fig:reducedbg_obs}}
\end{figure*}

From the values in Table~\ref{table:performance_metrics}, we see that while the ANN models trained on the data simulated with the observed background or with the reduced background yield only slightly poorer performance on the observed data as compared to that on the simulated spectra, the model trained on spectra without background performs very poorly on the observed spectra. Comparing the performance of these models with respect to the net counts and the background contribution of the observed spectra, shown in Fig.~\ref{fig:reducedbg_obs}, gives us more insight into this behavior.

We see that the performance of the ANN model trained on the spectra without background contribution degrades steeply at low net counts ($\lesssim 1000$) or when the ratio of background-to-net counts is high ($\gtrsim 0.02$). This is because the contribution of background is high in this regime. This background noise can be confused with emission lines, leading to many AGN being classified as stars. Looking at the performance of the ML models trained with background (observed and/or reduced background) with respect to the background-to-net count ratio also allows us to 
understand 
the slightly poorer overall accuracy on the observed data. Comparing Fig.~\ref{subfig:bg} and Fig.~\ref{subfig:bg_simulated}, we 
see that the the observed AGN have much higher background than the simulated AGN. Both plots show that the accuracy is greater than 90\% for background counts $\lesssim 5\%$ of the net counts from the source.

\subsection{Using ML to detect extragalactic sources in COUP}

Using the ANN model trained on the data simulated with the reduced background, we try to identify the extragalactic sources detected in the COUP catalog. Among the sources marked as extragalactic in \citet{Getman_2005b}, 63 sources have net counts greater than 100. Our ANN model is able to identify $\sim$40 of these 63 extragalactic sources. This corresponds to a recall of $\sim$63\% which is much lower than the recall on the CDFS spectra. This poorer performance is because most AGN in the COUP catalog are heavily absorbed and have a high background (60\% of extragalactic sources have background-to-net count ratio $\gtrsim$0.3, and 84\% of extragalactic sources have background-to-net count ratio $\gtrsim$0.1).

\subsection{Changing response of Chandra}
\label{sec:changing_response}
The response function of {\it Chandra} has been changing over the years, 
as the effective area at soft energies degrades. 
Thus the same star could have different emission line strengths  (especially $<$ 2 keV) across the years. We generate 10,000 artificial spectra of stars and AGN each, with the ACIS-I aimpoint responses  from the {\it Chandra} Cycle 25 call for proposals to check if our model could be applied if similar {\it Chandra}  observations were taken in 2023. We find that our model (trained on the COUP responses) can can classify the simulated stars having $>$100 net counts with a accuracy of ($84 \pm 3$)\%, retrieve ($77 \pm 7$)\% of AGN (recall), and identify ($90 \pm 3$)\% of the stars (TNR) i.e. the performance on identifying stars is only slightly decreased, but that for AGN decreases a lot. We will explore the training of ML models with the changing response of {\it Chandra} in our future works.

\section{Discussions}
\label{sec:discussions}
\subsection{Incorrect classification of Sources}
\begin{figure}
    \centering
    \includegraphics[width=\columnwidth]{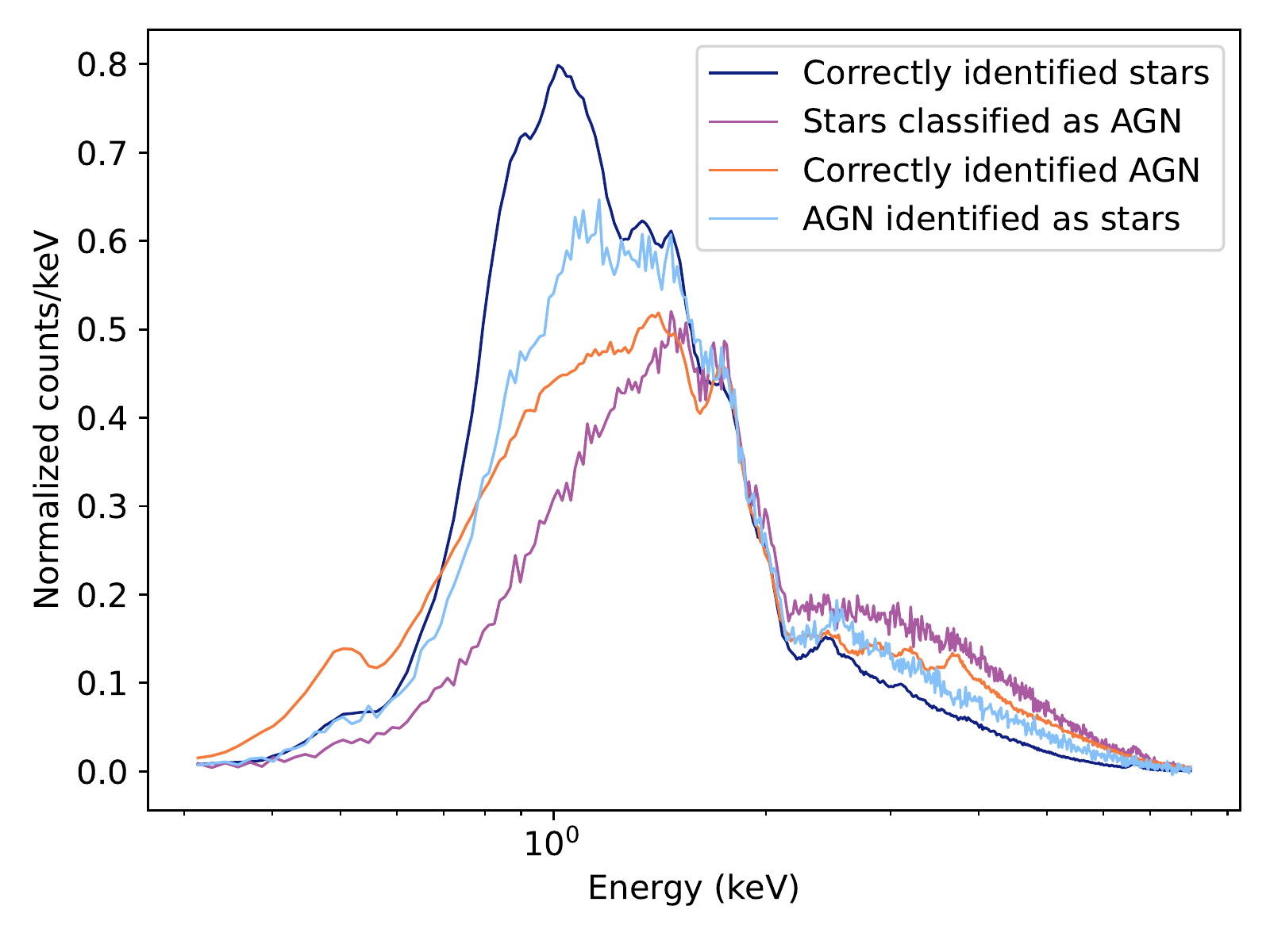}
    \figcaption{Mean normalized spectra of correctly identified stars and AGN, compared to those that are not classified properly. Notice that the stars that have been identified as AGN do not show strong emission lines, are heavily absorbed, and are harder than the stars that have been correctly identified. \label{fig:meanspec_correct_wrong_classified}}
\end{figure}

From our results, we notice that the trained ANN model uses a combination of emission-line strengths (interpreted from Figs.~\ref{subfig:perf_nh}, \ref{subfig:perf_kT} \& \ref{fig:reducedbg_weights}) and hardness of the spectra (Fig.~\ref{subfig:perf_gamma}). Fig.~\ref{fig:meanspec_correct_wrong_classified} shows the mean spectra of stars and AGN that have been correctly/incorrectly identified. From the figure, we notice that incorrectly classified stars show stronger absorption, show fainter emission lines, and/or are harder than correctly identified stars, i.e., their spectra are not line-dominated. 
AGN that have been identified as stars seem to show higher absorption and/or are softer than correctly identified AGN.

We further explore the mis-classification of softer AGN. The poor recall of AGN for $\Gamma > 2.0$ is because these AGN are affected more by increasing $N_H$, i.e. for $N_H \lesssim 10^{21}$ cm$^{-2}$, all AGN have similar recall ($\sim$98--99\%), but at $N_H \sim 10^{22}$ cm$^{-2}$, the recall for AGN with $\Gamma > 2.0$ decreases to $\sim$72\% while the recall for hard AGN with $\Gamma < 2.0$ is $\sim$96\%. This can be explained as classification of AGN with hardness similar to stars will mainly be based on identification of soft X-ray emission lines, which can be confusing at high absorption. Our sample of soft AGN with $\Gamma > 2.0$ need a net count $\gtrsim 3000$ photons, or a strong Fe-line with equivalent width $\gtrsim 1500$ eV, to be detected with a recall of $> 90\%$. The improvement in the detection of AGN with the increasing fraction of continuum soft-emission also explains why our performance is better on AGN with soft component and Compton thick AGN.

\subsection{Applications}
Based on our results, our algorithm works very well on sources with sufficient X-ray photons to enable the detection of Fe-L,  Mg and Si lines against the background and Poisson noise. Based on the distribution of weights (Fig.~\ref{fig:reducedbg_bgcounts}) and the decreased accuracy of classification on stars with $kT > 2$ keV, we see that the trained ANN picks out the energies corresponding to emission lines and identifies line-dominated X-ray spectra. Thus, ANNs could provide an efficient and practical approach to classify serendipitous X-ray sources.

X-ray emission from 
most isolated NSs
in our Galaxy can be fit with a power-law model with photon index $\Gamma \approx$ 1--2, a blackbody of temperature 0.05--0.3 keV, or a combination of these, and do not show any lines \citep{Pavlov_2002}. 
Those which are dominated by the power-law will then be similar to AGN, without the Fe-K lines. 
Based on Fig.~\ref{subfig:fe_line}, we expect that our model should be able to identify such NSs from chromospherically active stars in our Galaxy with a good recall (80--90\% if they follow a similar distribution of net counts, column density, and  background contribution). 
Our results indicate that the additional presence of softer components (e.g. a blackbody) does not affect our performance. 
Many 
neutron star X-ray spectra are dominated by a blackbody-like 
 thermal component. Discriminating between blackbody-like spectra and soft thermal plasma spectra (e.g. APEC models with temperatures $<$2 keV) will be even easier, as these low-temperature plasma spectra are even more dominated by lines than the harder spectra. 

X-ray binaries have spectra similar to those of AGN in the energy range of 0.1-10 keV. Thus our results can also be applied to identifying X-ray binaries. However, distinguishing X-ray binaries from AGN based on their spectra alone would be more challenging, and we would need to use the information from location of the source (X-ray binaries are primarily found along the Galactic plane and globular clusters), X-ray variability and multiwavelength observations to properly differentiate X-ray binaries and AGN. The presence of a fluorescent Fe line will also be important for distinguishing AGN at significant redshift from XRBs. Our results show that ANNs can detect the Fe-line irrespective of the redshift.  

Another application of this method would be in identifying millisecond pulsars (MSPs) in highly absorbed regions like the Galactic center and bulge (where soft X-ray sources are obscured). MSPs have  hard power-law spectra ($\Gamma \sim 1$) from the intrabinary shock between the NS and its companion, or magnetospheric emission. Identifying the Fe-K line would allow us to distinguish MSPs from other accreting hard X-ray sources like cataclysmic variables. 

The X-ray spectra of SNRs can also be modeled by thermal plasma components with high metal abundance. Young SNRs (where the reverse shock has not reached the core) usually have a cooler component with $kT < 2 keV$, and strong Mg and Si lines. Thus ANN models that can pick out these lines will be able to differentiate the young SNRs in distant galaxies from low-luminosity AGN (see for instance \citealt{Hebbar_2019}). 

\subsection{Current and Future missions}
In this work, we have focused on {\it Chandra} observations with similar response functions. Our results in \S\ref{sec:changing_response} indicate that we need a more detailed modeling of the responses during our training before we can combine the spectra from multiple {\it Chandra} observations to increase the total number of counts. This will enable us to classify the faint sources more accurately.

{\it XMM-Newton} has $\approx$ 5--10 times the effective area of {\it Chandra} at 1 keV (depending on the year), but also has a higher background for point-source extraction. Thus for individual observations we would only expect similar or a slight improvement in our performance. However, {\it XMM-Newton}'s soft energy response is more stable than {\it Chandra}'s, implying that spectra from multiple observations of the faint sources can be readily combined to increase the sensitivity in detecting emission lines. The {\it eRosita} probe also has similar characteristics as {\it XMM-Newton} but has observed a wider portion of the sky, and we can expect similar results.

Our results also point to the characteristics of future missions that could be ideal for our work. The Resolve instrument onboard {\it XRISM}, set to be launched later this year, will have an excellent spectral resolution allowing  detection of  abundant emission lines in the spectra of X-ray sources. However, its effective area in the soft energies is lower than {\it Chandra}, and it has a higher background (for point source spectroscopy). Thus, studying faint serendipitous sources will be difficult with XRISM. Proposed X-ray missions like the AXIS-probe with $\sim$10 times the effective area of {\it Chandra} and excellent angular resolution throughout the field will allow for the detection of higher counts for the faint sources, and allow classifying them accurately. The higher effective area will also lead to the detection of more serendipitous sources making machine learning methods much more efficient than traditional methods. X-ray missions like the Line Emission Mapper and Athena X-IFU would be the most ideal for our work. Their large effective area in the soft-energy X-rays ($< 2$ keV) and spectral resolution of a few electron-volts will allow for the detection of individual emission lines from elements with different ionization states, even in faint sources.

\section{Conclusion}
\label{sec:conclusion}

We discussed the application of a neural network model in distinguishing {\it Chandra} X-ray spectra of AGN in CDFS and stars in the Orion nebula cluster. We are able to achieve accuracy, recall and precision of $\sim$92\% on the simulated spectra and $\sim$91\% on the observed spectra. The algorithm is most efficient when the net counts in the 0.3--8.0 keV regime are $\gtrsim$ 200, background contribution is $\lesssim 5\%$, the stars have absorption column densities $N_H < 10^{22}$ cm$^{-2}$, $kT \lesssim 2$ keV, and the AGN are hard with power-law index $\Gamma \leq 2$ and have a strong Fe-K emission line with equivalent width $\gtrsim$ 0.5 keV.

We also tested the robustness of our method with the changing soft energy response of {\it Chandra} ACIS, and found that the performance of our model, trained on the COUP and CDFS responses, decreased on simulated spectra from responses in the Cycle 25 {\it Chandra} Call for proposals, especially in identifying AGN. Thus, combining multiple observations of faint sources would need more detailed modelling of the {\it Chandra} responses. Applying these algorithms to {\it XMM-Newton} observations will allow us to utilize its larger effective area and near-constant response to increase the signal-to-noise ratio for faint sources.

In this paper, we have only used the X-ray spectra for classifying the source. For sources with known optical/radio counterparts, we can use their position, variability, and multiwavelength properties in addition to the output of our machine learning methods (in place of commonly-used hardness ratios) to improve the accuracy of the classification.

X-ray catalogs have a unique property that while hundreds and thousands of sources have available spectra, only a few of them are labelled. Such  datasets allow the use of semi-supervised learning such as auto-encoders. These algorithms use clustering methods (i.e., use the distribution of features in the unlabelled data), representation of the data into smaller dimensions, etc., along with the information from the classified/labelled sources to efficiently train the ML model. Using these kinds of training algorithms would allow us to study the data without the need to simulate artificial spectra.

We plan 
to extend this method to distinguish and identify other kinds of X-ray spectra, like those of neutron stars and SNRs, from stars and AGN. We would also like to check the performance of our model on sources that have slightly different properties than the ones used in training. These methods will be especially useful with future higher-spectral-resolution X-ray data from upcoming microcalorimeter missions, such as 
Athena.

\section*{Acknowledgements}

The authors thank Dr. Abram Hindle for very constructive suggestions. 
COH is supported by NSERC
Discovery Grant RGPIN-2016-04602. This work has made use of data obtained
from the Chandra Data Archive and the Chandra Source Catalog, version 2.0.

\section*{Data Availability}

The COUP source list and the membership catalogues used to extract the properties of the stars are available at the VizieR online databases \citet{Getman_cat, Getmanb_cat}. The catalogue of AGN in CDFS can also be found at the VizieR online catalogue \citet{Tozzi_2006_cat}.  
The {\it Chandra} data themselves are available via the {\it Chandra} Source Catalog, \url{https://cxc.cfa.harvard.edu/csc/}, and/or the {\it Chandra} Data Archive \url{https://cxc.cfa.harvard.edu/cda/}.
The spectra of simulated stars and their properties are available upon reasonable request to the first author.

\bibliography{references}{}
\bibliographystyle{aasjournal}



\appendix

\section{Typical spectra of simulated stars and AGN}

\begin{figure}
    \centering
    \includegraphics[width=0.5\columnwidth]{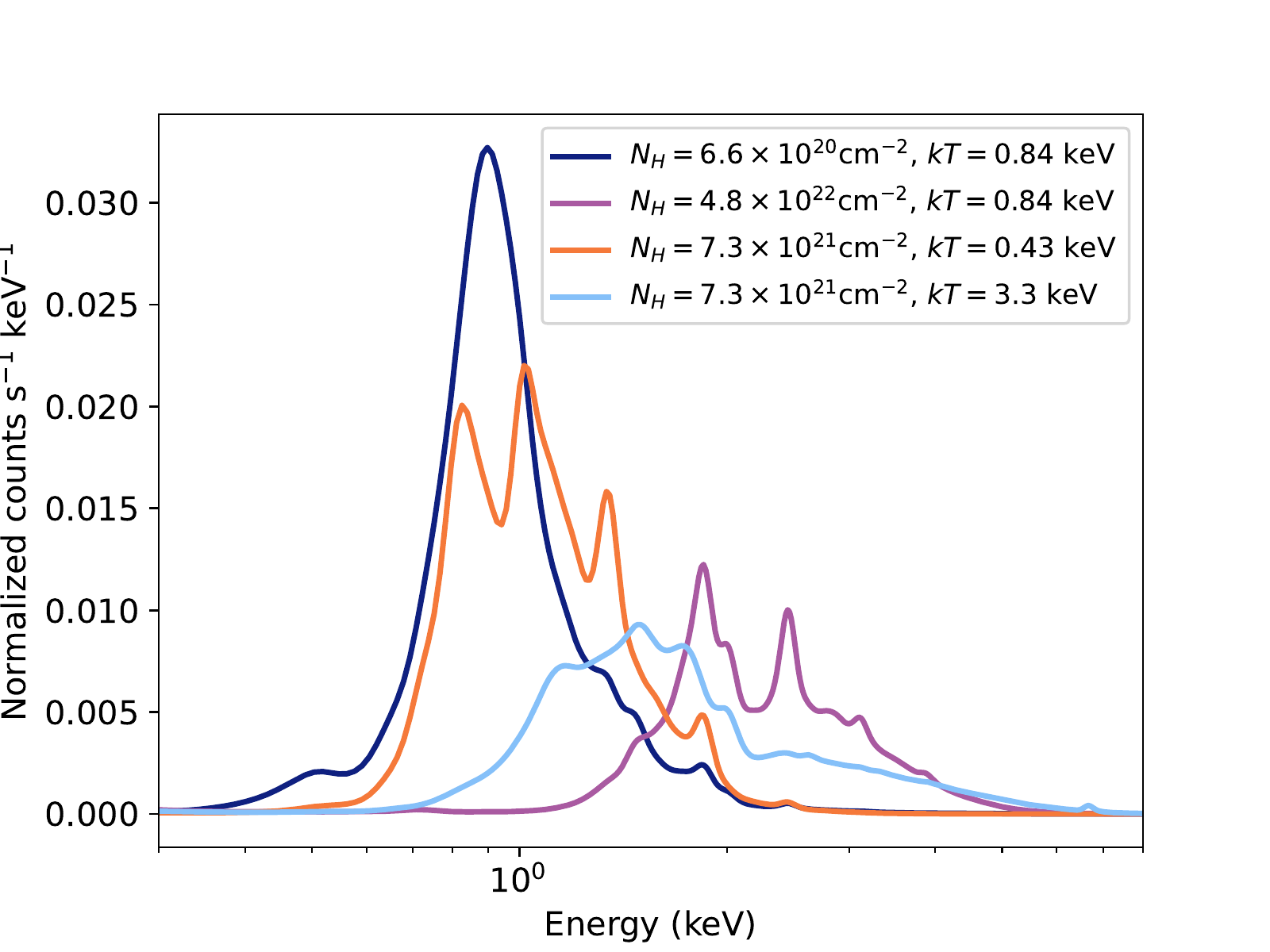}
    \caption{Normalized model spectra of stars for 90-percentile intervals in $N_H$ and $kT$. The spectra have been normalized such that the sum of counts in all energy channels is one. These spectra also do not account for the Poisson error due to low counts. For low $N_H$, we see a dominant Fe-L/Ne-K emission line, which disappears for higher $N_H$ (we only see Si and S lines for $N_H = 4.8 \times 10^{22}$). Similarly at lower $kT$, we clearly see Fe-L, Ne-K, Mg-K and Si-K lines, but these aren't dominant at higher temperatures.}
    \label{fig:star_typ_spectra}
\end{figure}

We show the spectra of stars simulated by us with respect to the variation in the absorption column density, $N_H$, and plasma temperature, $kT$ in Fig~\ref{fig:star_typ_spectra}. The values of $N_H$ and $kT$ shown correspond to the 90-percentile intervals. We use the model \texttt{tbabs*apec} to generate the spectra. We notice that while the Fe-L/Ne-K lines are clearly detected for stars with low-$N_H$, they are absorbed in the stars with very high $N_H$, leading to more pronounced Si-K and S-K lines (this is because we normalize the spectra by dividing each channel by the net counts). Similarly, as $kT$ increases the spectra become less line-dominated.

\begin{figure}
    \centering
    \includegraphics[width=0.5\columnwidth]{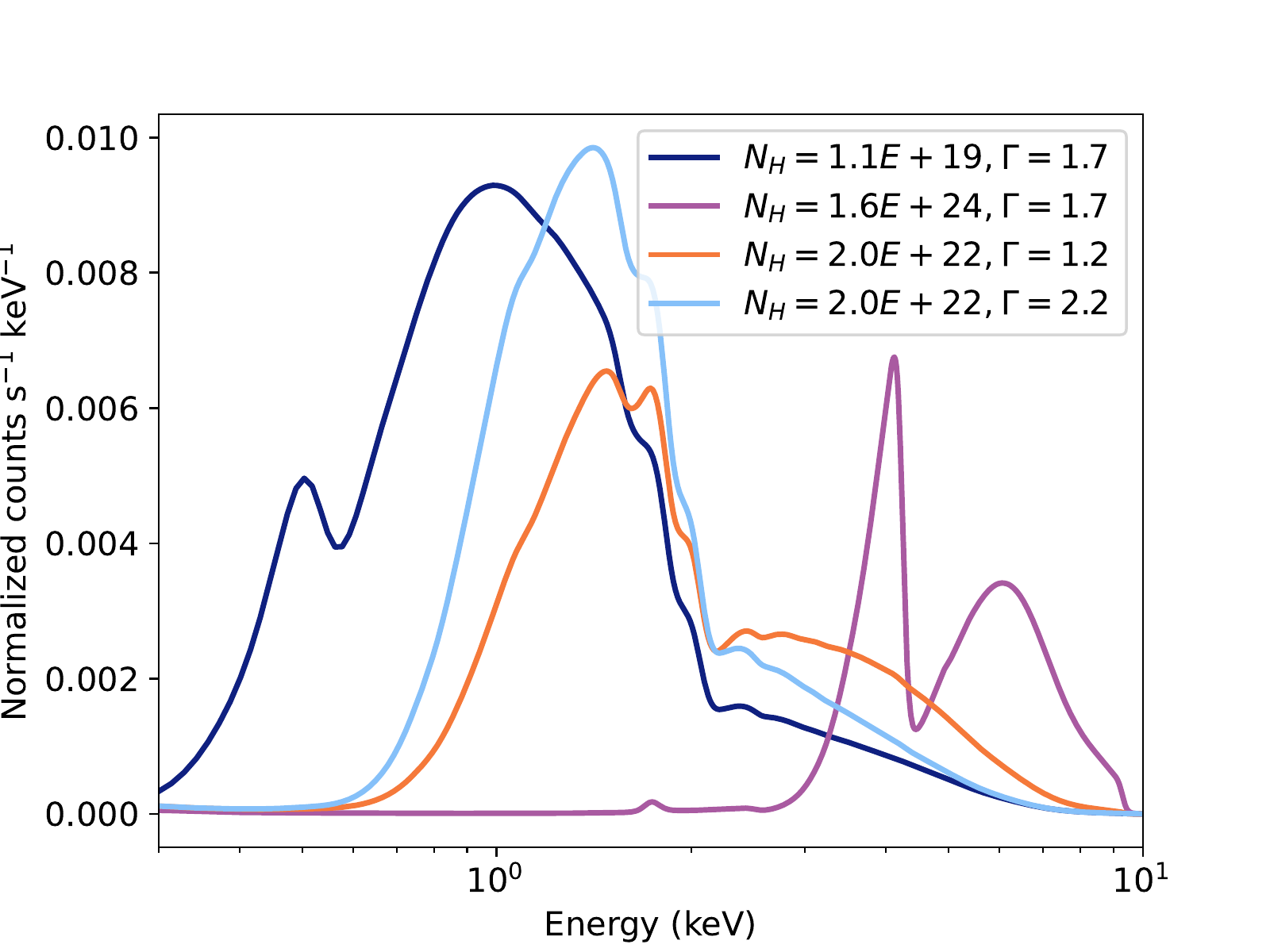}
    \caption{Normalized spectra of AGN for 90-percentile intervals in $N_H$, and $\Gamma$. We fix the redshift to 0.7 (most probable value in the distribution of $z$). The features in the spectra are mainly due to the effective area of {\it Chandra} ACIS and absorption edges. In this figure, we only show AGN of type \texttt{tbabs*ztbabs*pegpwrlw}. Some AGN also have an additional soft component and an Fe-K line. 4\% of the AGN are better represented by a \texttt{pexrav} model.}
    \label{fig:agn_typ_spectra}
\end{figure}

We also show the spectra of our simulated AGN with changing $N_H$ and $\Gamma$ in Fig~\ref{fig:agn_typ_spectra}. The values of $N_H$ and $\Gamma$ correspond to the 5th-percentile and 95th-percentile values. We use the model \texttt{tbabs*ztbabs*pegpwrlw} to generate the spectra. The spectra of AGN remain continuum dominated throughout the range of $N_H$ and $\Gamma$. Most of the features that we see in the AGN spectra are due to a  combination of the absorption edges and effective area of the Chandra ACIS detector.

\section{TensorFlow model of the ANN}
We use Keras and TensorFlow modules in python to model the ANN.
\begin{lstlisting}[language=Python]
ann_model = tf.keras.Sequential([
    layers.Dense(
    8, activation=`relu',
    kernel_regularizer=regularizers.l1(
        0.001)),
    layers.Dense(2, activation='softmax')])
# Using 2 output nodes allows changing the 
# threshold for the two classes separately
# and extend this to multi-class
# classification, if needed.

loss_fn = tf.keras.losses.BinaryCrossentropy(
    from_logits=False)  
# We notice setting `from_logits=False' 
# more compatible with the calculation of
# keras metrics and does not affect our
# performance.

ann_model.compile(optimizer=`adam',
                  loss=loss_fn)
\end{lstlisting}

\section{Choosing number of nodes, L1 value, and the size of training set}

\begin{figure}
    \centering
    \includegraphics[width=0.45\columnwidth]{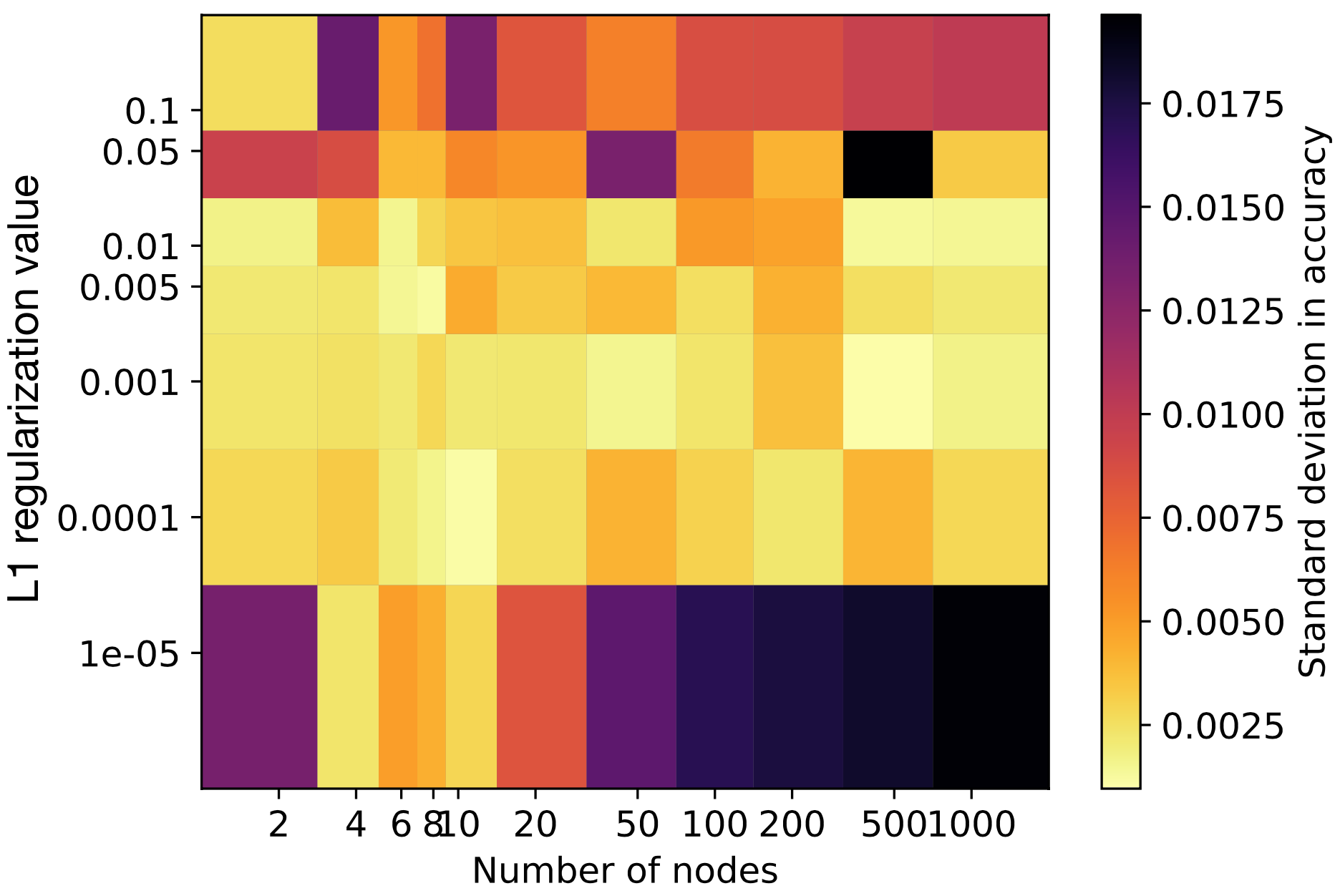}
    \includegraphics[width=0.45\columnwidth]{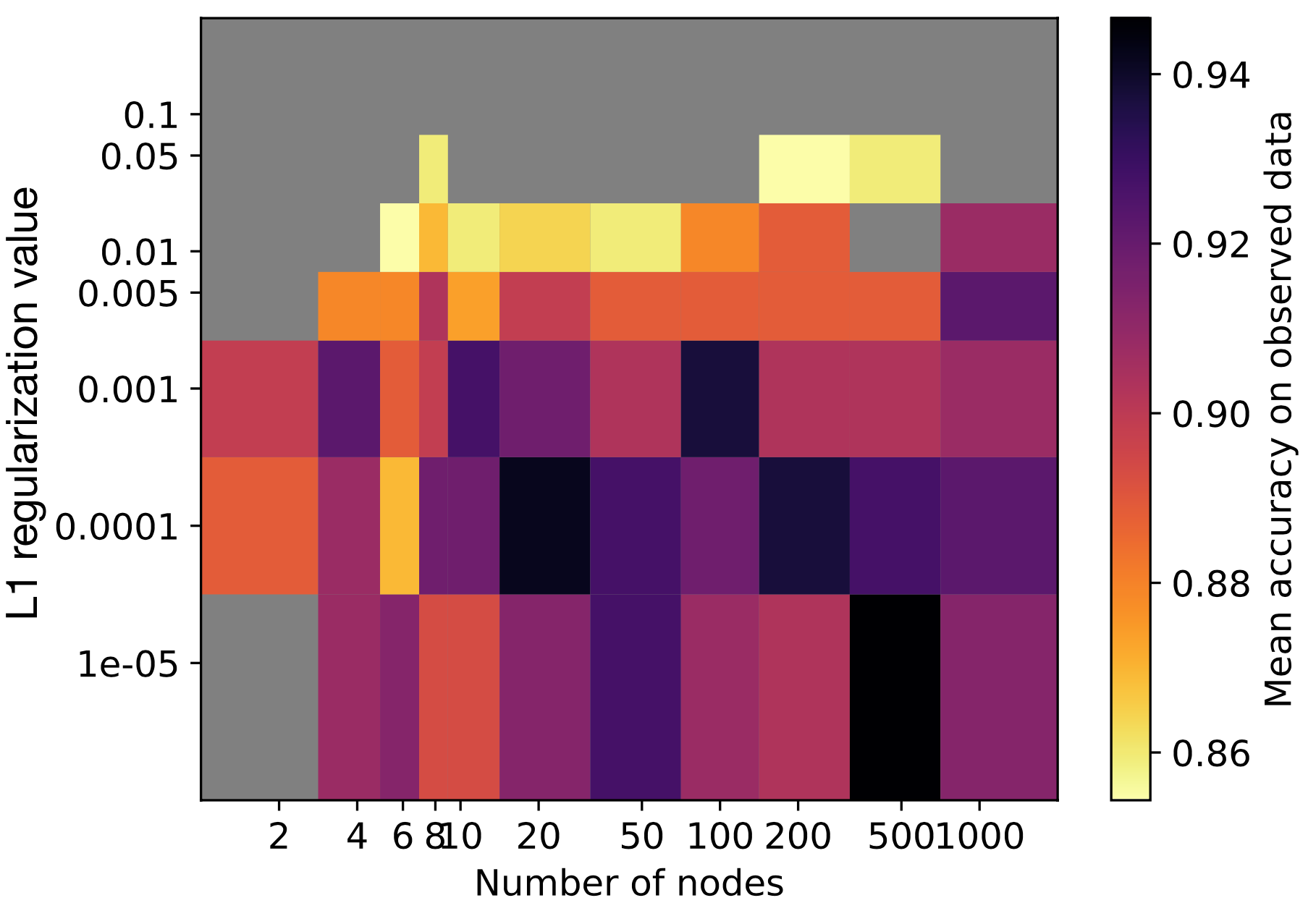}
    \caption{({\it Left}) Mean accuracy on the observed spectra and the ({\it Right}) standard deviation in accuracy on the simulated spectra for different numbers of nodes and the L1 regularization values, $\lambda$s. We notice that while the accuracy seems to increase with the number of nodes and decreasing $\lambda$, the standard deviation also increases, making these models less robust when the data has low signal-to-noise ratios.}
    \label{fig:hyperparametrize_accuracy}
\end{figure}

\begin{figure}
    \centering
    \includegraphics[width=0.5\columnwidth]{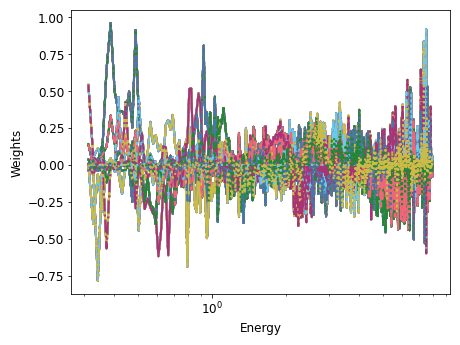}
    \caption{Best fit weights for the ANN model with 20 nodes in the hidden layer and L1 regularization value, $\lambda=0.0001$. We see that these weights are more variable than those in Fig.~\ref{fig:reducedbg_weights}. }
    \label{fig:1e4_20nodes_weights}
\end{figure}

We also hyper-parameterize our ANN model for the number of nodes and the value of the L1 regularization. For this purpose, we train and test the ANN model on the simulated data with reduced background, and also test the performance on the observed spectra (thus ensuring that our model is robust for higher noise and background). We use a 5-fold cross-validation in our training process. The standard deviation in the accuracy is calculated using the values of accuracy in the cross-validation process. In the top plot of Fig.~\ref{fig:hyperparametrize_accuracy} 
we show the accuracy of the ANN algorithm  on the observed spectra for a range of L1 values and numbers of nodes. We also show the standard deviation in the accuracy values during the cross-validation in the bottom plot of Fig.~\ref{fig:hyperparametrize_accuracy}. We see that the accuracy of classification changes only slightly, given the standard deviations, for any number of nodes greater than 10 and any L1 value over $10^{-3}$. However, from Fig.~\ref{fig:1e4_20nodes_weights}, we notice that for an L1 value of $\lambda \lesssim 10^{-4}$, the distribution of weights becomes highly variable, i.e. while the higher values of weights still correspond to emission lines, they seem to pick out lines smaller than the resolving power of the CCD detectors. This can happen as a result of over-fitting, and might not yield good results for data with higher noise. Therefore, we choose $\lambda = 0.001$ for our model. For this value of $\lambda$, the accuracy does not change for number of nodes $\gtrsim 10$. Therefore, we choose 10 nodes and $\lambda = 0.001$ for our ANN model.

\begin{figure}
    \centering
    \includegraphics[width=0.5\textwidth]{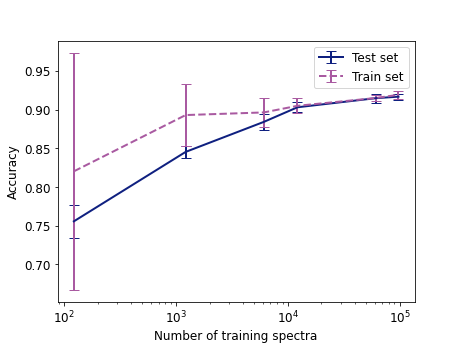}
    \caption{Change in classification accuracy with the number of spectra used for training. The accuracy seems to saturate at $\sim 90\%$ at $\sim 10^{5}$ training points.}
    \label{fig:acc_vs_train_no}
\end{figure}

Fig.~\ref{fig:acc_vs_train_no} plots our the performance with the size of the training data. For small number of data sets, the model over-fits the data during training and we get poor performance on the test set. The accuracy seems to saturate at $\sim 90$\% (within error-bars) for $10^5$ data points in the training spectra.

\end{document}